\documentclass[fleqn,usenatbib]{mnras}

\usepackage{newtxtext,newtxmath}

\usepackage[T1]{fontenc}

\DeclareRobustCommand{\VAN}[3]{#2}
\let\VANthebibliography\thebibliography
\def\thebibliography{\DeclareRobustCommand{\VAN}[3]{##3}\VANthebibliography}

\usepackage{graphicx}	
\usepackage{amsmath}	
\usepackage{subcaption}
\usepackage{caption}
\usepackage[dvipsnames]{xcolor}
\usepackage{tikz}       
\usepackage{pagecolor}
\usepackage{multirow}
\usepackage{siunitx}
\usepackage{pdflscape}

\newcommand{\revb}{}
\newcommand{\revbb}{}

\title[Diverse morphologies of FRII-lows]{The diverse morphologies and evolution of low-luminosity edge-brightened radio galaxies}

\author[B. Barkus et al.]{
B. Barkus,$^{1,2}$\thanks{E-mail: bonny.barkus@outlook.com}
J. H. Croston,$^{2}$
B. Mingo,$^{1,2}$
M. J. Hardcastle,$^{1}$
G. G\"urkan,$^{1,4}$
V. H. Mahatma$^{3,5}$
\\
$^{1}$Centre for Astrophysics Research, University of Hertfordshire, College Lane, Hatfield, AL10 9AB, UK\\
$^{2}$School of Physical Sciences, The Open University, Walton Hall, Milton Keynes, MK7 6AA, UK\\
$^{3}$Cavendish Laboratory, University of Cambridge, 19 JJ Thomson Avenue, Cambridge CB3 0HE, UK \\
$^{4}$CSIRO Space and Astronomy, ATNF, PO Box 1130, Bentley, WA 6102, Australia \\
$^{5}$Kavli Institute for Cosmology, University of Cambridge, Madingley Road, Cambridge CB3 0HA, UK\\
}

\date{Accepted XXX. Received YYY; in original form ZZZ}

\pubyear{\the\year{}}

\begin{document}
\label{firstpage}
\pagerange{\pageref{firstpage}--\pageref{lastpage}}
\maketitle

\begin{abstract}

Fanaroff-Riley class I (FRI) radio galaxies show centre-brightened emission from disrupted lower power jets, while traditionally more luminous class II (FRIIs), are edge-brightened, with relativistic jets terminating in hotspots. Population studies of radio-loud AGN (RLAGN) with low frequency, deep, wide-field surveys have revealed FRII-like radio structures at lower luminosities. We present the first high-resolution morphological investigation of a representative LOFAR-selected sample of low-luminosity FRIIs, to determine whether this new population is physically distinct from traditional high-luminosity FRIIs. Using new $1.5$-GHz Jansky Very Large Array (VLA) observations for a sample of 19 low-luminosity FRIIs, from the LOFAR Two Metre Sky Survey Data Release 1 (LoTSS DR1), with luminosities up to three orders of magnitude lower than the typical FR break ($L_{150} = 10^{26}$ W Hz$^{-1}$). We examine the compact features and perform spectral index analysis to identify hotspots, cores and signatures of restarting or remnant activity. We find a higher prevalence of cores and a comparable number of hotspots in the low-luminosity FRII sample compared to a randomly-selected sample of luminous ($L_{150}>10^{26}$ W Hz$^{-1}$) FRIIs selected from the same parent LOFAR sample. Approximately 32 per cent of low-luminosity FRIIs show restarting or remnant behaviour, while $\sim 32$ per cent are active FRIIs with compact hotspots. Our results show that FRII source dynamics occur at low radio luminosities, but reinforce earlier conclusions that the population of low-luminosity edge-brightened RLAGN is highly diverse. Binary morphological classifications should be used cautiously as a first step towards more nuanced investigations of the complexity of jet life cycles and evolution.


\end{abstract}




\section{Introduction}
\label{Sec:Intro}


The \revb{kpc}-scale radio structures (morphologies) of radio-loud AGN (RLAGN) reflect the interaction between powerful jets launched from close to the central black hole of their host galaxy and their environment on a range of different physical scales. Traditionally, RLAGN have been classified, using radio maps that resolve their components, by morphology into two classes, Fanaroff-Riley \citep[FR;][]{fanaroff_morphology_1974} class I and II. FRIs have a centre-brightened radio appearance, in comparison with FRIIs, which are edge brightened and \revb{typically} terminate in hotspots (internal shock) \citep{meisenheimer_synchrotron_1989, bicknell_relativistic_1995, laing_dynamical_2002, tchekhovskoy_three-dimensional_2016}.  It is thought that this structural difference is caused by the interplay between jet kinetic power and environment density, with jets of similar power disrupting more easily in richer environments. FRI jets are observed to decelerate from relativistic speeds on kpc scales, whereas FRII jets remain relativistic throughout \citep{bicknell_relativistic_1995, kaiser_luminosity_2007}. \citet{ledlow_20_1996} provided observational evidence for this theory by showing that the FR luminosity break is dependent on host-galaxy magnitude.  They found that FRIs were likely to have higher radio luminosities in brighter host galaxies, where the interstellar medium is assumed to be denser.  However the comparatively small, flux \revb{density} limited samples, leading to selection effects, such as different redshift distributions and environments of the FRIs and FRIIs, has led to debate over whether this is true for the whole population \citep{best_radio_2009, lin_populations_2010, wing_galaxy_2011, singal_fanaroff-riley_2014, capetti_astrophysics_2017, shabala_role_2018, clews_radio-loud_2025}.  There has also been considerable debate about a relationship between accretion mode and jet morphology \citep{hardcastle_chandra_2007, hardcastle_active_2009, best_fundamental_2012, gendre_relation_2013, mingo_x-ray_2014, mingo_accretion_2022, ineson_link_2015, tadhunter_radio_2016, hardcastle_simulation-based_2018,chilufya_nature_2025-1}.

Until recently, most of our understanding of morphological comparisons came from surveys carried out with a high flux density limit, such as the 3CRR catalogue \citep{mackay_observations_1971, laing_bright_1983}, which were subsequently followed up with observations from other telescopes, such as the Karl G. Jansky Very Large Array (VLA).  A more comprehensive view of RLAGN is being built up from recent deep, wide-area surveys \citep{norris_data_2011, jarvis_meerkat_2016, shimwell_lofar_2019, hurley-walker_galactic_2017} by exploring the distant, faint, and low surface-brightness radio population.  These surveys are reducing the bias of earlier studies and have demonstrated a substantial overlap in luminosity between those sources with FRI and FRII morphology rather than a clear luminosity break \citep{best_radio_2009, miraghaei_morphological_2017, mingo_revisiting_2019}.  The new surveys are also revealing a more complex population of extended sources at low surface brightness and expanding our knowledge of \revb{remnant} sources, restarting sources and sources showing highly complex environmental interaction \citep[e.g.][]{brienza_lofar_2016, brienza_search_2017, brienza_non-thermal_2025, kapinska_radio_2017, mahatma_remnant_2018, mahatma_lotss_2019, mingo_revisiting_2019}. These results demonstrate that binary classification of RLAGN into FRI and FRII sources is an oversimplication that can neglect the underlying diversity of jet behaviour. 

Using the LOw Frequency ARray (LOFAR) Two Metre Sky Survey Data Release 1 (LoTSS DR1) at 150-MHz \citep{shimwell_lofar_2019} \citet{mingo_revisiting_2019} identified a population of edge-brightened (FRII-like) RLAGN with luminosities up to 3 orders of magnitude lower than the traditional FR break ($L_{150} = 10^{26}$ W Hz$^{-1}$).  \revbb{They concluded that these ‘FRII-lows’ are likely to include older luminous FRIIs fading from their peak (including young remnants with edge-brightened structure) and FRIIs hosted by lower mass galaxies (with active supersonic jets and terminal hotspots) at much lower radio luminosity than the traditional FRI/FRII break.}
A population of low-luminosity FRIIs raises many questions.  For example, if there exists a population of low-luminosity AGN with FRII source dynamics, then the models discussed above predict that they must have differing host/environment properties to FRIs at similar luminosities (e.g. lower ambient gas density).  This also raises questions as to whether low-luminosity FRIIs have similar or different triggers and distributions to the general low-luminosity population \citep{tadhunter_radio_2016, hardcastle_simulation-based_2018}. If this population of low-luminosity sources has true FRII dynamics, then we might expect the\revb{ir lobes to be less contaminated} by entrainment due to jet disruption\revb{ than FRIs of the same power}, which would have \revb{important} implications for jet power inference and environmental impact \citep{croston_particle_2018}.

If low-luminosity FRIIs do not have the same dynamics as high-luminosity FRIIs, then attempts to apply FR classifications to deep, low- to medium-resolution wide-area surveys \citep[e.g.][]{alhassan_first_2018} are likely to be misleading and lead to categories that are not well-linked to underlying physical behaviour.  Therefore, in order to accurately infer relationships between black-hole and galaxy evolution using large samples of RLAGN from current and upcoming surveys, an understanding of the structure and dynamics of low-luminosity FRIIs needs to be developed.

In this paper, we report on a high angular resolution follow-up programme to search for the presence of compact features, such as cores, jets, and \revb{hotspots} in a sample of LOFAR-selected low-luminosity FRIIs. We use the LoTSS DR1 images, combined with new, high resolution VLA images of a subsample of \citet{mingo_revisiting_2019}'s low-luminosity FRIIs, to determine the fraction of sources in this sample with compact features. We combine this information with an investigation into their spectral properties to explore the source evolution and underlying diversity of this population.

This paper is structured as follows: \autoref{Sec:Data} describes our new VLA datasets and summarizes the data reduction and imaging, with \autoref{sec:Images} then containing the details of the final images. \autoref{sec:SDyn} discusses the source dynamics, including the definition of compact features and their prevalence in our sample. \autoref{sec:SpI} discusses the spectral properties of each source.  Finally,  \autoref{sec:RandR} presents a discussion and interpretation of what the observational results tell us about the dynamics and evolution of the population. For this paper we have used a concordance cosmology with $H_0$ = 70 km s$^{-1}$ Mpc$^{-1}$, $\Omega_m$ = 0.3, and $\Omega_\Lambda$ = 0.7.

\section{Sample selection and data reduction}
\label{Sec:Data}

For this work we have selected a sample of low-luminosity FRIIs from \citet{mingo_revisiting_2019}, which we have followed up with the VLA.  The sample was selected from the preliminary images of the LoTSS DR1 dataset \citep{shimwell_lofar_2019, williams_lofar_2019}.  LoTSS is a 120 – 168-MHz continuum survey, and DR1 covered 424 deg$^2$ of the northern hemisphere, in the HETDEX field.  LoTSS has a resolution of 6 arcsec (1 pixel $=$ 1.5 arcsec), and 73 per cent of the 318,520 sources have optical host identifications from Pan-STARRS/WISE as reported by \citet{williams_lofar_2019}.  Morphological classification for LoTSS DR1 was carried  out using LoMorph\footnote{\url{https://github.com/bmingo/LoMorph/}} \citep{mingo_revisiting_2019}; which has a $>95$ per cent reliability for FRII classifications (compared to visual inspection) and is described in detail by \citet{mingo_revisiting_2019}.  To select a sample for VLA follow-up, we considered a parent sample with $S_{150} > 10$ mJy and angular size > 50 arcsec. We then selected all those classed as FRIIs where $L_{150} < 10^{25}$ W Hz$^{-1}$.  The luminosity cut was applied to ensure a selection of sources well below the traditional FR break.  This left a sample of 20 sources with luminosities ranging from $5 \times 10^{23}-10^{25}$ W Hz$^{-1}$. We note that the luminosities are based on distances obtained via spectroscopic redshifts for 75 per cent of the sample (including two of the sources with the lowest luminosities).  Photometric redshifts were used for the rest of the sample, and have uncertainties $\lesssim 0.1$ \citep{Duncan2019}.  This gives a sample which is representative of the full \citet{mingo_revisiting_2019} low-luminosity FRII population, including those which have $S_{150} < 10$ mJy but have luminosities and sizes in the same range as the targets.  The details of the nineteen sources observed with the VLA (one was not observed due to time constraints) can be found in \autoref{tab:samp_prop}.\revb{In this paper the full  source names will be given in the tables, figures and captions, but the shorter versions will be used throughout the text.}

\begin{table*}
    \centering
        \caption{The properties of the FRII-low sample. The columns are: \revb{LoTSS Source Name, }RA and DEC (in degrees), 150-MHz luminosity density, redshift, angular size, physical size, and host galaxy magnitude \revb{($K_s$)}. The luminosity, angular and linear size are those calculated by \citet{mingo_revisiting_2019} and the host galaxy $K_s$-magnitudes are those given by \citet{Duncan2019}. The final column highlights which images include observations with the C configuration.}
        \begin{tabular}{l S S S S S S S c}
        \hline
        \textbf{LoTSS Source Name} &  \textbf{RA} & \textbf{DEC} & \textbf{Luminosity} & \textbf{Redshift} & \textbf{Angular Size} & \textbf{Linear Size} & \textbf{Host Magnitude} & \textbf{C-Configuration} \\
         ~ & \revb{\textbf{(J2000)}} & \revb{\textbf{(J2000)}} &  \textbf{($\times 10^{24}$ W Hz$^{-1}$)} & ~ & \textbf{(arcsec)} & \textbf{(kpc)} & \textbf{($K_s$)} & \textbf{Observation} \\
        \hline \hline
        ILTJ105946.75+563136.4  &   164.945  &  56.527  &  8.4    &   0.181    &   61.8    &   189.7   &   -23.19  &   -   \\
        ILTJ112015.05+503254.9  &   170.063  &  50.549  &  4.4    &   0.281    &   76.7    &   329.6   &   -23.19  &   Yes \\
        ILTJ112654.44+540415.3  &   171.727  &  54.071  &  1.0    &   0.093    &   84.9    &   148.0   &   -23.12  &   -   \\
        ILTJ113626.52+501320.3  &   174.111  &  50.222  &  1.2    &   0.054    &   126.9   &   134.4   &   -22.13  &   Yes \\
        ILTJ114351.49+511712.6  &   175.965  &  51.287  &  5.1    &   0.157    &   128.9   &   353.1   &   -22.24  &   Yes \\
        ILTJ115011.27+534320.9  &   177.547  &  53.722  &  2.7    &   0.060    &   60.7    &   71.6    &   -22.23  &   -   \\
        ILTJ121623.58+524409.4  &   184.098  &  52.736  &  8.0    &   0.121    &   122.0   &   269.2   &   -23.74  &   Yes \\
        ILTJ130109.83+560623.4  &   195.291  &  56.107  &  9.7    &   0.269    &   110.4   &   459.9   &   -23.56  &   Yes \\
        ILTJ130605.63+555127.6  &   196.523  &  55.858  &  0.8    &   0.107    &   56.0    &   110.6   &   -23.36  &   -   \\
        ILTJ133217.44+484221.7  &   203.073  &  48.706  &  1.5    &   0.066    &   61.7    &   78.0    &   -20.77  &   Yes \\
        ILTJ133729.25+481822.3  &   204.372  &  48.306  &  5.8    &   0.119    &   67.2    &   145.5   &   -23.26  &   Yes \\
        ILTJ134315.98+553139.6  &   205.817  &  55.528  &  13.5   &   0.476    &   52.1    &   311.9   &   -24.29  &   -   \\
        ILTJ135152.95+521618.8  &   207.971  &  52.272  &  5.2    &   0.166    &   175.2   &   503.4   &   -23.33  &   Yes \\
        ILTJ135630.51+555245.1  &   209.127  &  55.879  &  12.2   &   0.360    &   107.1   &   544.8   &   -23.86  &   Yes \\
        ILTJ144644.12+492012.3  &   221.684  &  49.337  &  9.8    &   0.247    &   93.9    &   367.8   &   -23.74  &   Yes \\
        ILTJ144650.49+514625.3  &   221.710  &  51.774  &  12.0   &   0.240    &   102.9   &   394.9   &   -23.52  &   Yes \\
        ILTJ145759.29+490219.2  &   224.497  &  49.039  &  13.4   &   0.444    &   68.8    &   397.1   &   -23.72  &   Yes \\
        ILTJ145936.33+484219.8  &   224.901  &  48.706  &  12.0   &   0.206    &   118.2   &   403.6   &   -24.00  &   Yes \\
        ILTJ150827.77+541507.1  &   227.116  &  54.252  &  2.7    &   0.096    &   73.0    &   130.9   &   -22.84  &   -   \\
        \hline \hline 
        \textbf{Median}           &     ~      &   ~      &  5.8    &   0.17     &   84.9    &   311.9   &   -23.60   &    \\
        \hline
        \end{tabular}
    \label{tab:samp_prop} 
\end{table*}

The aim of the observing programme was to obtain high resolution VLA images to compare the morphological properties of the low-luminosity FRIIs with well-studied luminous FRIIs. The observations were designed to achieve sufficient sensitivity to image compact hotspots in all the FRII-lows, if present, and potentially jet features in some objects.  The detection (or non-detection) of \revb{hotspots} will therefore establish what fraction of our sample (and hence population) have true FRII source dynamics.

The VLA observations were taken in the L-Band (1 – 2 GHz) over 5 observing sessions (project ID 19A-151).  The A configuration sessions were on 2019 August 10th, and 2019 August 31st, the B configuration observations were taken on 2019 March 27th, and 2019 May 11th, and finally the C configuration observations were taken on 2019 May 19th.  Each session contains the flux density calibrator 3C 286, the same four secondary calibrators, and two scans on the 19 target sources (13 for the C configuration).  This gives a total of four or five \revb{sessions} on each target source. 

To carry out the data reduction the Common Astronomy Software Applications Version 5.7.0$-134$ \citep[CASA,][]{mcmullin_casa_2007} created by the National Radio Astronomy Observatory (NRAO) was used. The data reduction for the \revb{combined A and B configurations was carried out manually, and the addition of the C configuration was done using the standard CASA pipeline. Checks were carried out to ensure that the manual processing was comparable to the pipeline outcomes}. The flux density scale for the flux calibrator was determined using the \citet{perley_accurate_2017} 3C 286 model.  After calibration was completed \texttt{statwt} was used to find the weightings of each of the A, B, and C configuration data, based on their RMS, allowing them to be successfully combined for imaging. 

The A configuration provides a resolution of $\sim 1.3$ arcsec and was selected to highlight hotspots in the sources.  The B configuration, with a resolution of $\sim 4.3$ arcsec, complements this to sample the extended emission with a resolution more similar to LoTSS to enable spectral analysis. Finally the C configuration, with a resolution of 14 arcsec, was used to ensure \revb{optimal} sampling of the extended emission for the 13 sources that exceed 60 arcsec in size.

The final calibrated, combined datasets were imaged using \texttt{wsclean} \citep{offringa_optimized_2017}. The images were produced at a scale of 0.4 arcsec per pixel for a resolution of 1.3 arcsec using a multiscale, multi-frequency clean with a multiscale bias of 0.7, and a Briggs weighting \citep{Briggs1995} of $0.5$. The weighting of $0.5$ was chosen as it gave an improvement in the dynamic range of the images without loss of the hotspot detail. A second set of images were produced using a 6-arcsec Gaussian taper and a scale of 1.5 arcsec, to give a 6-arcsec resolution. These were generated to match the resolution of LoTSS for use in spectral comparisons.

\section{Sample Images}
\label{sec:Images}

Here we present and discuss the images of the FRII-lows, and define and discuss a comparison sample. 

\subsection{FRII-lows}
\label{subsec:FRII-lows}

The full set of FRII-low images are shown in \autoref{fig:LowImages} (figures made using AplPy: \citealt{robitaille_aplpy_2012,robitaille_aplpy_2019}) and their properties are given in \autoref{tab:image_prop}. When generating these images the pixel size was chosen to match the A configuration and to oversample the B and C configuration. This produced images at \revb{the full 1.3-arcsec resolution and allows us to observe the compact features such as hotspots and cores.} LoTSS 150-MHz images for the sample are shown for comparison. We note that the LoTSS images presented here were produced using the DR2 image processing \citep[][]{shimwell_lofar_2022}, which incorporated improvements relative to the DR1 images used for the original VLA sample selection. 

\begin{figure*}
    \centering
        \begin{minipage}{0.47\linewidth}
        \centering
            \includegraphics[width=\linewidth]{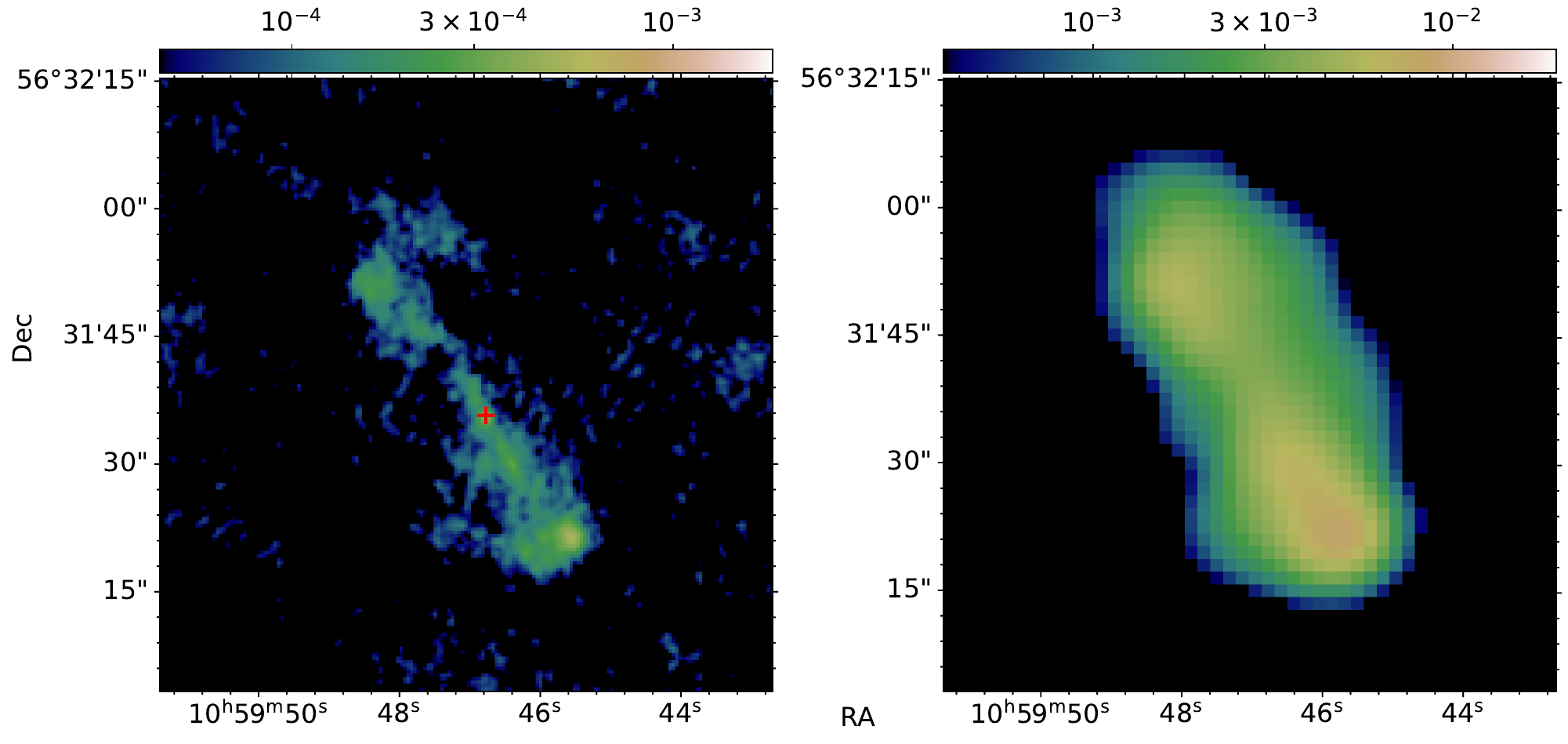}
        \subcaption{ILTJ105946.75+563136.4\label{fig:ErrExt}}
        \end{minipage}
        \hfill
        \begin{minipage}{0.47\linewidth}
        \centering
            \includegraphics[width=\linewidth]{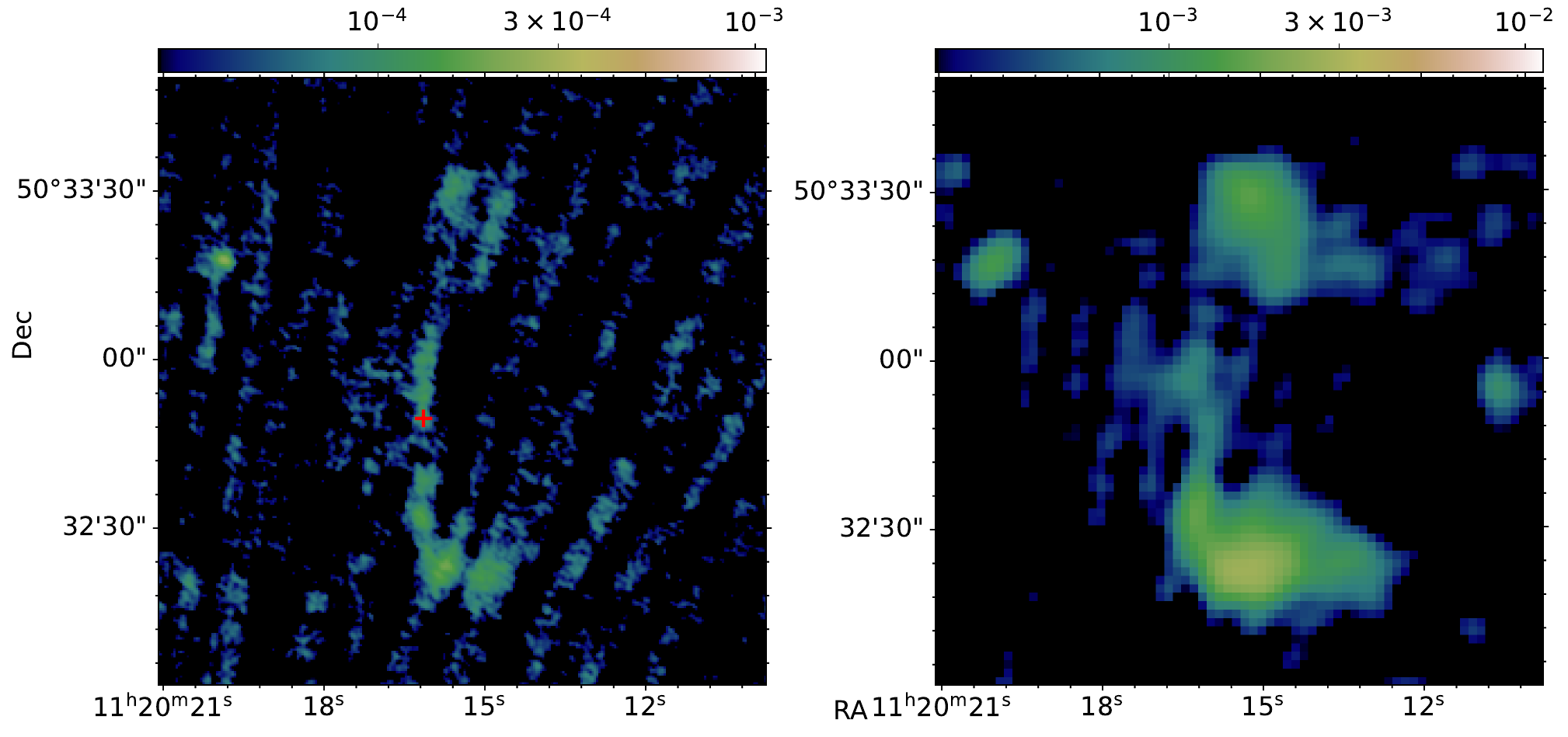}
        \subcaption{ILTJ112015.05+503254.9\label{fig:ErrExt}}
        \end{minipage}

    \vspace{0.07cm} 

    \centering
        \begin{minipage}{0.47\linewidth}
        \centering
            \includegraphics[width=\linewidth]{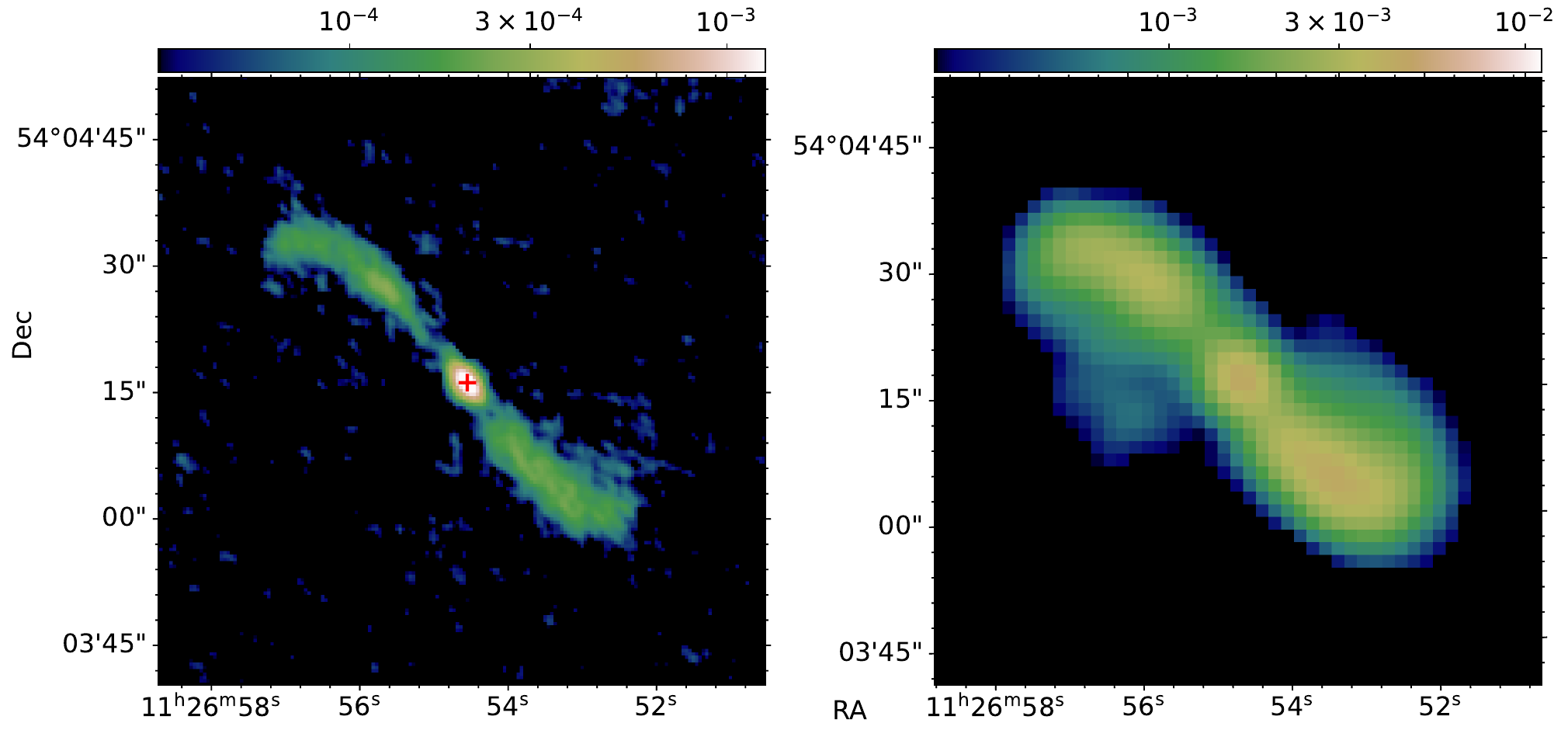}
        \subcaption{ILTJ112654.44+540415.3\label{fig:ErrExt}}
        \end{minipage}
        \hfill
        \begin{minipage}{0.47\linewidth}
        \centering
            \includegraphics[width=\linewidth]{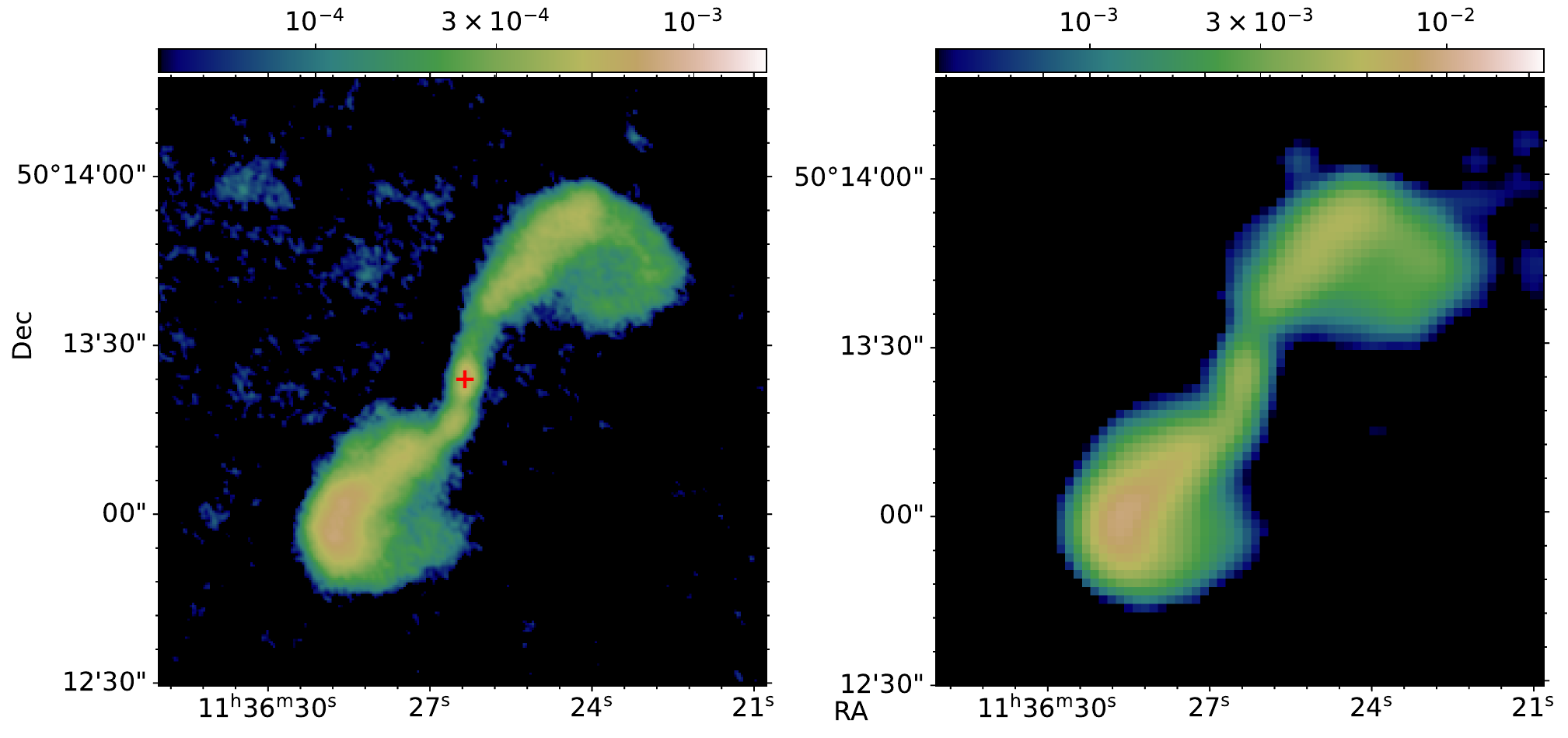}
        \subcaption{ILTJ113626.52+501320.3\label{fig:ErrExt}}
        \end{minipage}

    \vspace{0.07cm} 
    
    \centering
        \begin{minipage}{0.47\linewidth}
        \centering
            \includegraphics[width=\linewidth]{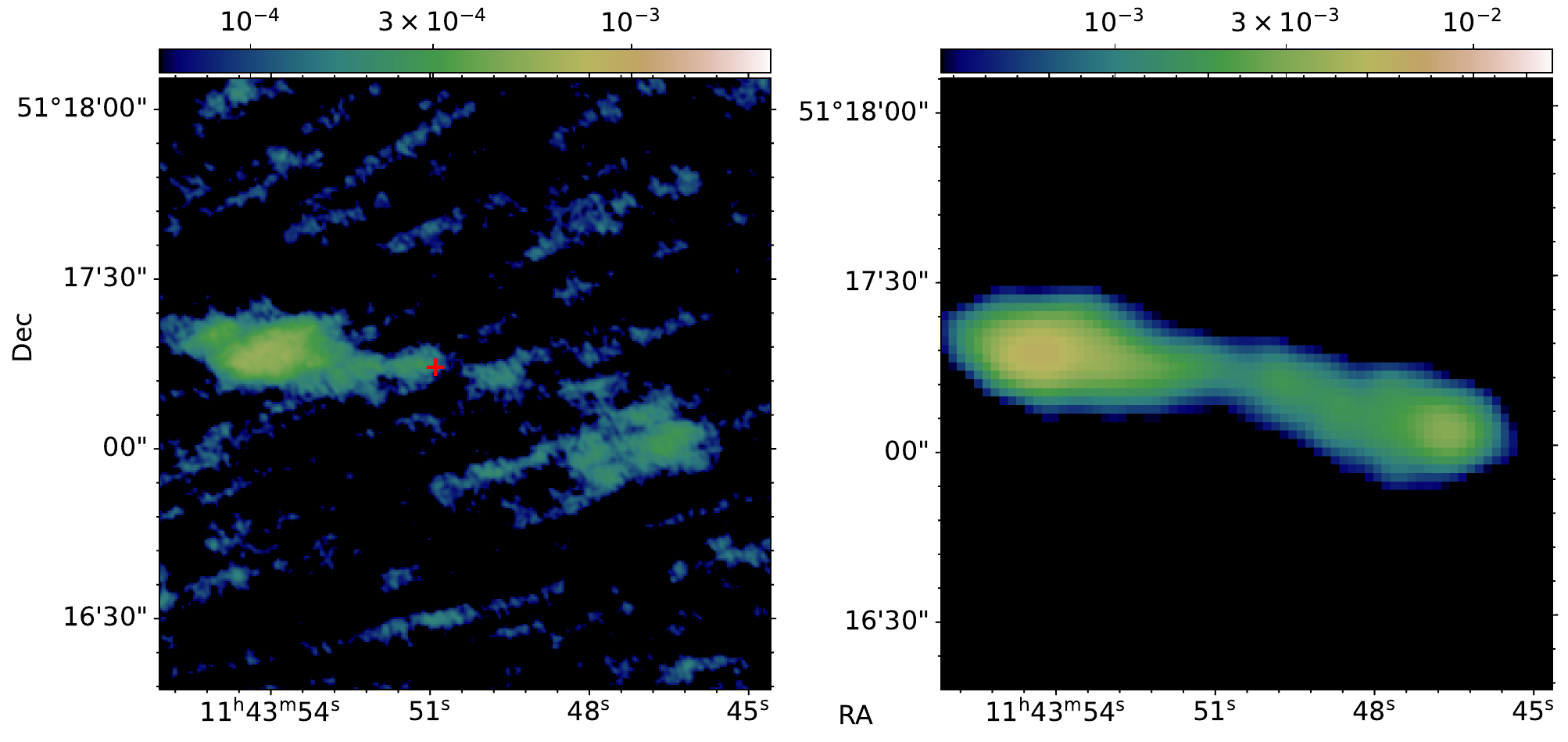}
        \subcaption{ILTJ114351.49+511712.6\label{fig:ErrExt}}
        \end{minipage}
        \hfill
        \begin{minipage}{0.47\linewidth}
        \centering
            \includegraphics[width=\linewidth]{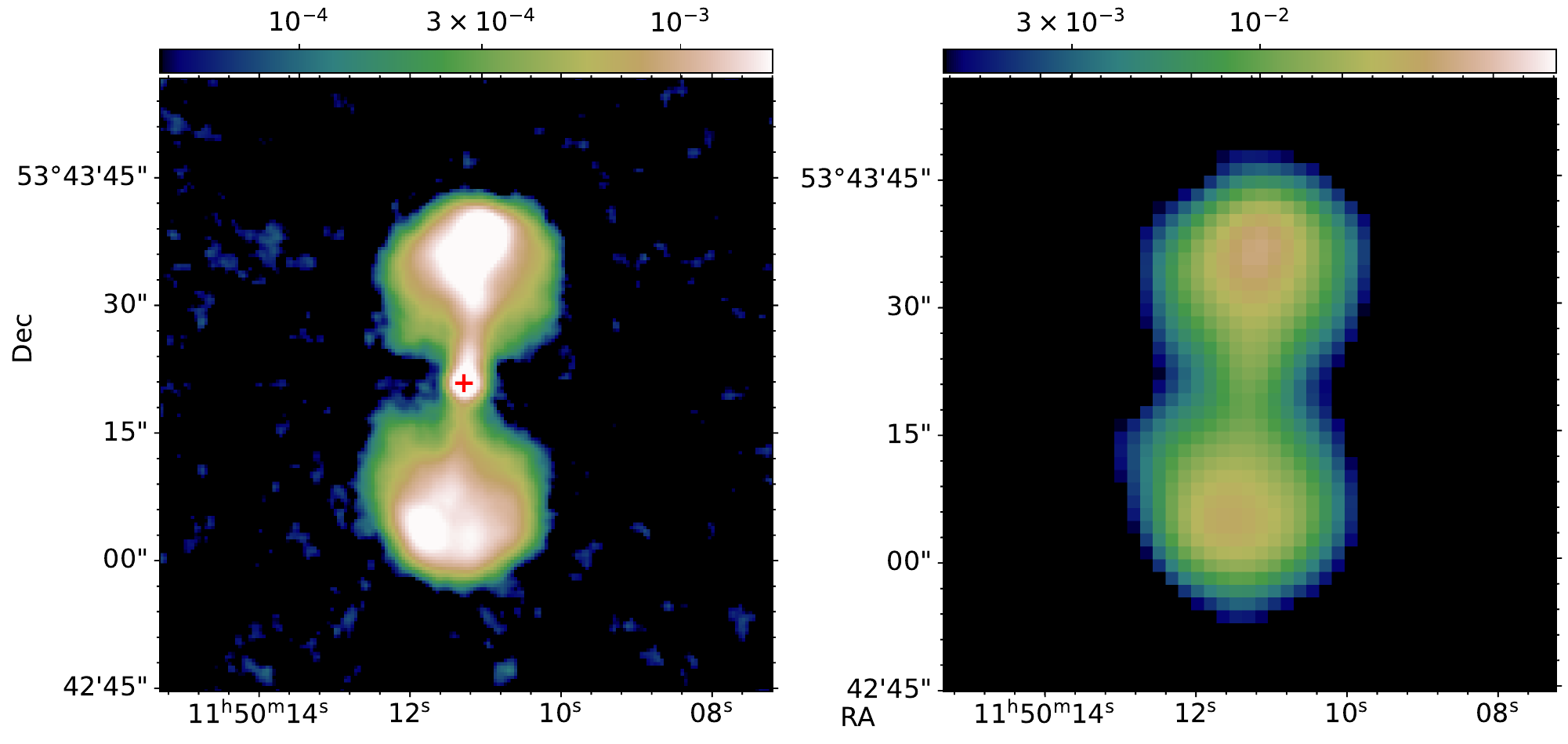}
        \subcaption{ILTJ115011.27+534320.9\label{fig:ErrExt}}
        \end{minipage}

    \vspace{0.07cm} 

    \centering
        \begin{minipage}{0.47\linewidth}
        \centering
            \includegraphics[width=\linewidth]{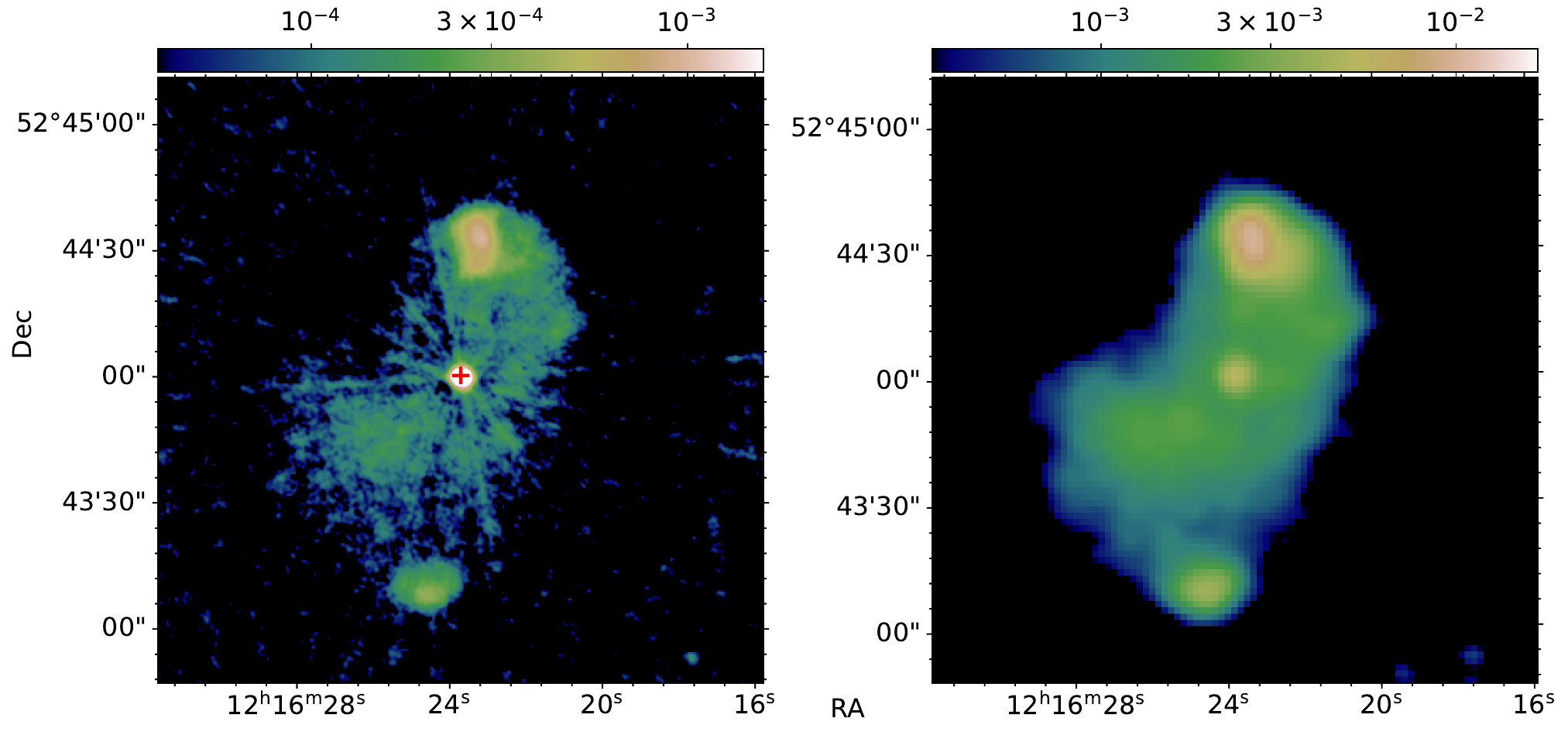}
        \subcaption{ILTJ121623.58+524409.4\label{fig:ErrExt}}
        \end{minipage}
        \hfill
        \begin{minipage}{0.47\linewidth}
        \centering
            \includegraphics[width=\linewidth]{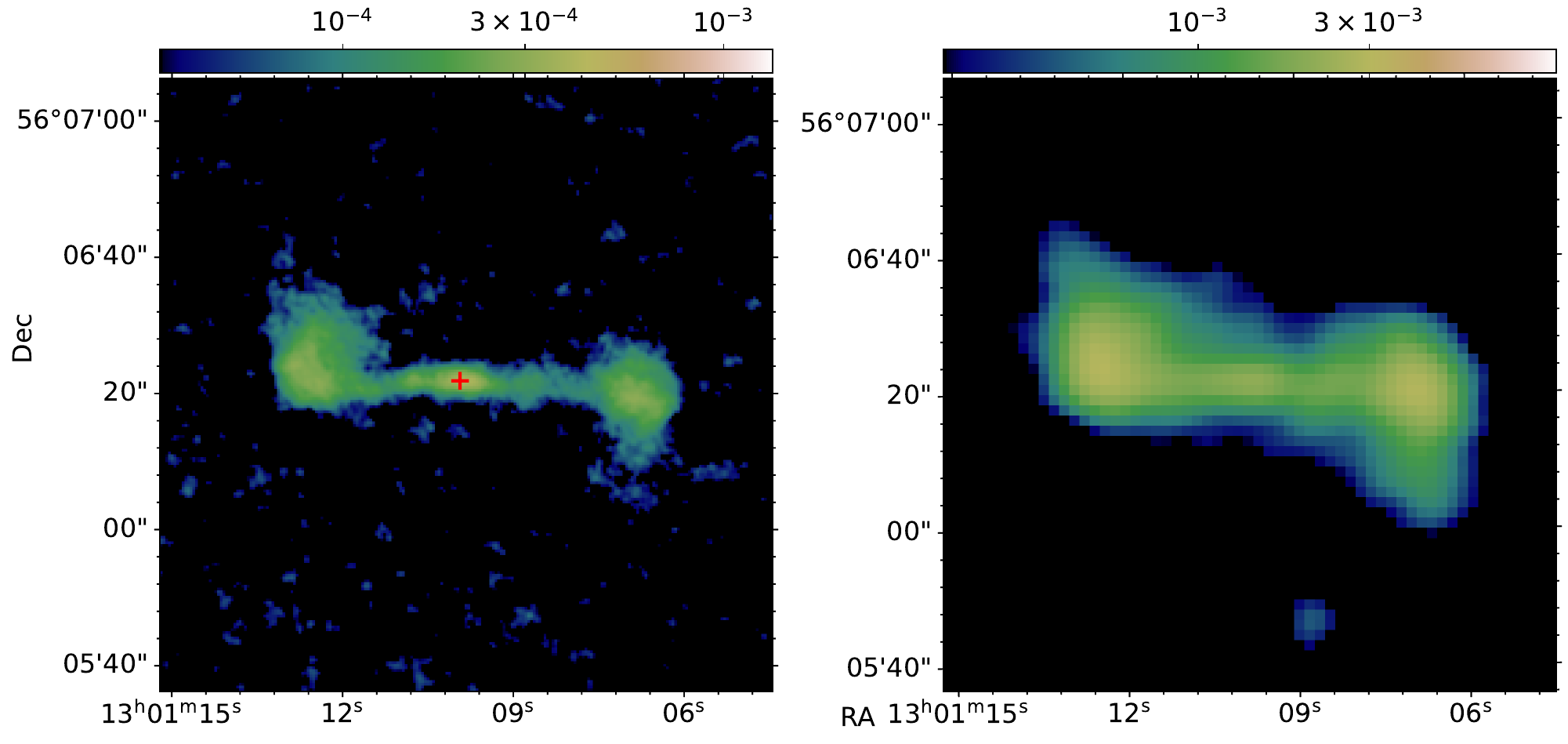}
        \subcaption{ILTJ130109.83+560623.4\label{fig:ErrExt}}
        \end{minipage}
    
    \vspace{0.07cm} 

    \centering
        \begin{minipage}{0.47\linewidth}
        \centering
            \includegraphics[width=\linewidth]{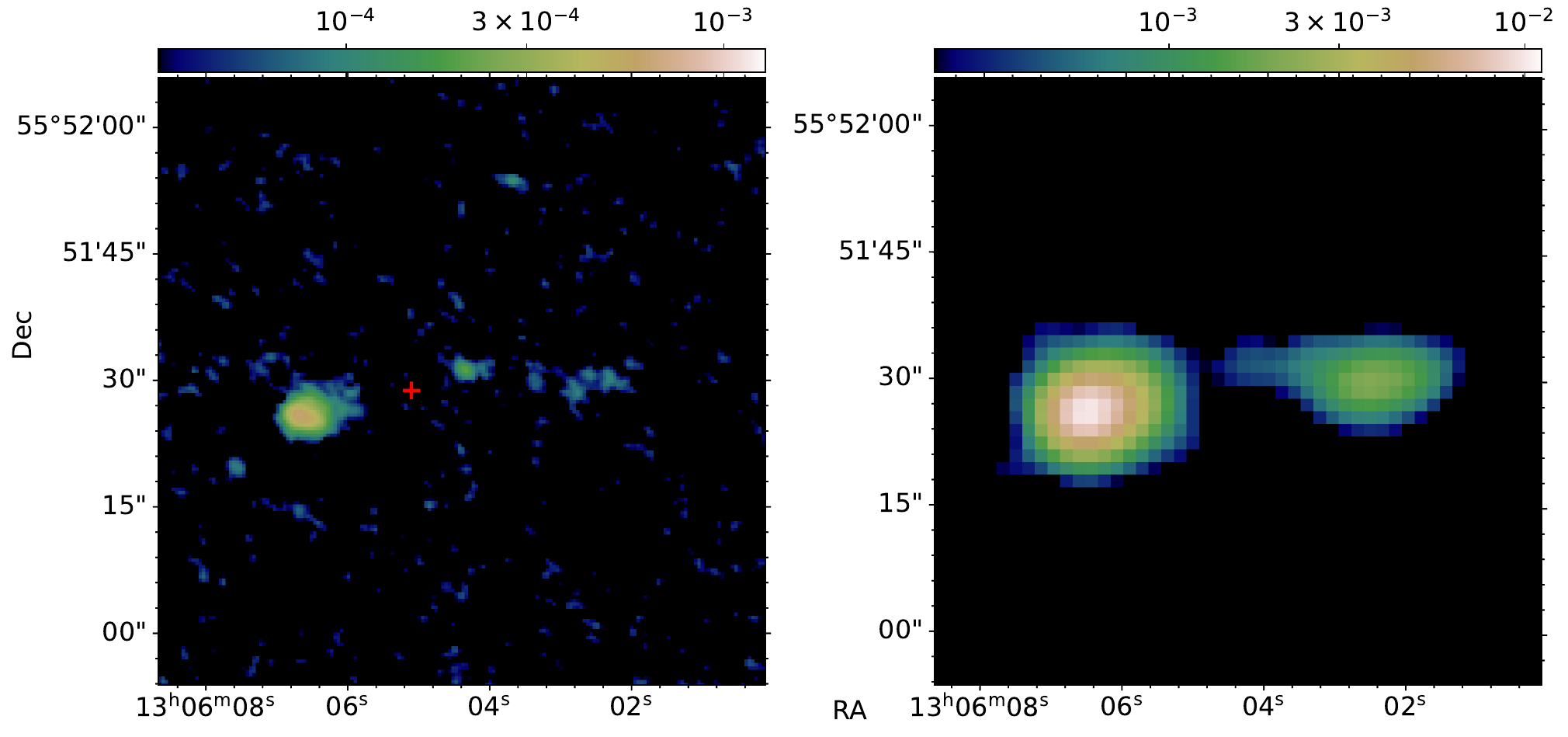}
        \subcaption{ILTJ130605.63+555127.6\label{fig:ErrExt}}
        \end{minipage}
        \hfill
        \begin{minipage}{0.47\linewidth}
        \centering
            \includegraphics[width=\linewidth]{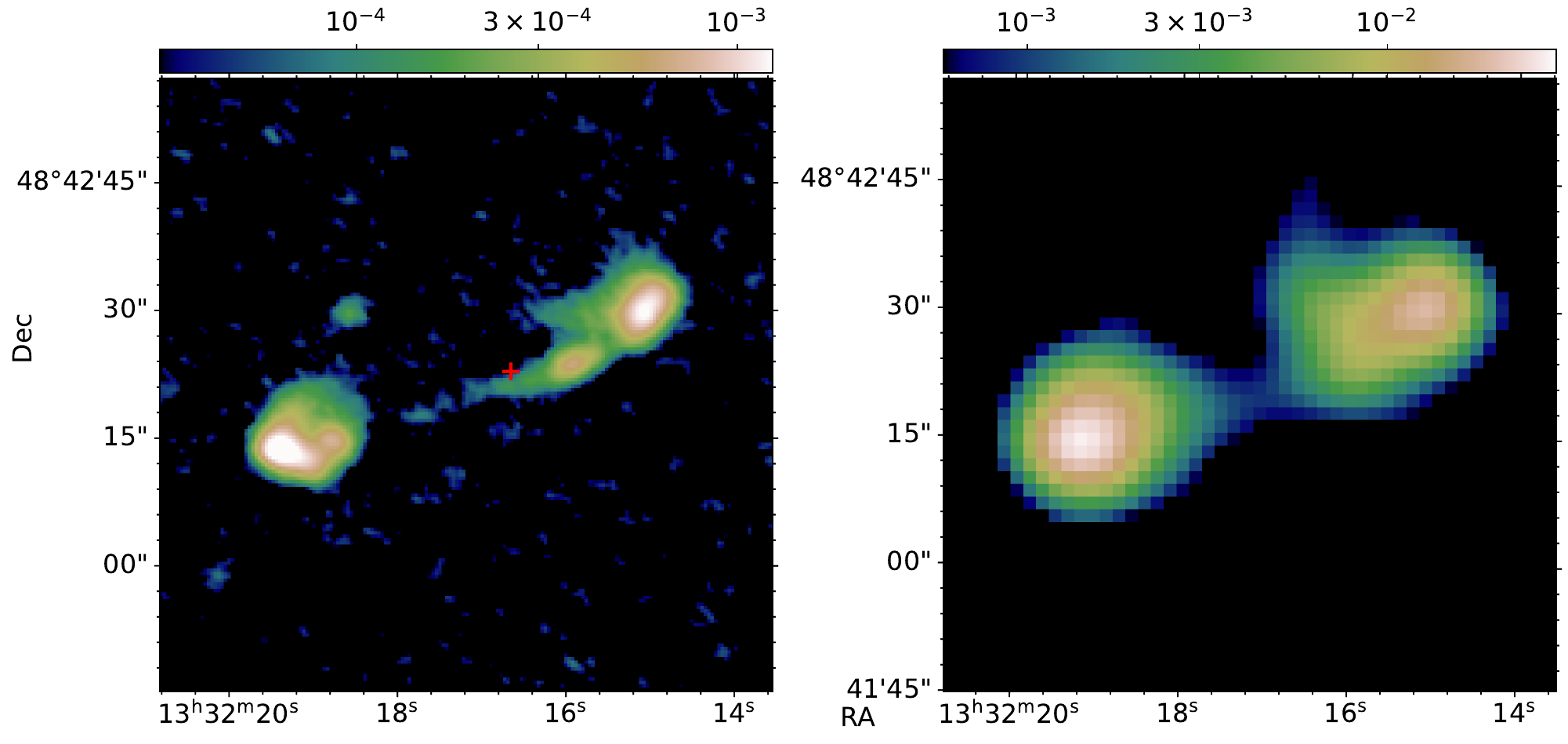}
        \subcaption{ILTJ133217.44+484221.7\label{fig:ErrExt}}
        \end{minipage}
    
\end{figure*}    

\begin{figure*}\ContinuedFloat
    \centering
        \begin{minipage}{0.47\linewidth}
        \centering
            \includegraphics[width=\linewidth]{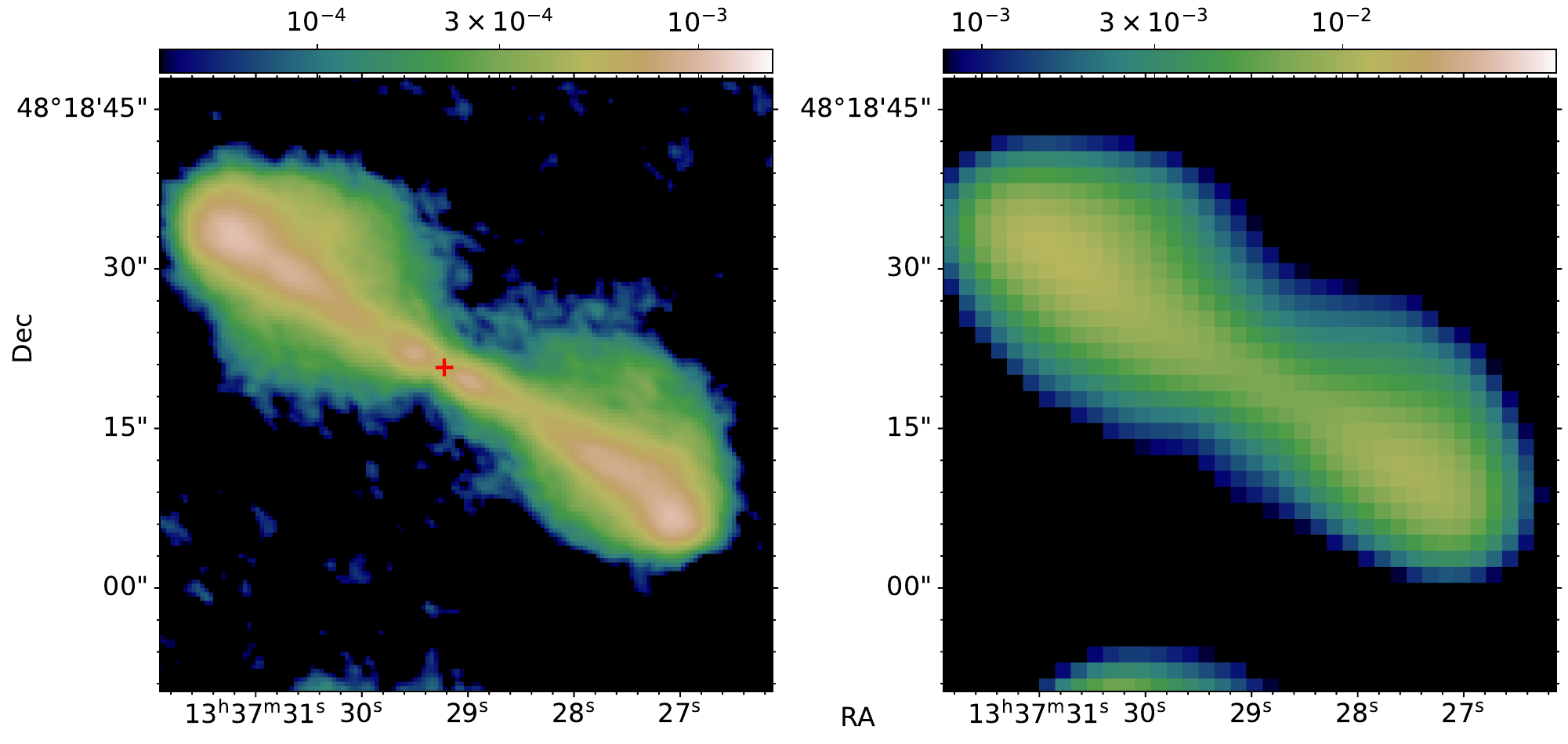}
        \subcaption{ILTJ133729.25+481822.3\label{fig:ErrExt}}
        \end{minipage}
        \hfill
        \begin{minipage}{0.47\linewidth}
        \centering
            \includegraphics[width=\linewidth]{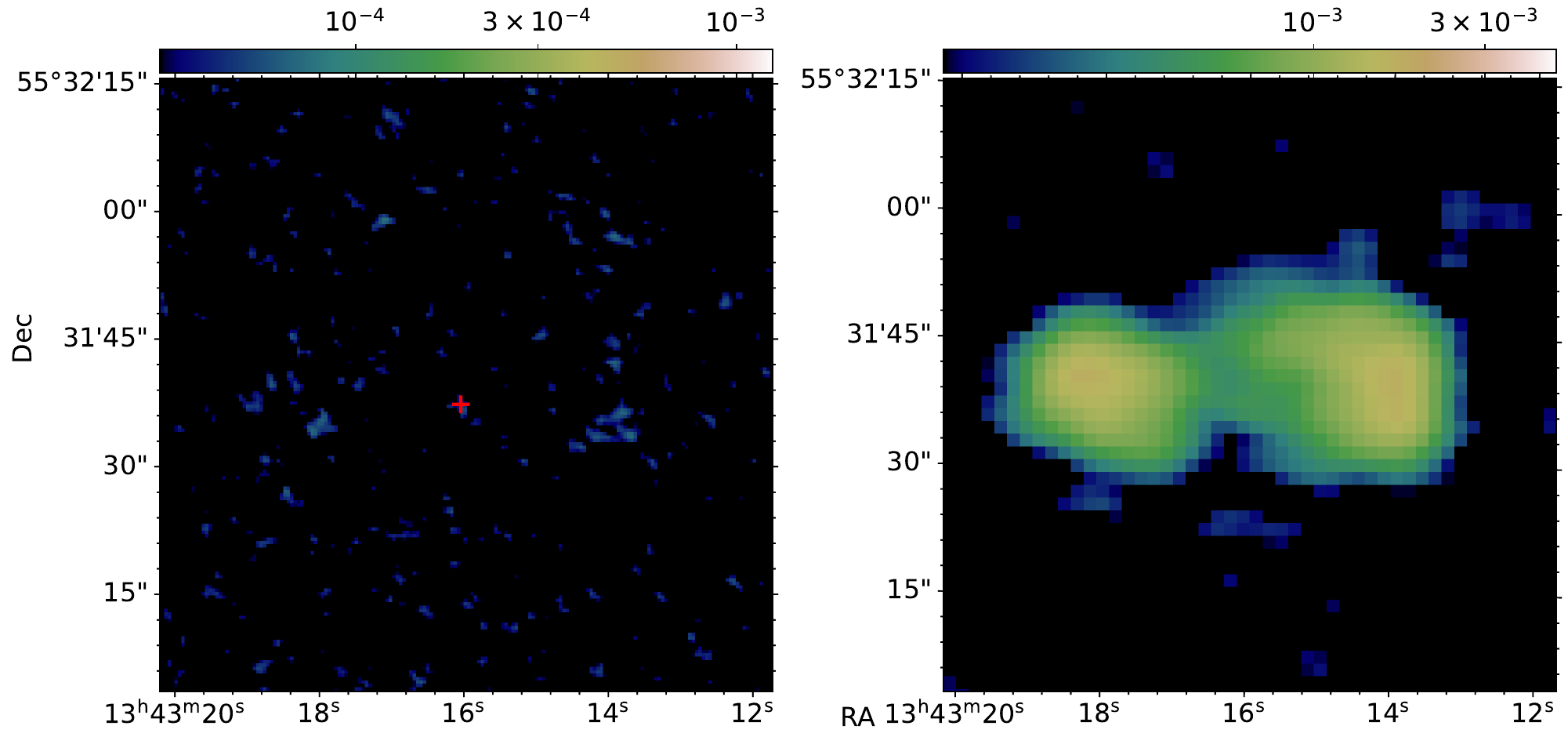}
        \subcaption{ILTJ134315.98+553139.6$^+$\label{fig:ErrExt}}
        \end{minipage}

    \vspace{0.07cm} 

    \centering
        \begin{minipage}{0.47\linewidth}
        \centering
            \includegraphics[width=\linewidth]{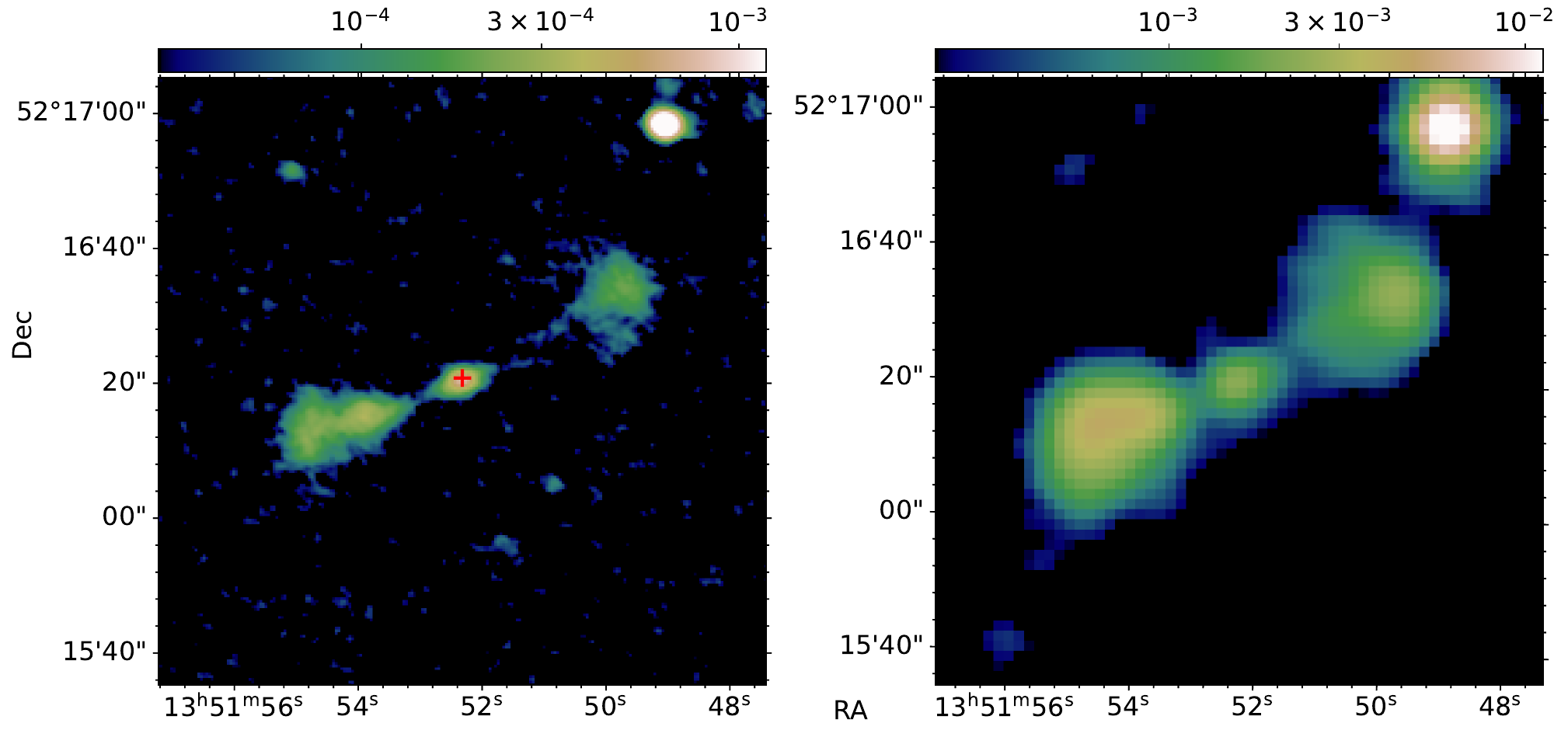}
        \subcaption{ILTJ135152.95+521618.8\label{fig:ErrExt}}
        \end{minipage}
        \hfill
        \begin{minipage}{0.47\linewidth}
        \centering
            \includegraphics[width=\linewidth]{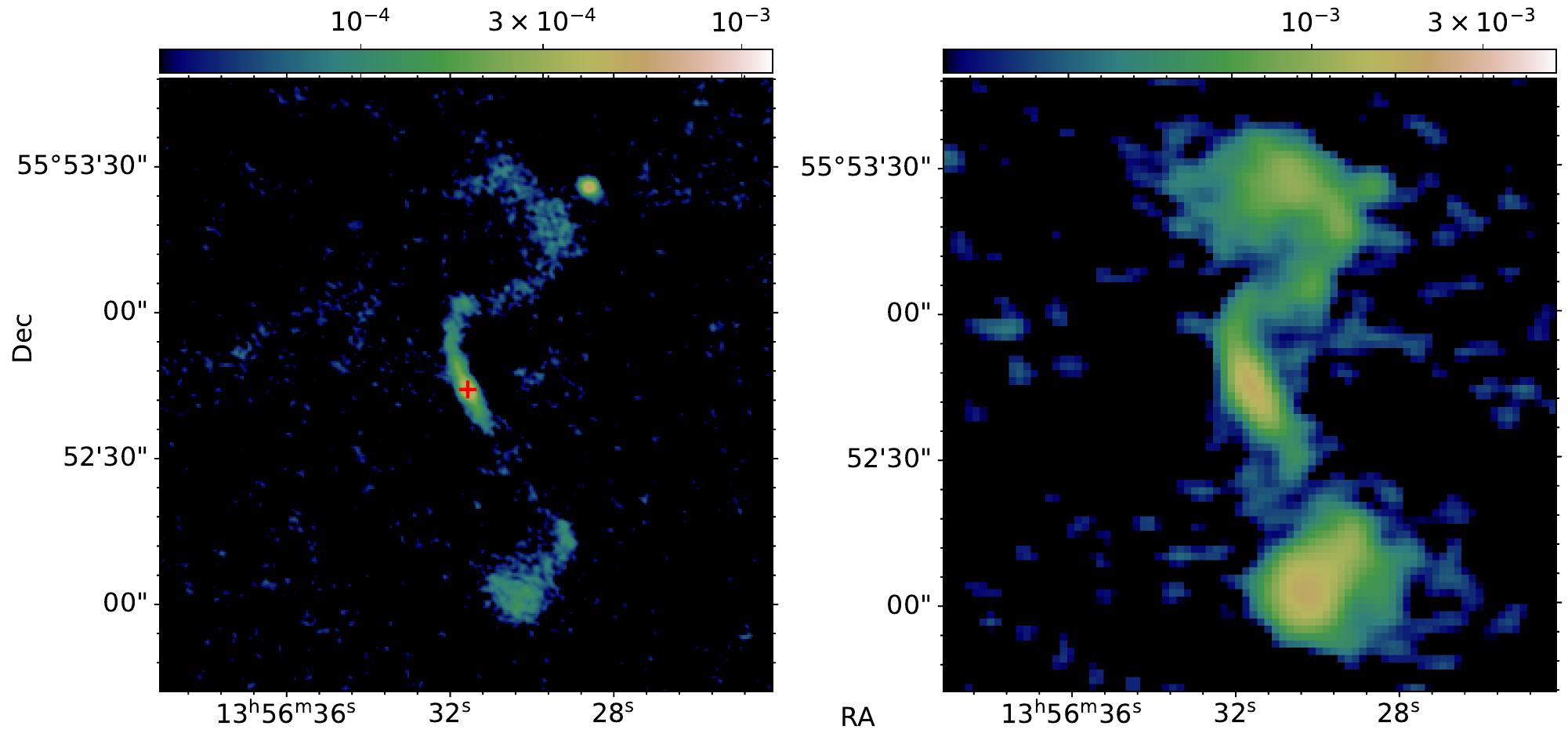}
        \subcaption{ILTJ135630.51+555245.1\label{fig:ErrExt}}
        \end{minipage}

    \vspace{0.07cm} 

    \centering
        \begin{minipage}{0.47\linewidth}
        \centering
            \includegraphics[width=\linewidth]{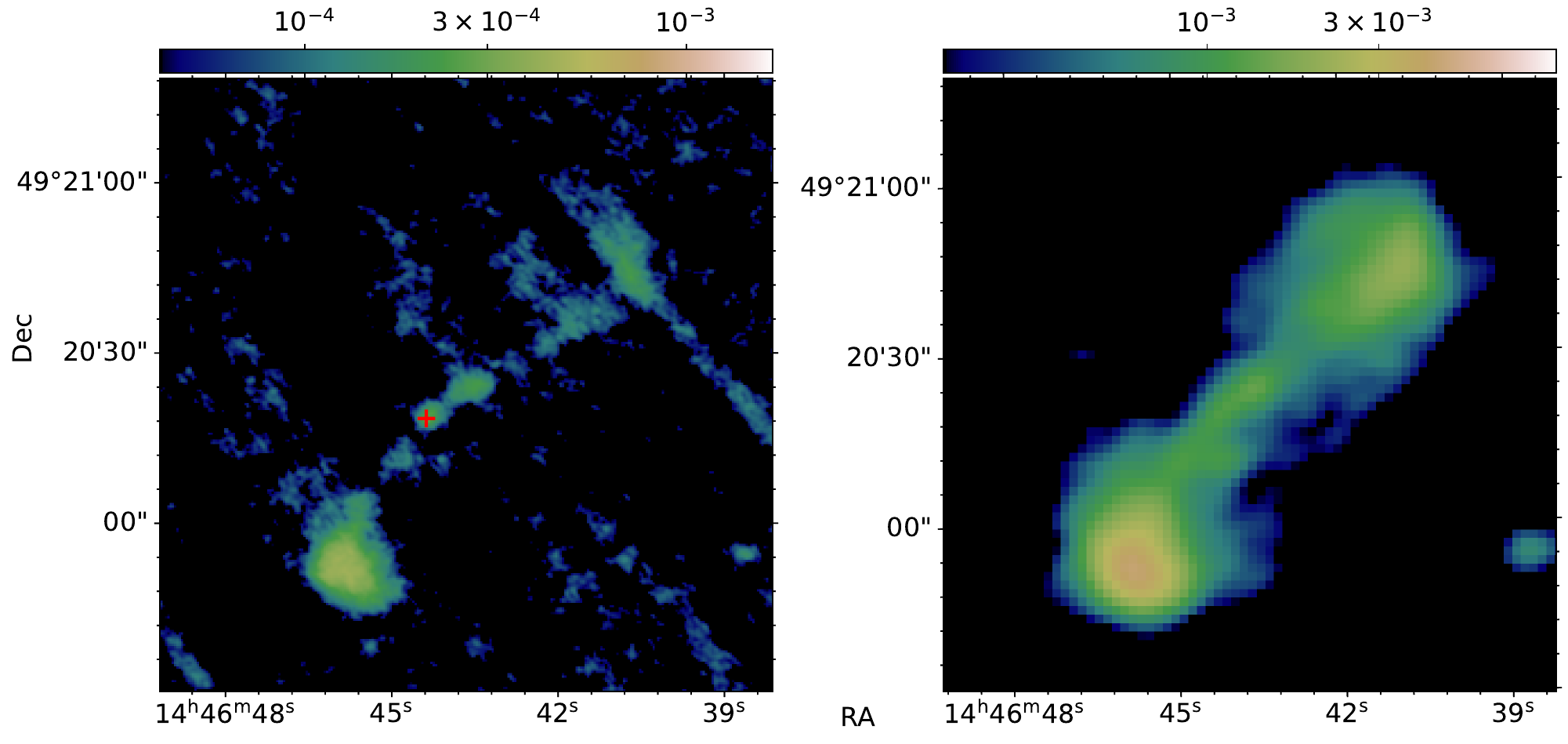}
        \subcaption{ILTJ144644.12+492012.3\label{fig:ErrExt}}
        \end{minipage}
        \hfill
        \begin{minipage}{0.47\linewidth}
        \centering
            \includegraphics[width=\linewidth]{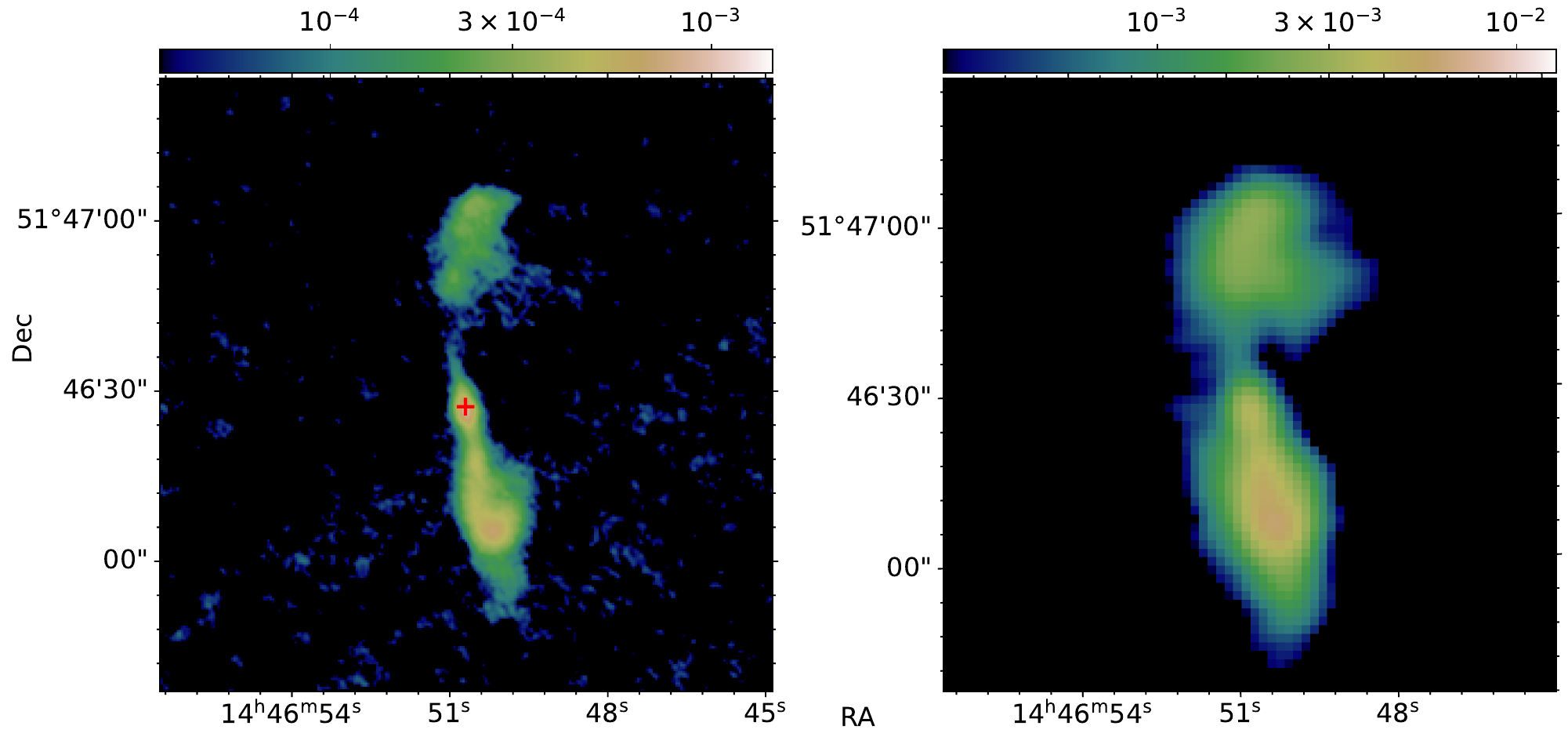}
        \subcaption{ILTJ144650.49+514625.3\label{fig:ErrExt}}
        \end{minipage}

    \vspace{0.07cm} 
    \centering
        \begin{minipage}{0.47\linewidth}
        \centering
            \includegraphics[width=\linewidth]{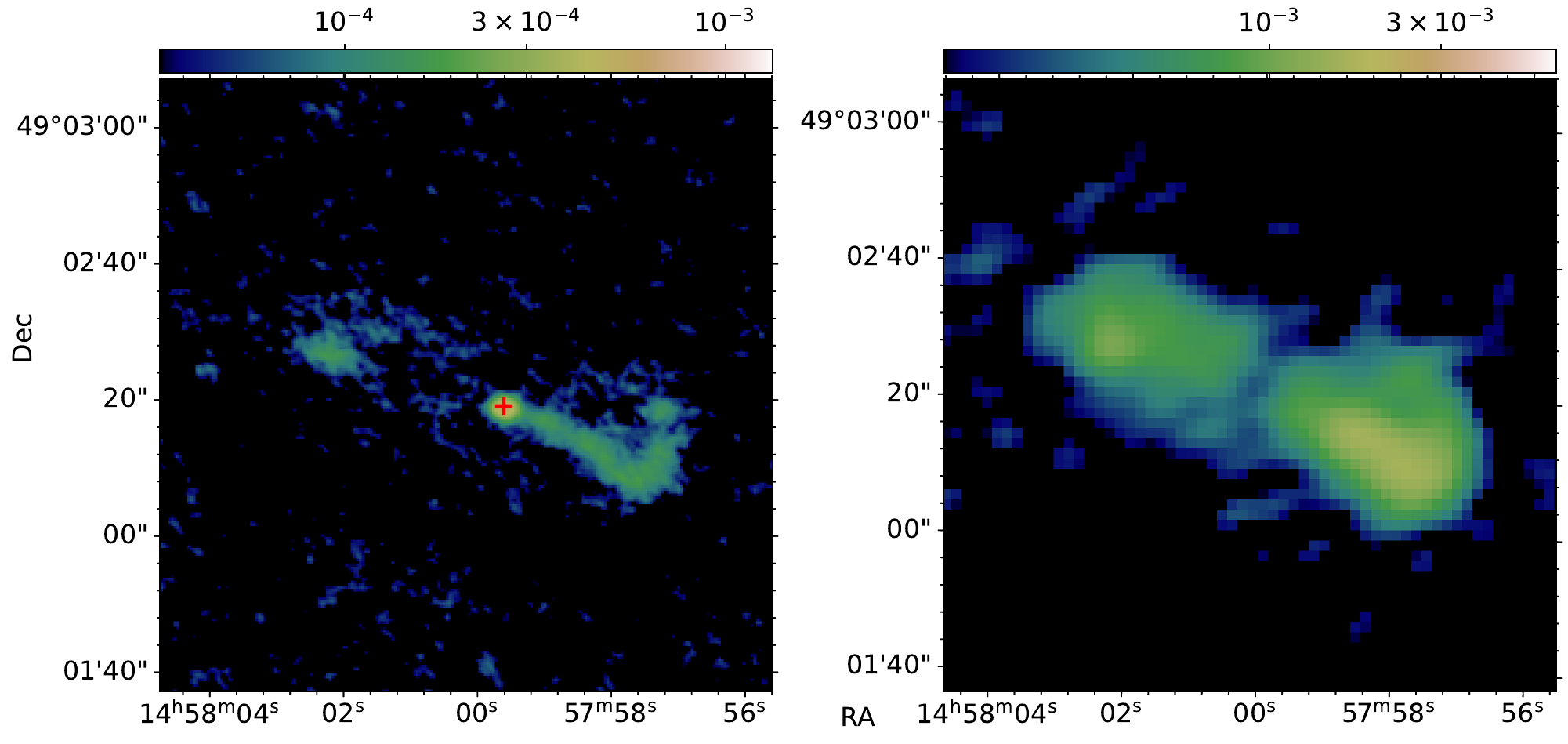}
        \subcaption{ILTJ145759.29+490219.2\label{fig:ErrExt}}
        \end{minipage}
        \hfill
        \begin{minipage}{0.47\linewidth}
        \centering
            \includegraphics[width=\linewidth]{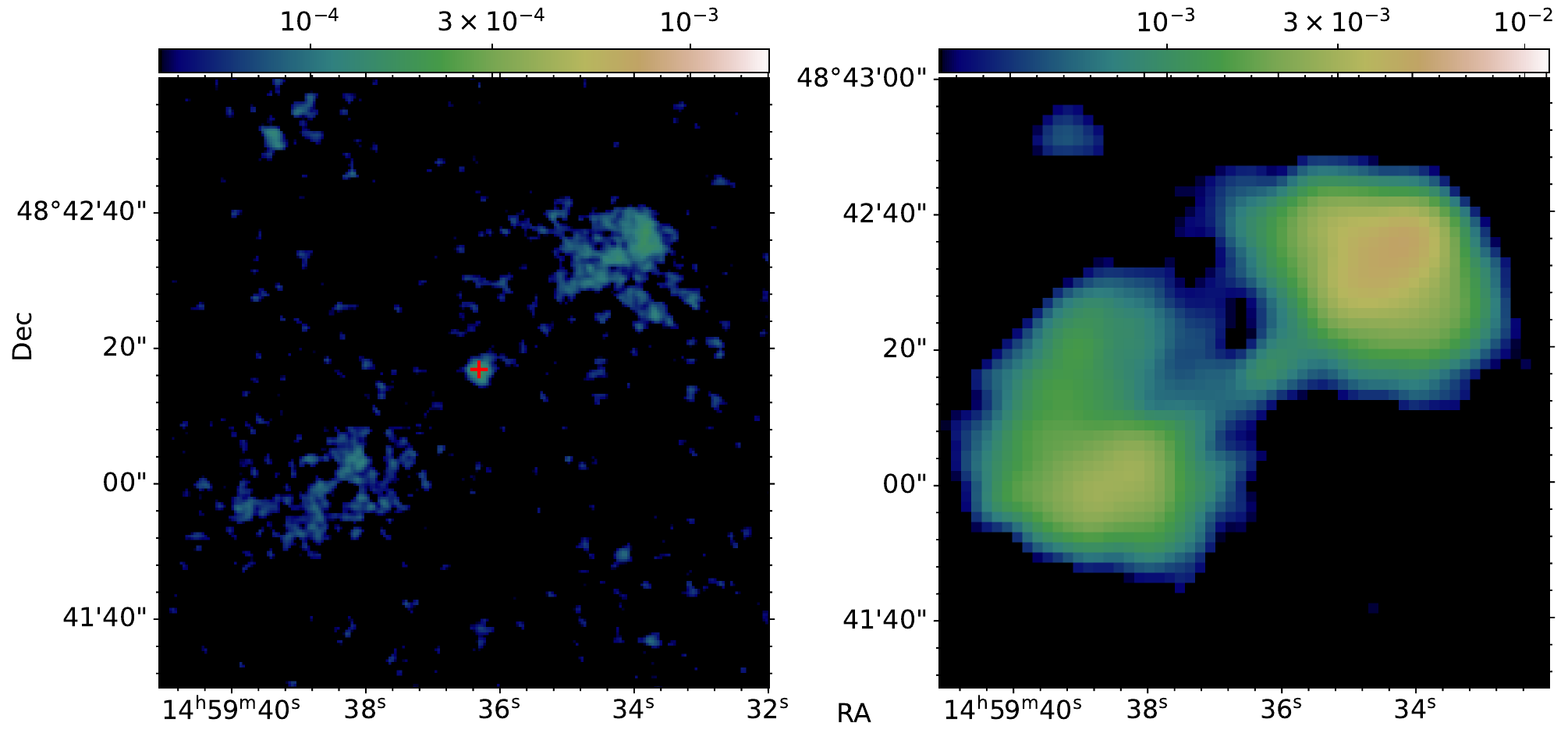}
        \subcaption{ILTJ145936.33+484219.8\label{fig:ErrExt}}
        \end{minipage}

    \vspace{0.07cm} 

    \centering
        \begin{minipage}{0.47\linewidth}
        \centering
            \includegraphics[width=\linewidth]{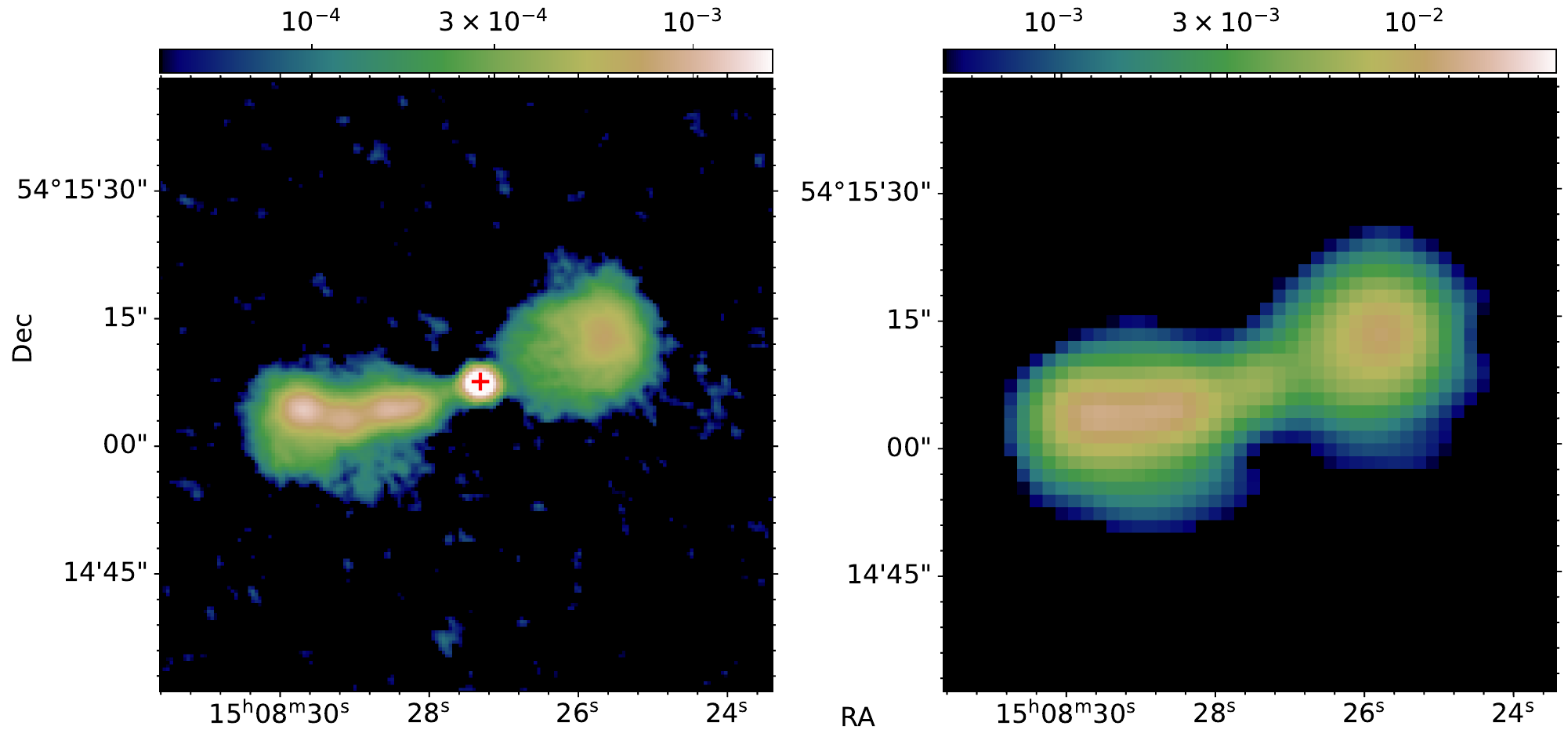}
        \subcaption{ILTJ150827.77+541507.1\label{fig:ErrExt}}
        \end{minipage}

    \caption{The images of the FRII-lows. These are given in two sets of two columns, in the left hand column are the 1.3-arcsec VLA images, made of the combination of the A, B, and C observations, and in the right hand column is the original LOFAR 6-arcsec image. Each pair from (a) through to (s) matches up with the source order given in \autoref{tab:image_prop}. The red cross is the position of the optical host, and the \revb{source indicated with a + is described in the text. The colour bars show the flux density in \revbb{mJy/beam.}}}
    \label{fig:LowImages} 
\end{figure*}

Out of the 19 sources, \revb{one did not yield a scientifically usable detection of the source above 5$\times$RMS (this has been marked with a + in \autoref{tab:image_prop} and \autoref{fig:LowImages}).} 

From the observational status summary for 2019A\footnote{\url{https://science.nrao.edu/facilities/vla/docs/manuals/oss2019A/performance/sensitivity}} the theoretical thermal noise for each image has been calculated.
The theoretical noise is given in the fourth column of \autoref{tab:image_prop} and can be compared with the observed noise levels in the fifth column. For all sources self-calibration is theoretically possible; however, although some images (such as ILTJ112015.05 and ILTJ121623.58) may benefit from self-calibration, in general we did not find significant improvement from self-calibration. The maps have a noise level within 1.5 to 3.2 times the theoretical noise. The peak flux \revb{density} value in the image and off source noise were obtained using \texttt{imstat}, and are used to calculate the dynamic range, which varies between approximately 160 to 5900.


\begin{table*}
    \centering
        \caption{Details of the FRII-Low final images. The first two columns give the LoTSS source name and the calibrator(s) used for the VLA observations. The following columns contain the image statistics for the combined 1.5 GHz VLA observations. These include the total time spent observing the source, both the theoretical and observed RMS levels, and the peak flux \revb{density} of the 1.3 arcsec image. The final two columns give the dynamic range and the ratio of the RMS levels (observed/theoretical). $^*$ Source ILTJ114351.49+511712.6 had a different calibrator for the A Configuration (J1219+4829) and the B and C Configurations (J1035+5628). \revb{$^+$ The source with a faint detection in the VLA as described in the text.}}
        \begin{tabular}{l l r r r r r r}
        \hline
        \textbf{LoTSS Source Name} & \textbf{Calibrator} & \textbf{Time on} & \textbf{Theoretical Noise} & \textbf{Observed RMS} & \textbf{Peak Flux \revb{density}} & \textbf{Dynamic} & \textbf{RMS} \\
         ~ &  ~ & \textbf{Source (s)} & \textbf{(Jy/beam)} & \textbf{(Jy/beam)} & \textbf{(Jy/beam)} & \textbf{Range} & \textbf{Ratio} \\
        \hline \hline
        ILTJ105946.75+563136.4      & J1035+5628 & 2314 & $10.4 \times 10^{-6}$ & $2.2 \times 10^{-5}$ & $135.2 \times 10^{-3}$ & 5927.5 & 2.2 \\
        ILTJ112015.05+503254.9      & J1035+5628 & 3380 & $8.6 \times 10^{-6}$  & $1.3 \times 10^{-5}$ & $57.7 \times 10^{-3}$  & 4328.5 & 1.6 \\
        ILTJ112654.44+540415.3      & J1035+5628 & 2555 & $9.9 \times 10^{-6}$  & $1.6 \times 10^{-5}$ & $43.6 \times 10^{-3}$  & 2760.6 & 1.6 \\
        ILTJ113626.52+501320.3      & J1035+5628 & 2899 & $9.3 \times 10^{-6}$  & $1.9 \times 10^{-5}$ & $4.1 \times 10^{-3}$   & 212.9  & 2.1 \\  
        \multirow{2}*{ILTJ114351.49+511712.6}  & J1035+5628$^*$ & \multirow{2}*{3098} & \multirow{2}*{$9.0 \times 10^{-6}$} & \multirow{2}*{$2.9 \times 10^{-5}$} &   \multirow{2}*{ $131.3 \times 10^{-3}$}       & \multirow{2}*{4516.2} & \multirow{2}*{3.2} \\
                                      & J1219+4829$^*$ &      &  &  &    &   \\
        ILTJ115011.27+534320.9      & J1219+4829 & 2276 & $10.4 \times 10^{-6}$ & $2.2 \times 10^{-5}$ & $15.2 \times 10^{-3}$  & 695.1  & 2.1 \\
        ILTJ121623.58+524409.4      & J1219+4829 & 3005 & $9.1 \times 10^{-6}$  & $2.0 \times 10^{-5}$ & $10.3 \times 10^{-3}$  & 519.0  & 2.2 \\
        ILTJ130109.83+560623.4      & J1219+4829 & 3240 & $8.8 \times 10^{-6}$  & $1.7 \times 10^{-5}$ & $7.9 \times 10^{-3}$   & 468.1  & 1.9 \\
        ILTJ130605.63+555127.6      & J1349+5341 & 2494 & $10.0 \times 10^{-6}$ & $1.6 \times 10^{-5}$ & $4.2 \times 10^{-3}$   & 258.6  & 1.6 \\
        ILTJ133217.44+484221.7      & J1349+5341 & 2944 & $9.2 \times 10^{-6}$  & $1.5 \times 10^{-5}$ & $2.5 \times 10^{-3}$   & 162.9  & 1.7 \\
        ILTJ133729.25+481822.3      & J1349+5341 & 2848 & $9.3 \times 10^{-6}$  & $2.0 \times 10^{-5}$ & $50.0 \times 10^{-3}$  & 2554.6 & 2.1 \\
        ILTJ134315.98+553139.6$^+$  & J1349+5341 & 2460 & $10.4 \times 10^{-6}$ & $1.6 \times 10^{-5}$ & $8.5 \times 10^{-3}$   & 546.8  & 1.5 \\
        ILTJ135152.95+521618.8      & J1349+5341 & 3060 & $9.0 \times 10^{-6}$  & $1.5 \times 10^{-5}$ & $4.6 \times 10^{-3}$   & 310.6  & 1.6 \\
        ILTJ135630.51+555245.1      & J1349+5341 & 3179 & $8.8 \times 10^{-6}$  & $1.5 \times 10^{-5}$ & $15.0 \times 10^{-3}$  & 999.2  & 1.7 \\
        ILTJ144644.12+492012.3      & J1549+5038 & 3252 & $8.7 \times 10^{-6}$  & $2.1 \times 10^{-5}$ & $62.7 \times 10^{-3}$  & 2976.7 & 2.4 \\
        ILTJ144650.49+514625.3      & J1549+5038 & 3033 & $9.0 \times 10^{-6}$  & $1.8 \times 10^{-5}$ & $3.0 \times 10^{-3}$   & 167.5  & 2.0 \\
        ILTJ145759.29+490219.2      & J1549+5038 & 3126 & $8.9 \times 10^{-6}$  & $1.7 \times 10^{-5}$ & $13.5 \times 10^{-3}$  & 814.3  & 1.9 \\
        ILTJ145936.33+484219.8      & J1549+5038 & 3234 & $8.8 \times 10^{-6}$  & $2.0 \times 10^{-5}$ & $46.2 \times 10^{-3}$  & 2288.3 & 2.3 \\
        ILTJ150827.77+541507.1      & J1549+5038 & 2272 & $10.4 \times 10^{-6}$ & $2.0 \times 10^{-5}$ & $4.3 \times 10^{-3}$   & 211.2  & 1.9 \\
        \hline
        \end{tabular}
    \label{tab:image_prop} 
\end{table*}

\subsection{FRII-highs}
\label{subsec:FRII-highs}

In order to establish whether the FRII-low population is physically distinct from the more traditional FRII population at high luminosity (hereafter FRII-highs) we defined a sample of ten high-luminosity FRIIs. This is a comparison group, representative of well-studied luminous FRIIs, i.e. the population that have formed the basis of studies of FRII physics, dynamics, and energetics to date. The sample of ten FRII-highs were selected at random from \citet{mingo_revisiting_2019}'s sources classified as FRIIs with $L_{150}>10^{26}$ W Hz$^{-1}$ and make up $\sim$ 7.5 per cent of the total number of FRII-highs in \citet{mingo_revisiting_2019}'s sample. 

\revb{Appendix \ref{App:Features} provides the 6-arcsec LoTSS DR2 images of the FRII-highs, along with their corresponding FIRST images. For reference throughout the following sections: these FRII-highs have 20$\sigma$ contours included on the FIRST images. Appendix \ref{App:Features} also includes a set of images of the FRII-low sources \revbb{with structures above the 20$\sigma$ contour threshold}.}

\subsection{Comparing populations}
\label{subsec:CompPP}

\autoref{tab:medians} compares the median values of key properties for each sample group and their corresponding parent populations. The comparison shows the expected difference in median luminosity between the FRII-lows ($5.8 \times 10^{24}$ W Hz$^{-1}$) and the FRII-highs ($179.6 \times 10^{24}$ W Hz$^{-1}$), and also shows that there is only a small difference in the host galaxy magnitudes, with the FRII-highs having slightly brighter host galaxies. The FRII-lows have a median magnitude of \revb{$K_s = -23.26$} compared to the FRII-highs at -23.97. \revb{\autoref{fig:CompPlots} shows that t}he FRII-lows are smaller in both their angular and linear sizes compared to the FRII-highs, and they also cover a lower redshift range. Although the LOFAR data span a wide luminosity range at low to medium redshifts, on average low-luminosity sources are expected to have lower redshifts than higher-luminosity sources.

\begin{table*}
    \centering
    \caption{The medians, standard deviation, and \revb{Kolmogorov-Smirnov (KS) test} $p$-values of the key properties of the FRII-lows and -highs and their corresponding parent populations \revb{(size given in brackets)}. Both samples are selected from the catalogue by \citet{mingo_revisiting_2019}, as described in the main text.}
    \begin{tabular}{l r r r r r r r r r r}
        \hline
        ~ & \multicolumn{5}{c}{\textbf{FRII-lows}} & \multicolumn{5}{c}{\textbf{FRII-highs}}  \\
        \hline 
        ~  & \multicolumn{2}{c}{\textbf{Sample \revb{(19)}}} & \multicolumn{2}{c}{\textbf{Population \revb{(120)}}}  &  ~  & \multicolumn{2}{c}{\textbf{Sample \revb{(10)}}} & \multicolumn{2}{c}{\textbf{Population \revb{(131)}}} & ~ \\
        \textbf{Property}  & \textbf{median}  &  \textbf{stdev}  & \textbf{median}  &  \textbf{stdev}  & \textbf{\textit{p}} & \textbf{median}  &  \textbf{stdev}  & \textbf{median}  &  \textbf{stdev}  & \textbf{\textit{p}} \\ 
        \hline\hline
        \textbf{\rule{0pt}{2.5ex}\revb{$L_{150}$} Luminosity (10$^{24}$ W Hz$^{-1}$)} & 5.8     &  4.4    &  6.2      &  5.7    &  \revb{\textit{0.52}}  &  179.6   &  558.1  &  301.5   &  1323.1  &  \textit{0.61}  \\
        \textbf{Redshift}                           & 0.17    &  0.12   &  0.38     &  0.16   &  \revb{$\mathit{3.8\times10^{-5}}$} &  0.49    &  0.15   &  0.57    &  0.15    &  \textit{0.52}  \\
        \textbf{Angular size (arcsec)}              & 84.9    &  31.8   &  56.5     &  40.5   &  $\mathit{2.2\times10^{-3}}$  &  95.5    &  44.0   &  85.2    &  330.4   &  \textit{0.31}  \\
        \textbf{Linear size (kpc)}                  & 311.9   &  147.2  &  168.5    &  78.0   &  \textit{0.093}               &  587.2   &  221.7  &  531.7   &  1277.1  &  \textit{0.63}  \\
        \textbf{Magnitude (\revb{$K_s$})}           & -23.26  &  0.90   &  -23.63   &  0.73   &  \textit{0.076}               &  -23.97  &  0.72   &  -24.14  &  0.66    &  \textit{0.59}  \\
        \hline
    \end{tabular}
    \label{tab:medians}
\end{table*}

\begin{figure*}
    \centering
    \begin{minipage}{0.9\textwidth}
        \centering
        \includegraphics[width=0.45\linewidth]{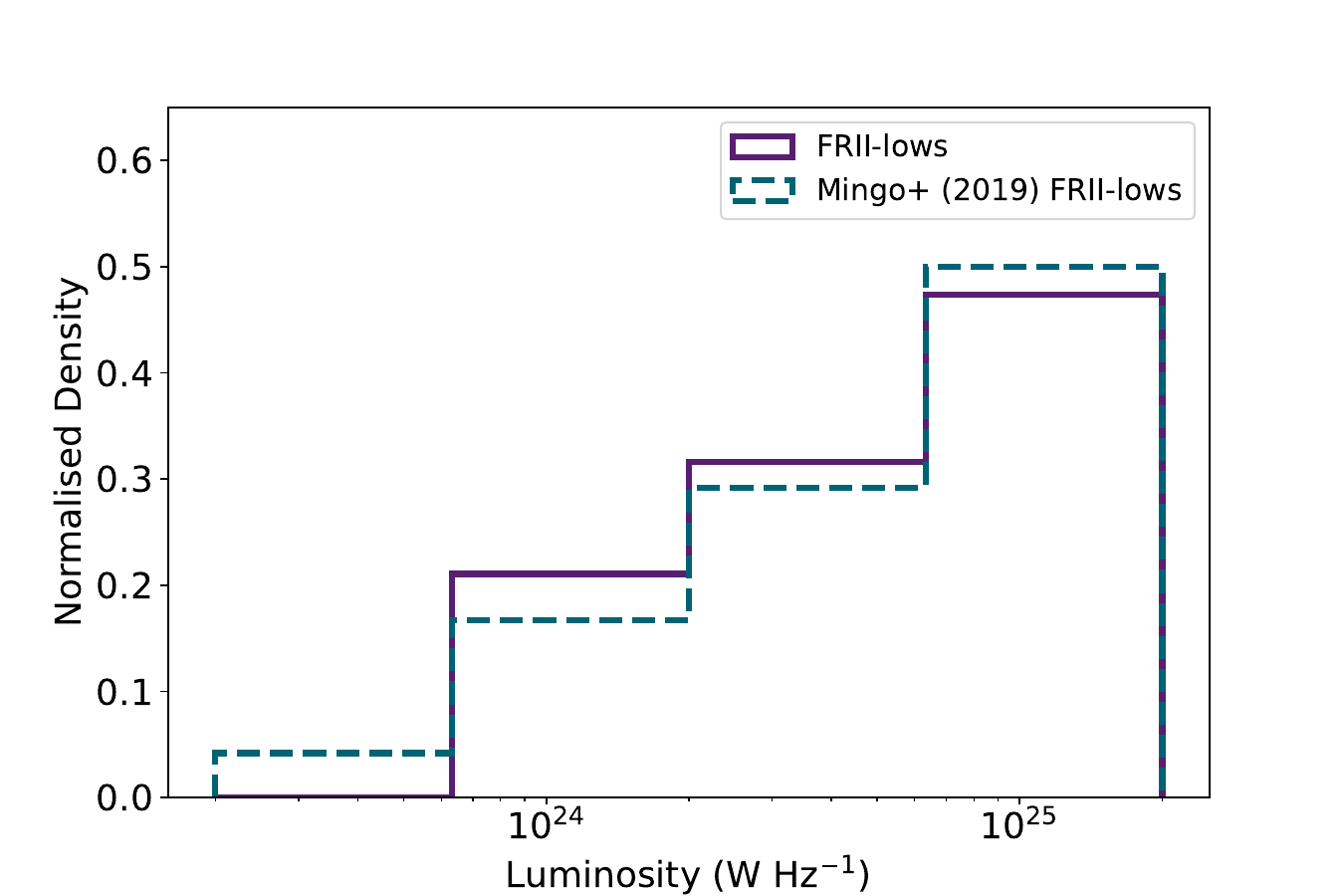}
        \includegraphics[width=0.45\linewidth]{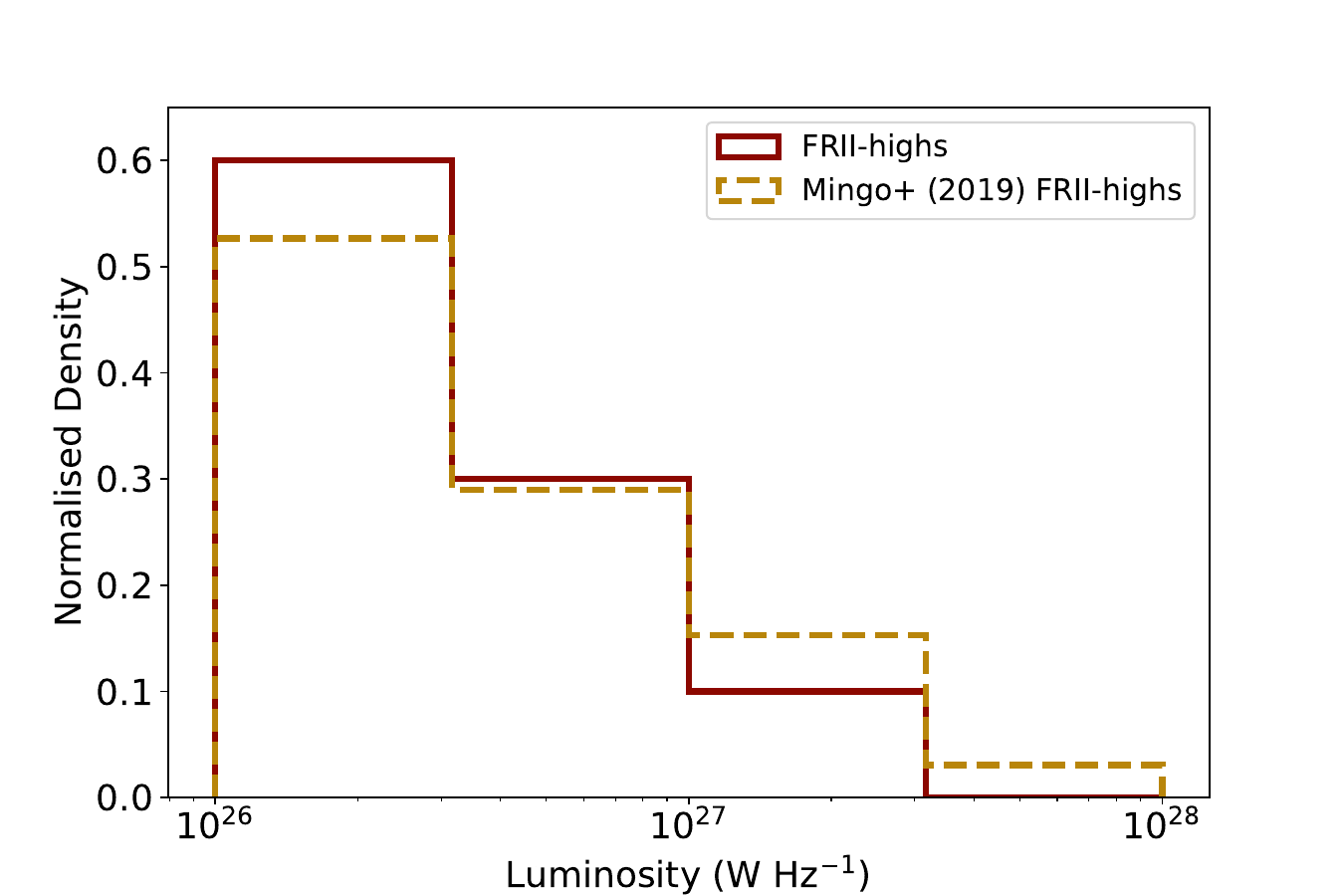}
        \subcaption{Luminosity (W Hz$^{-1}$)}
    \end{minipage}

    \vspace{0.09cm}

    \begin{minipage}{0.9\textwidth}
        \centering
        \includegraphics[width=0.45\linewidth]{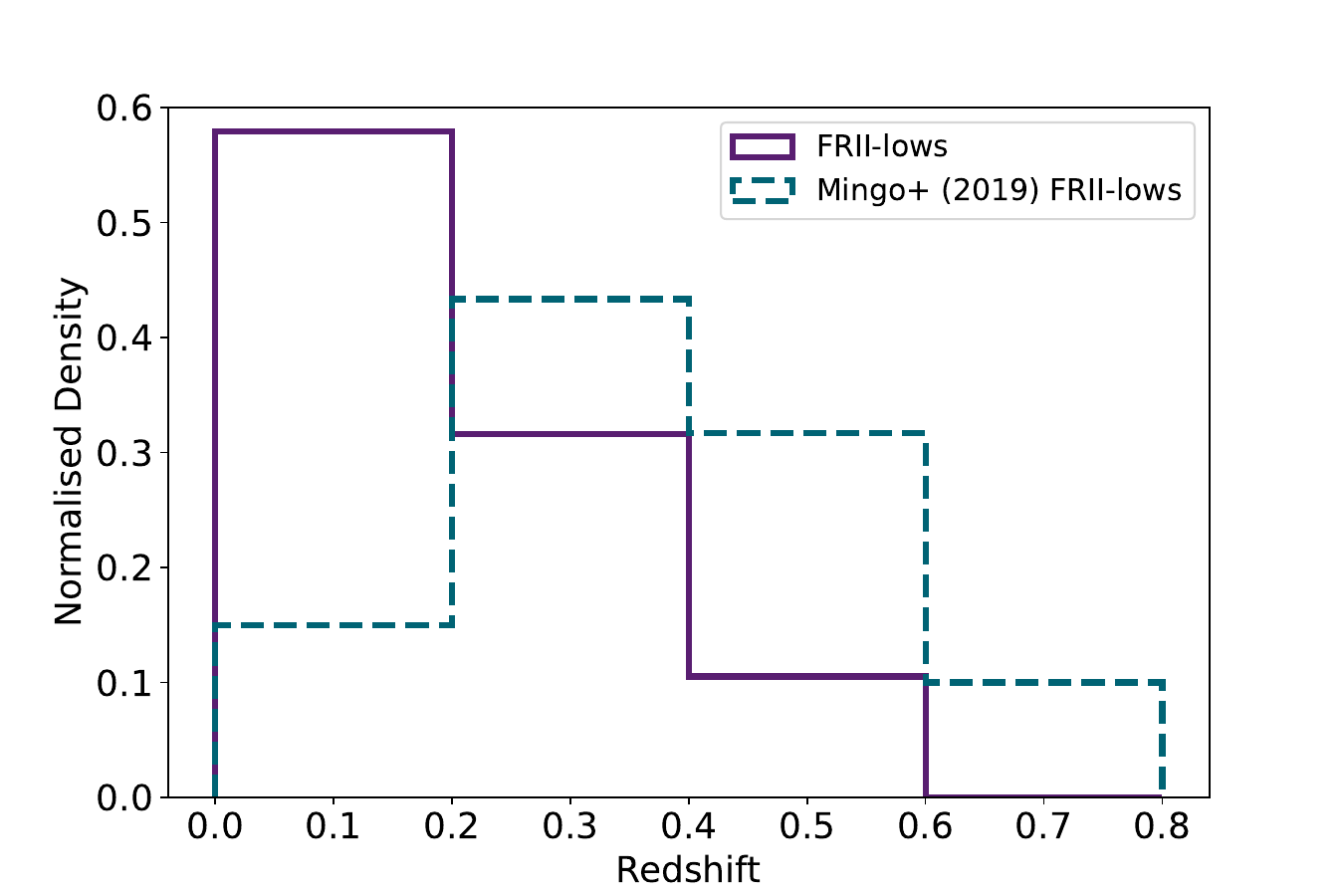}
        \includegraphics[width=0.45\linewidth]{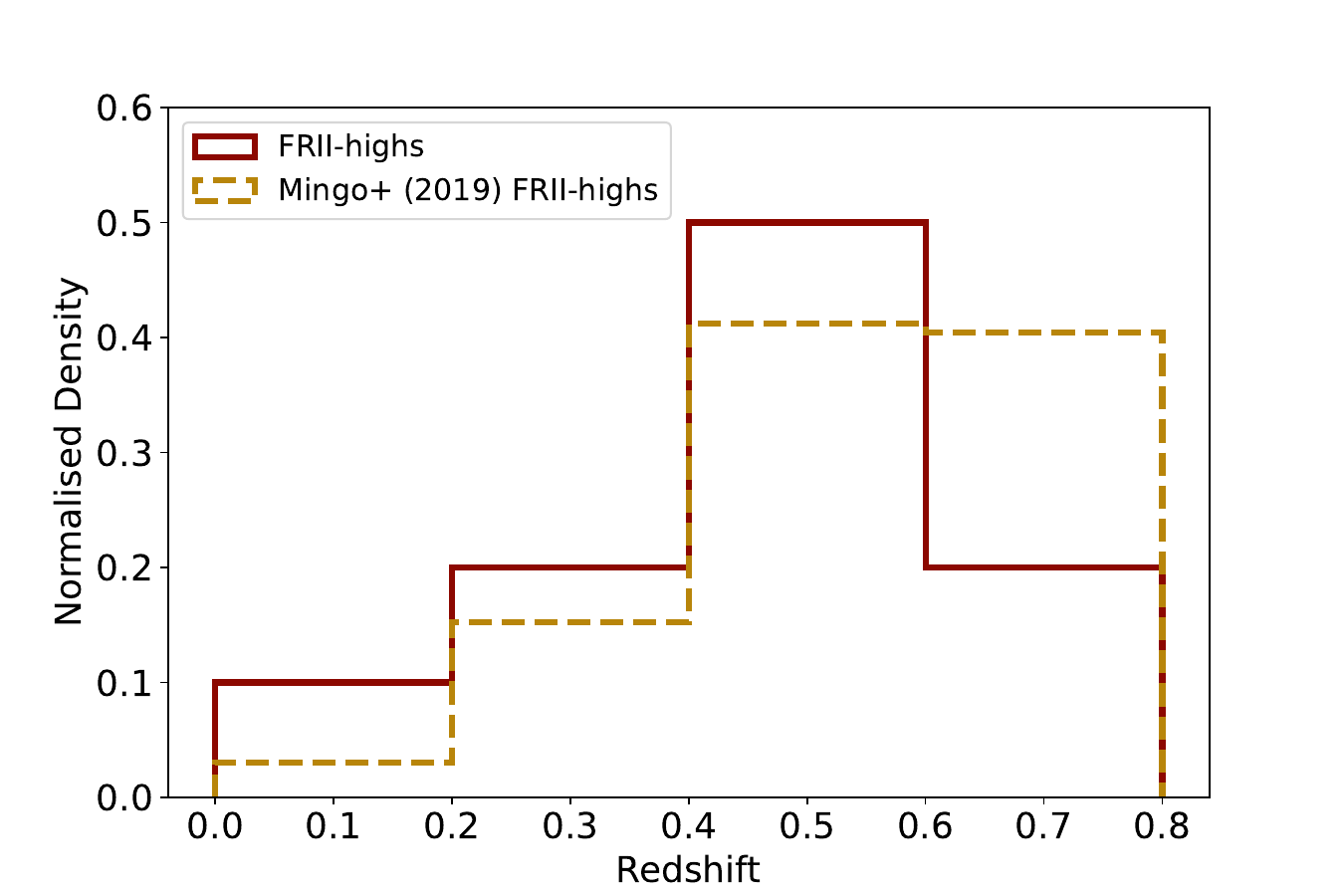}
        \subcaption{Redshift}
    \end{minipage}

    \vspace{0.09cm}

    \begin{minipage}{0.9\textwidth}
        \centering
        \includegraphics[width=0.45\linewidth]{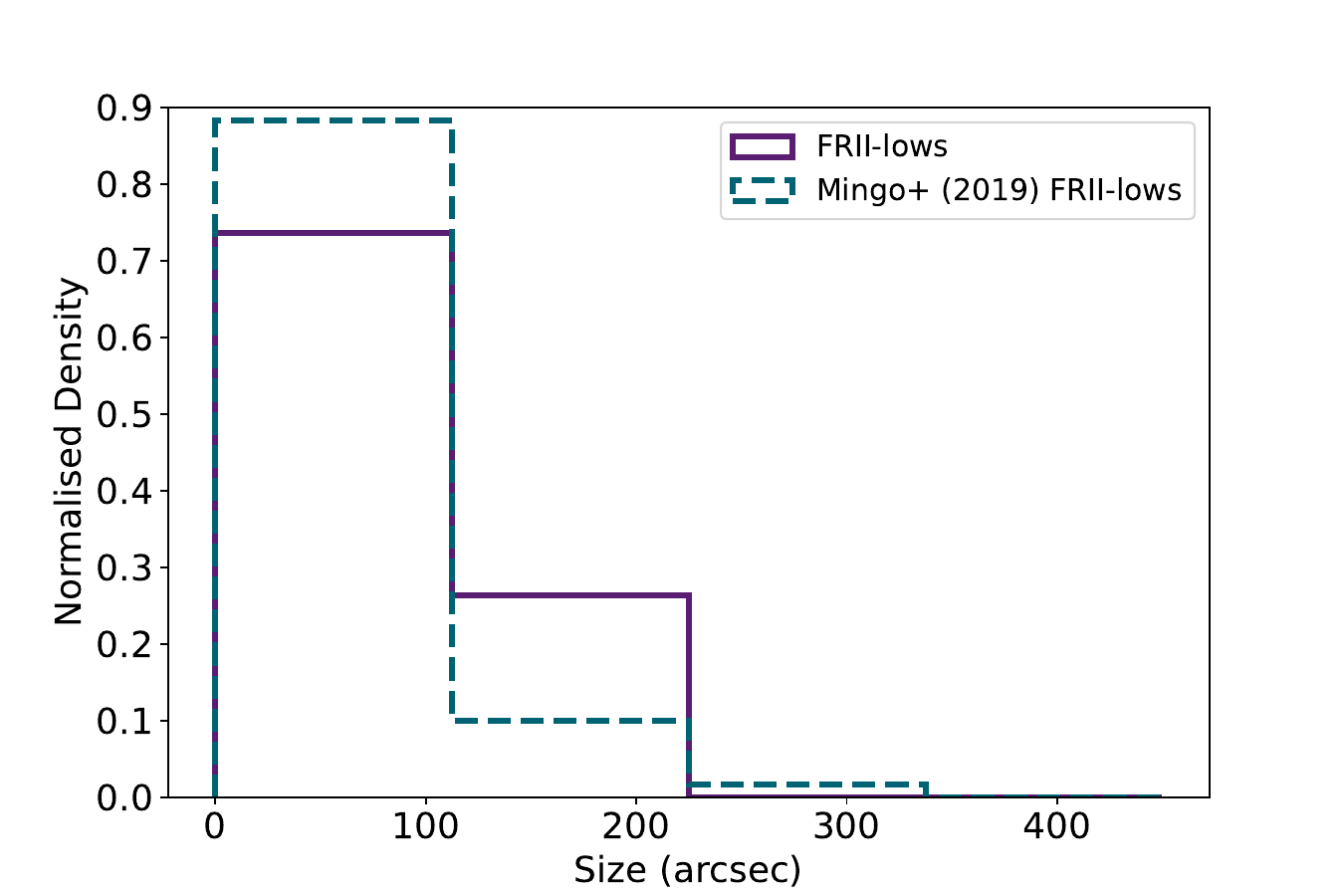}
        \includegraphics[width=0.45\linewidth]{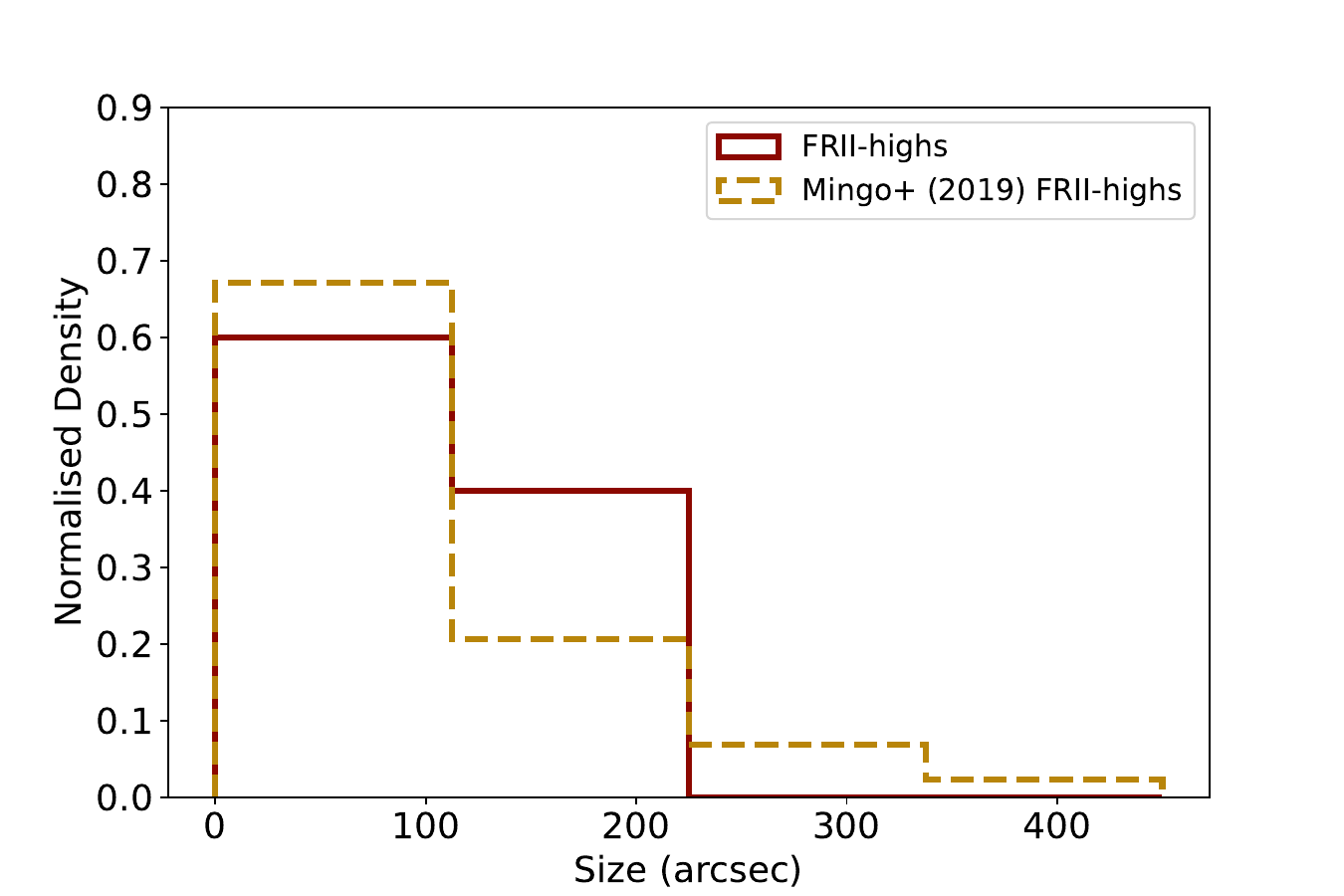}
        \subcaption{Angular size (arcsec)}
    \end{minipage}

    \vspace{0.09cm}
    
    \begin{minipage}{0.9\textwidth}
        \centering
        \includegraphics[width=0.45\linewidth]{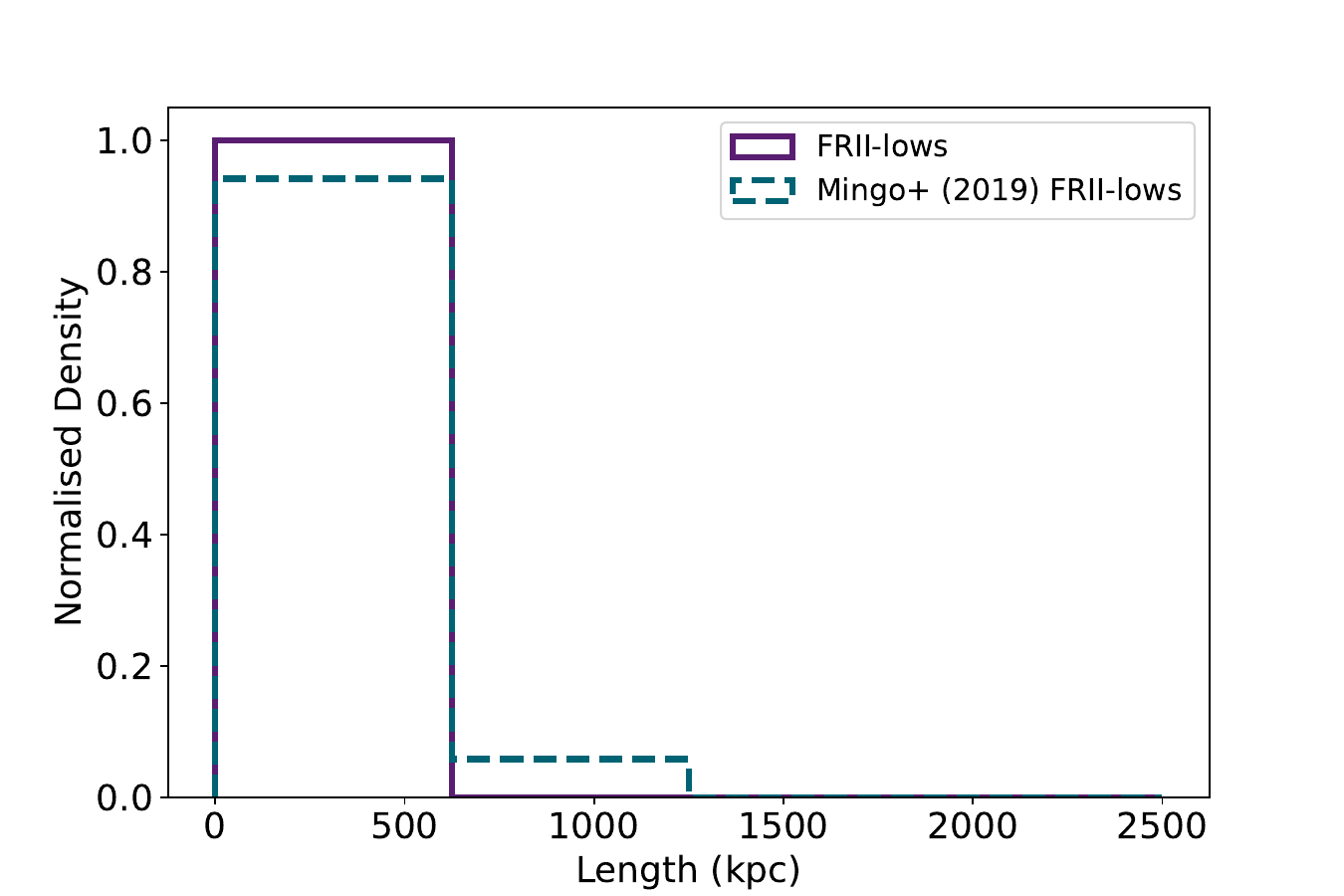}
        \includegraphics[width=0.45\linewidth]{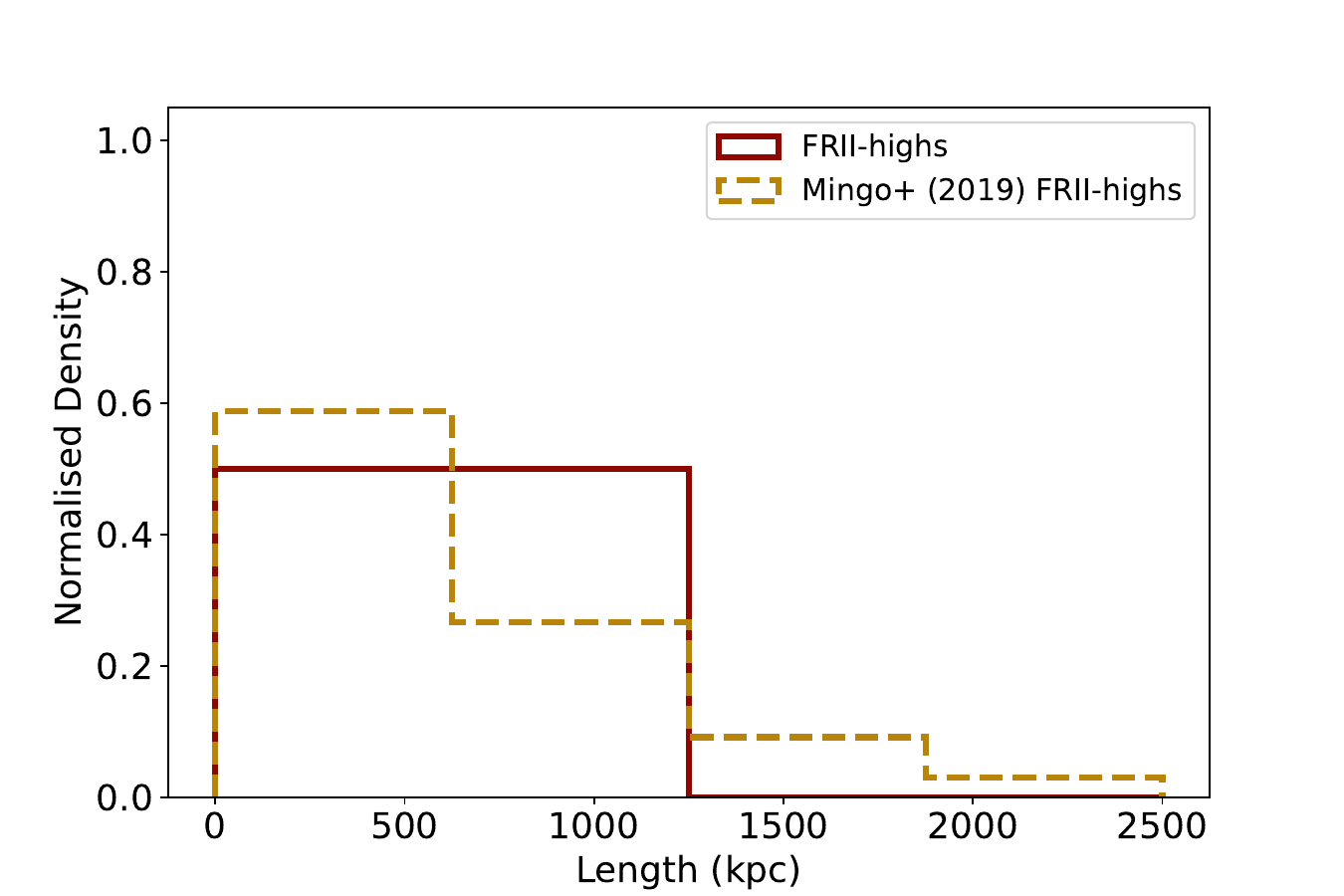}
        \subcaption{Linear size (kpc)}
    \end{minipage}

\end{figure*}

\begin{figure*}\ContinuedFloat

    \begin{minipage}{0.9\textwidth}
        \centering
        \includegraphics[width=0.45\linewidth]{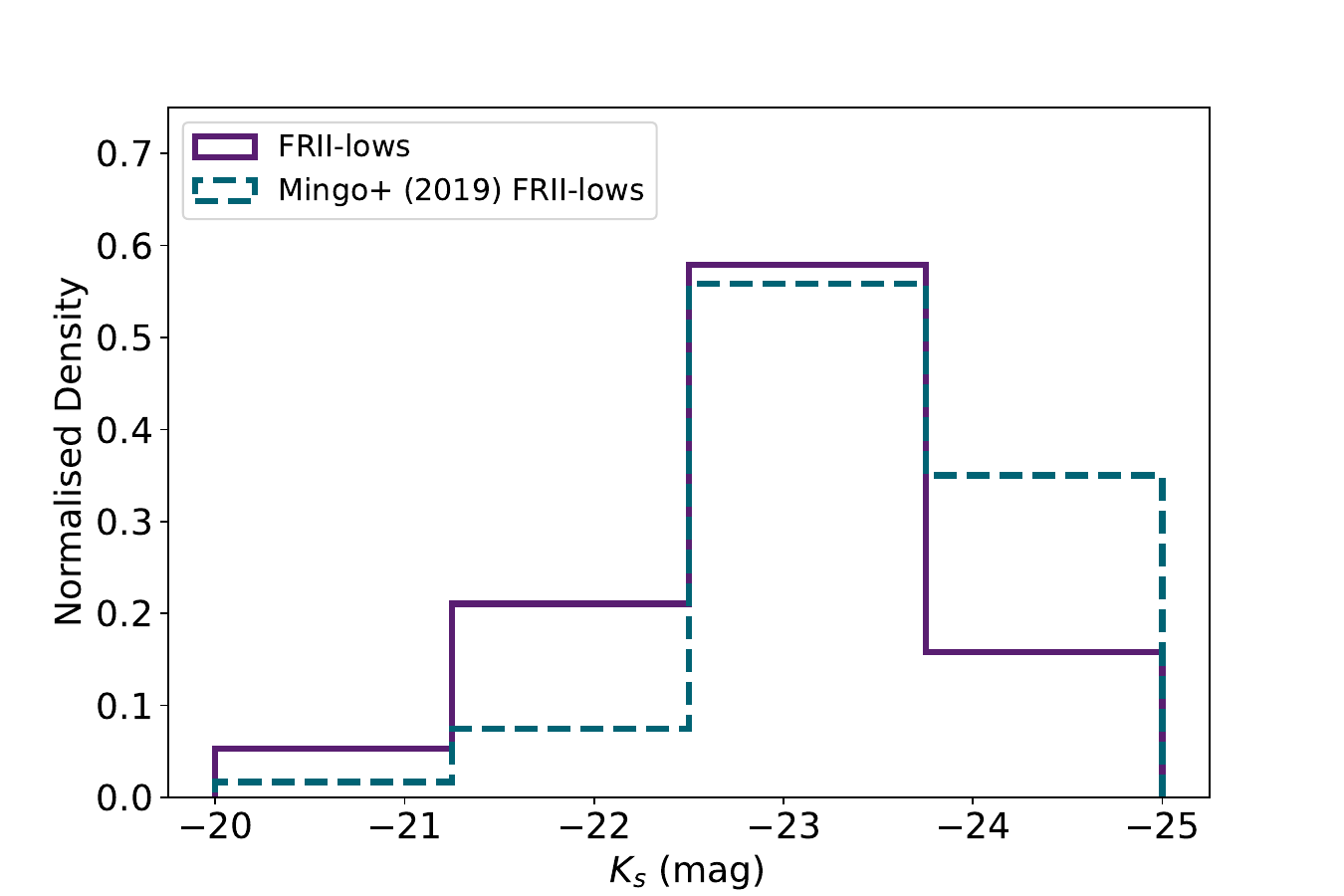}
        \includegraphics[width=0.45\linewidth]{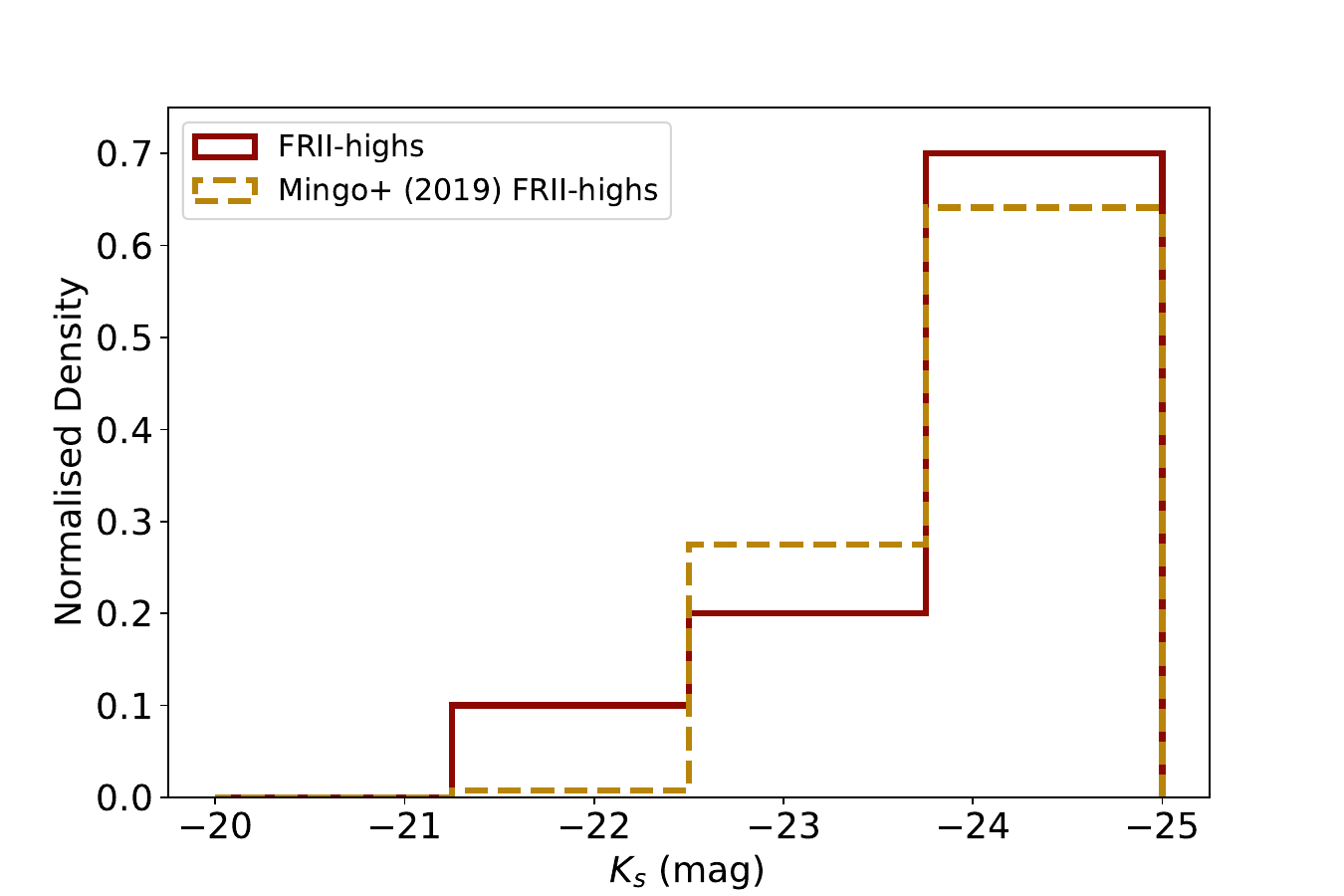}
        \subcaption{Magnitude \revb{($K_s$)}}
    \end{minipage}

    \caption{The comparison histograms of the different properties of the FRII-low and high samples, and their corresponding parent populations. The left-hand column shows the comparison of the FRII-lows and the right-hand column the comparison of the FRII-highs. \revb{The properties shown are: (a) luminosity (W Hz$^{-1}$), (b) redshift, (c) angular size (arcsec), (d) linear size (kpc), and (e) magnitude ($K_s$).}}
    \label{fig:CompPlots}
\end{figure*}

For each of these key properties \autoref{tab:medians} shows that each sample is broadly representative of their corresponding parent population. We have calculated the $p$-value from a two sample Kolmogorov-Smirnov (KS) test, given in \autoref{tab:medians}, along with the medians and standard deviations. In almost all cases the $p$-values demonstrate no significant difference between the sample and their parent population, except for redshift and angular size for the FRII-lows. \revb{The difference in angular size distributions arises from imposing a minimum angular size of $> 50$ arcsec to ensure sufficient resolution in the new data for structural analysis and reliable identification of compact features. This cut preferentially favours lower redshifts.} Apart from this selection-driven effect, the chosen samples match the parent populations well.

\section{Source dynamics}
\label{sec:SDyn}

In order to investigate the source dynamics of the FRII-lows, and see how they compare to those of the FRII-highs, the higher frequency and higher resolution VLA images of the FRII-lows were used to identify compact features, and compared with the VLA Faint Images of the Radio Sky at Twenty-Centimeters (FIRST) \citep{becker_first_1995} images of the FRII-highs. The FIRST images were chosen because of their good match in resolution to our L-band images from the VLA (the sources in the new FRII-low subsample are too faint to obtain useful information from FIRST). To carry out our comparison we have used a similar definition to that of \citet{mullin_observed_2008} to classify compact features in the FRII-low and FRII-high subsamples, where hotspots are described as features which (i) are not part of a jet, (ii) are smaller than 10 per cent of the main axis length of the source, and (iii) have a peak brightness over ten times the off-source noise. 
\revb{Compared to their definition, w}e have chosen to use \revbb{contours at $20\times$RMS level} to determine bright regions. As our sample consists of fainter sources than those of \citet{mullin_observed_2008}, we experimented with the threshold level for defining compact features and found that the $20\sigma$ level was the optimal choice across the sample to isolate obvious hotspot and core features, and to prevent any artefacts from being selected. 

If an identified feature is less than 10 per cent of the length of the source then it is considered a compact feature. If this compact feature is at a distance from the core > 50 per cent of the source length from the centre, then we classify it as a hotspot. If it is centrally located, it is classified as a core feature. Those sources with core features are discussed further in \autoref{sec:RandR}, in the context of discussion of potential remnant and restarting behaviour. \revb{\citet{mullin_observed_2008} also define a jet as being at least 4 times as long as it is wide, and as a narrow ridge running through more diffuse emission, or a narrow feature in the inner part entering more extended emission in the outer part.}

\subsection{Locating compact features}
\label{subsec:LocComFea}

Compact features were identified in both the VLA images of the FRII-lows and the FIRST images of the FRII-highs. \revb{\autoref{tab:lowscompact} and \autoref{tab:highscompact} summarise whether the compact features can be classified as hotspots, cores, or possible jet features.} Almost all of \revb{the VLA-detected} sources were found to include regions over 20 times the RMS in brightness. 

\begin{table}
    \centering
    \caption{The occurrence of compact features in the FRII-low sample. The columns indicate \revb{whether any compact features are} present in each source, and whether they are a hotspot, core, or possible jet. \revb{The cores in ILTJ133729.25 and }ILTJ150011.27 \revb{are} discussed in the text and marked with (), \revb{likewise the hotspot in ILTJ114351.49 is discussed in the text and highlighted with an $^*$.} \revb{The low detection source ILTJ134315.98 is marked with a + and is discussed in an earlier section.} The final row gives the percentage of each feature. The uncertainties of the percentages are calculated as described in the text.} 
    \begin{tabular}{l c c c}
    \hline
    \textbf{Source Name} & \textbf{Cores} & \textbf{Hotspots} & \textbf{Jets} \\
    \hline \hline
    \revb{ILTJ105946.75+563136.4}   &\revb{-}        &\revb{$\checkmark$} &\revb{-}   \\
    \revb{ILTJ112015.05+503254.9}   &\revb{-}        &\revb{-}        &\revb{-}       \\
    ILTJ112654.44\revb{+540415.3}   & $\checkmark$   & -              & -             \\
    ILTJ113626.52\revb{+501320.3}   & $\checkmark$   & -              & $\checkmark$  \\
    \revb{ILTJ114351.49+511712.6}   &\revb{-}        &\revb{$\checkmark^*$} &\revb{-} \\
    ILTJ115011.27\revb{+534320.9}   & $(\checkmark)$ & -              & $(\checkmark)$\\
    ILTJ121623.58\revb{+254409.4}   & $\checkmark$   & $\checkmark$   & -             \\
    ILTJ130109.83\revb{+560623.4}   & $\checkmark$   & -              & -             \\
    ILTJ130605.63\revb{+555127.6}   & -              & $\checkmark$   & -             \\
    ILTJ133217.44\revb{+484221.7}   & -              & -              & -             \\
    ILTJ133729.25\revb{+481822.3}   & $(\checkmark)$ & -              & $(\checkmark)$\\
    \revb{ILTJ134315.98+553139.6$^+$}   &\revb{-}    &\revb{-}        &\revb{-}       \\
    ILTJ135152.95\revb{+521618.8}   & $\checkmark$   & $\checkmark$   & -             \\
    ILTJ135630.51\revb{+555245.1}   & $\checkmark$   & -              & -             \\
    \revb{ILTJ144644.12+492012.3}   &\revb{-}        &\revb{$\checkmark$} &\revb{-}   \\
    ILTJ144650.49\revb{+514625.3}   & $\checkmark$   & -              & $\checkmark$  \\
    ILTJ145759.29\revb{+490219.2}   & $\checkmark$   & -              & -             \\
    \revb{ILTJ145936.33+484219.8}   &\revb{-}        &\revb{-}        &\revb{-}       \\
    ILTJ150827.77\revb{+541507.1}   & $\checkmark$   & -              & $\checkmark$  \\
    \hline
    &&&\\[-1em]
    \textbf{Percentage of sample} & $47.4^{+11.2}_{-11.0}$ & $31.6^{+11.3}_{-9.5}$ & $15.8^{+10.1}_{-6.8}$ \\
    &&&\\[-1em]
    \hline
    \end{tabular}
    \label{tab:lowscompact}
\end{table}

\begin{table}
    \centering
    \caption{The occurrence of the compact features in the FRII-high sample. The columns indicate which compact feature is present in each source, and whether they are a hotspot, core, or possible jet. The final row gives the percentage of each feature. The uncertainties of the percentages are calculated as described in the text.} 
    \begin{tabular}{l c c c}
    \hline
    \textbf{Source Name} & \textbf{Cores} & \textbf{Hotspots} & \textbf{Jets} \\
    \hline \hline
    J104706\revb{+534419}   & -   & $\checkmark$ & -  \\
    J111529\revb{+560039}   & -   & -              & -  \\
    J114559\revb{+512528}   & -   & $\checkmark$   & -  \\
    J121849\revb{+502617}   & -   & -              & -  \\
    J125804\revb{+541702}   & -   & -              & -  \\
    J131508\revb{+455846}   & -   & $\checkmark$   & -  \\
    J134837\revb{+470800}   & -   & -              & -  \\
    J142420\revb{+455837}   & -   & $\checkmark$   & -  \\
    J145352\revb{+500406}   & -   & -              & -  \\
    J151830\revb{+515817}   & -   & -              & -  \\
    \hline
    &&&\\[-1em]
    \textbf{Percentage of sample} & $0.0^{+10.7}_{-00.0}$ & $40.0^{+15.6}_{-13.8}$   & $0.0^{+10.7}_{-00.0}$ \\
    &&&\\[-1em]
    \hline
    \end{tabular}
    \label{tab:highscompact}
\end{table}

\revb{In ILTJ114351.49 the 20$\sigma$ contours highlighted an additional small region alongside the small hotspot, and it is not certain if this is a possible double hotspot or a random fluctuation in the flux density level. This source has been indicated in \autoref{tab:lowscompact} with an $^*$.} Possible jet features occurred in three sources: ILTJ113626.52, ILTJ144650.49, and ILTJ150827.77. These features can even be extending from possible core features. The maps of ILTJ115011.27 and ILTJ133729.25 are dynamic range limited, causing the 20 sigma contours to not fully capture the features of the source. Visually, for ILTJ115011.27, there is a compact core and a jet feature at least in the northern lobe, and in ILTJ133729.25 there could be jet features. The \revb{cores of both have} been characterised later and used further, with the caveat that there may be contamination of the cores from the possible extension into the jet features. These compact features have been marked with a bracket in \autoref{tab:lowscompact}; in the interests of consistency, they have not been included in the systematic comparison between the FRII-lows and FRII-highs.
ILTJ133217.44 is still an edge-brightened source; however, both the furthest regions are too large for hotspots, and the inner region is too close to the host location to meet the criteria for a hotspot.

\revb{For the FIRST FRII-high images, six} out of the ten sources have detections over twenty times the image RMS. Possibly due to a worse signal-to-noise ratio (SNR) in FIRST than for LoTSS and the VLA images, and a lower resolution in the FIRST images, some of the regions are larger than 10 per cent of the source size. Therefore, although by-eye the sources are clearly edge-brightened, the structures are not small enough to be compact by the definition used here.

\subsection{Comparing compact features}
\label{subsec:ComComFea}

\autoref{tab:comparecompact} gives the summary percentages from \autoref{tab:lowscompact} and \autoref{tab:highscompact} for each compact feature within the images that have significant detections. The errors on the percentages are calculated using the Agresti-Coull \citep{AgrestiCoull1998} method using a 68 per cent confidence interval. Both samples are small in size, leading to large uncertainties on the percentages. \revb{In the FRII-low sample there were four sources with no detected compact features, giving a detection rate for compact structures of $78.9_{-10.7}^{+7.9}$ per cent. For the FRII-highs, four have detected compact features, giving a detection rate of $40.0_{-13.8}^{+15.6}$ per cent. The FIRST resolution is not sufficient to apply the 10 per cent of source length criterion, and so to ensure that the hotspot definition can be applied to all of the FIRST sample, we have relaxed this criterion to 20 per cent.} However, it is important to note that all of the compact features, including hotspots, identified in the higher-resolution VLA FRII-low sample also meet the original 10 per cent criterion of \citet{mullin_observed_2008}. \citet{mullin_observed_2008} also note that a lower resolution leads to a larger hotspot size, and only one of their 98 3CRR FRIIs lacked any hotspot features. Comparing the FRII-lows and -highs shows that compact features are very common in the FRII-lows, and that \revb{the presence of hotspots is comparable in the FRII-lows to the FRII-highs.} \revb{If we apply this relaxed criterion to the VLA images tapered to match LoTSS and FIRST resolution then the presence of hotspots remains comparable in both samples.}

The FRII-lows show a higher prevalence of compact core features. These features were less distinct in the original LoTSS images due to the lower resolution (and frequency). However, in the VLA images a mix of core features are present, including compact, unresolved cores. These cores are present in $47.4_{-11.0}^{+11.2}$ per cent of the images with detections. In the FRII-highs there are no sources that have a FIRST-detected cores. We checked the predicted core flux \revb{density} for the FRII-high sources to see if their core prominence (see \autoref{subsec:CP}) was the same as the median for the FRII-lows, and found that we would have been able to detect similarly prominent cores if they were present. 

\revb{Using the same detection criteria (>5 RMS), the VLA FRII-low sample, it is clear that the LoTSS FRII-high sample have typical hotspot morphology} and there is only a single source with a possible core. Even considering the uncertainties of a sample having no cores, the FRII-lows show a higher prevalence of cores over the FRII-highs. ILTJ115011.27 does not show a core using the definition given earlier as it is too bright. However, by eye it has a compact core component. This has not been included in the comparison and does not affect the conclusions about core prevalence, but it is used later in \autoref{sec:RandR} when assessing remnants and restarters.

As previously mentioned, three sources or $15.8_{-6.8}^{+10.1}$ per cent of the FRII-lows have possible jet features. There are none in the FRII-high sample, but FIRST's resolution and sensitivity makes jet detection challenging. The VLA morphologies for the three FRII-lows with jets could be examples where the morphological classification of a source can change at different frequencies (for example where inner jet features are flatter spectrum than lobes). Having a brighter central region, and in the case of ILTJ150827.77 a clear variation between lobes, would change the classification from FRII at low frequency to hybrid or an FRI when observed at somewhat higher frequency.

\begin{table}
    \centering
    \caption{A comparison of the percentage occurrence of compact features in the FRII-lows and -highs. The calculation of the uncertainties is described in \autoref{subsec:ComComFea}.}
    \begin{tabular}{l c c c}
    \hline
    ~ & \textbf{Cores} & \textbf{Hotspots} & \textbf{Jets} \\
    \hline
    \hline
    &&&\\[-1em]
    \textbf{FRII-lows} & \revb{$47.4^{+11.2}_{-11.0}$} & \revb{$31.6^{+11.3}_{-9.5}$} & \revb{$15.8^{+10.1}_{-6.8}$} \\
    &&&\\[-1em]
    \hline
    &&&\\[-1em]
    \textbf{FRII-highs} & $0.0^{+32.1}_{-00.0}$ & \revb{$40^{+15.6}_{-13.8}$}   & $0.0^{+32.1}_{-00.0}$ \\
    ~&~&~&~\\[-0.8em]
    \hline
    \end{tabular}
    \label{tab:comparecompact}
\end{table}

\section{Spectral Indices}
\label{sec:SpI}

Spectral index analysis is useful for distinguishing between active and remnant sources where the jet has turned off. As remnant behaviour is one of the possible scenarios for the origin of FRII-lows, we explored this further using the available spectral information. We chose to compare the sample of FRII-lows to typical active radio galaxies by calculating their integrated spectral indices ($\alpha$, where radio flux density $S_{\nu} \propto \nu^{-\alpha}$). \citet{mingo_revisiting_2019} demonstrated that over half of their FRII-low sample for which spectral information was available have integrated spectral indices from LoTSS to NVSS between $\sim$0.7 and 1. To carry out a spectral index analysis we generated regions for each source based on the 3$\sigma$ contours on the LoTSS images. Using these regions, we then used the \texttt{RADIOFLUX}\footnote{\url{https://github.com/mhardcastle/radioflux}} tool to calculate the total flux \revb{density within in contours on the} VLA images that were reimaged at 6-arcsec resolution, matched to LoTSS. 
$\alpha_{150}^{1500}$ was calculated for the sample using the total flux densities from LoTSS and the VLA, and the results are given in \autoref{tab:SpI}.
The uncertainties on the integrated spectral indices are calculated using the standard error propagation formula, where the flux density calibration errors are $\sim$5 per cent for the VLA L-band\footnote{\url{https://science.nrao.edu/facilities/vla/docs/manuals/oss2019A/performance/fdscale}} and $\sim$ 20 per cent for LoTSS \citep{shimwell_lofar_2019}. As these values are constant this gives an uncertainty of $\pm$0.1 for all of the spectral indices.



Of the 19 sources, \revb{one image, ILTJ134315.98, had very faint detection from the VLA observation.  This faint detected structure became better defined in the tapered 6 arcsec image.} When calculating $\alpha_{150}^{1500}$ using the total flux density within the LoTSS 3$\sigma$ contours overlaid on this images, the value can only be a lower limit. We \revb{placed lower and upper bounds} on the 1.5-GHz flux \revb{density} based on the detected flux \revb{density} level within the source region, and this detected level plus 3 times the RMS (this is indicated in \autoref{tab:SpI} with a range for the spectral index).

The remaining sources have integrated spectral indices that lie in the range of $\sim$0.5 to 1.0, which is typical for active radio galaxies. As per \citet{mingo_revisiting_2019}'s findings using NVSS for the parent FRII-highs, over half of the sample have an $\alpha$ value in the range of 0.7 to 1.0, matching the typical values expected for FRII-highs from the literature.

\autoref{tab:SpI} also shows the integrated spectral indices for the cores and lobes of the sources.  The core region was determined by placing a 3 arcsec radius circle over the optical host location to coincide with the beam size and resolution of LoTSS and the tapered images. The \revb{total flux density in this} region was then subtracted from the total flux \revb{density in 3$\sigma$ region} to give the \revb{total} lobe flux \revb{density}. The three sources with the highest core spectral index values (ILTJ113626.52, ILTJ133729.25, and ILTJ144650.49) have been highlighted in bold in \autoref{tab:SpI}. These values are steeper than both the lobe and total spectral index for each source, and also unusually steep for active radio galaxy cores. This is discussed further in \autoref{sec:RandR}.

\begin{table}
    \centering
    \caption{The integrated spectral indices of the sample of FRII-lows. The columns show the integrated spectral indices for the whole source, the lobes, and the cores, calculated from LoTSS to the tapered VLA images as described in \autoref{sec:SpI}.  As described in the text the uncertainty on these values is calculated as $\pm$0.1, and the three bold entries highlight the sources with steeper cores. \revb{ILTJ134315.98+553139.6 has a marginal detection (as described in the text, and is indicated with a $^+$), and a range of possible spectral indices}.}
    \begin{tabular}{l c c c}
        \hline
        &&&\\[-1em]
        \textbf{Source} & $\alpha_{150 MHz}^{1500 MHz}$ & \textbf{Lobe SpI} & \textbf{Core SpI} \\
        &&&\\[-1em]
        \hline \hline
        ILTJ105946.75\revb{+563136.4}     & \revb{0.74} & -     & -     \\
        ILTJ112015.05\revb{+503254.9}     & \revb{0.80} & -     & -     \\
        ILTJ112654.44\revb{+540415.3}     & 0.5        & 0.57  & 0.20  \\
        \textbf{ILTJ113626.52}\revb{+501320.3}     & \textbf{0.6 }       & \textbf{0.59}  & \textbf{0.64}  \\
        ILTJ114351.49\revb{+511712.6}     & \revb{0.72} & -     & -     \\
        ILTJ115011.27\revb{+534320.9}     & 0.5        & 0.55  & 0.34  \\
        ILTJ121623.58\revb{+524409.4}     & 0.6        & 0.58  & 0.09  \\
        ILTJ130109.83\revb{+560623.4}     & 0.7        & 0.71  & 0.63  \\
        ILTJ130605.63\revb{+555127.6}     & 0.9        & -     & -     \\
        ILTJ133217.44\revb{+484221.7}     & 1.0        & -     & -     \\
       \textbf{ILTJ133729.25}\revb{+481822.3}     & \textbf{0.6}        & \textbf{0.61}  & \textbf{0.75}  \\
        ILTJ134315.98\revb{+553139.6}$^+$ & 1.24 - 1.54 & -     & -     \\
        ILTJ135152.95\revb{+521618.8}     & 0.7        & 0.68  & 0.54  \\ 
        ILTJ135630.51\revb{+555245.1}     & 0.7        & 0.67  & 0.53  \\   
        ILTJ144644.12\revb{+492012.3}     & \revb{0.71} & -     & -     \\ 
        \textbf{ILTJ144650.49}\revb{+514625.3}     & \textbf{0.6}        & \textbf{0.65}  & \textbf{0.70}  \\   
        ILTJ145759.29\revb{+490219.2}     & 0.6        & 0.65  & 0.33  \\  
        ILTJ145936.33\revb{+484219.8}     & \revb{1.34} & -     & -     \\ 
        ILTJ150827.77\revb{+541507.1}     & 0.6        & 0.59  & 0.20  \\  
        \hline
    \end{tabular}
    \label{tab:SpI}
\end{table}

\section{Evolution of sources}
\label{sec:RandR}

\citet{mingo_revisiting_2019} suggest that a possible explanation for the origin of some of the FRII-lows is that they are part of an older, fading population, i.e., they are in the remnant or restarting phase of their life cycle. During this period the typical indicators of active radio galaxies, such as cores, jets and hotspots are expected to have disappeared or to be highly diminished in brightness, and the expanding and ageing emission in the extended lobes will cause them to dissipate and fade. It can be hard to identify whether a source should be considered a remnant radio galaxy candidate. Initially individual cases were studied \citep[e.g.,][]{cordey_ic_1987, brienza_lofar_2016, randriamanakoto_j16155452_2020} or samples of candidates selected on a single criterion \citep[e.g.,][]{parma_search_2007, saripalli_atlbs_2012}. As our sample is from LoTSS we have chosen to investigate their remnant nature using a method similar to that of \citet{jurlin_multi-frequency_2021}, \revb{whose criteria are based on a combination of studies by e.g. \citet{brienza_search_2017}, \citet{godfrey_population_2017}, \citet{jurlin_life_2020}, and \citet{shabala_duty_2020}, and are complementary to the work of e.g. \citet{morganti_combining_2021}.}

Four criteria are considered in these studies when assessing whether a source is a remnant candidate. These are:

\begin{itemize}
    \item Ultra-steep integrated spectral index
    \item Spectral curvature criterion $\ge$ 0.5
    \item Lower than expected core prominence for active radio galaxies
    \item Relaxed morphology
\end{itemize}
Each criterion is covered in detail in the following sections.

\subsection{Ultra-steep spectrum sources}
\label{subsec:USS}

Ultra-steep spectrum (USS) sources are defined as having indices steeper than $\sim$1.2 \citep{komissarov_relic_1994}. There is no replenishment of new electrons, and the spectrum reflects the losses from radiation and expansion. It is common to see a steeper spectrum above 1.5-GHz due to the higher radiative loss rate of higher energy electrons, but when it is also present at the lowest frequencies it is a sign of a fairly old electron population. Therefore, a presence of a steeper spectrum at lower frequencies indicates an aged source. As such an USS is often the criterion by which a remnant candidate is chosen.

\begin{figure}
    \centering
    \includegraphics[width=\linewidth]{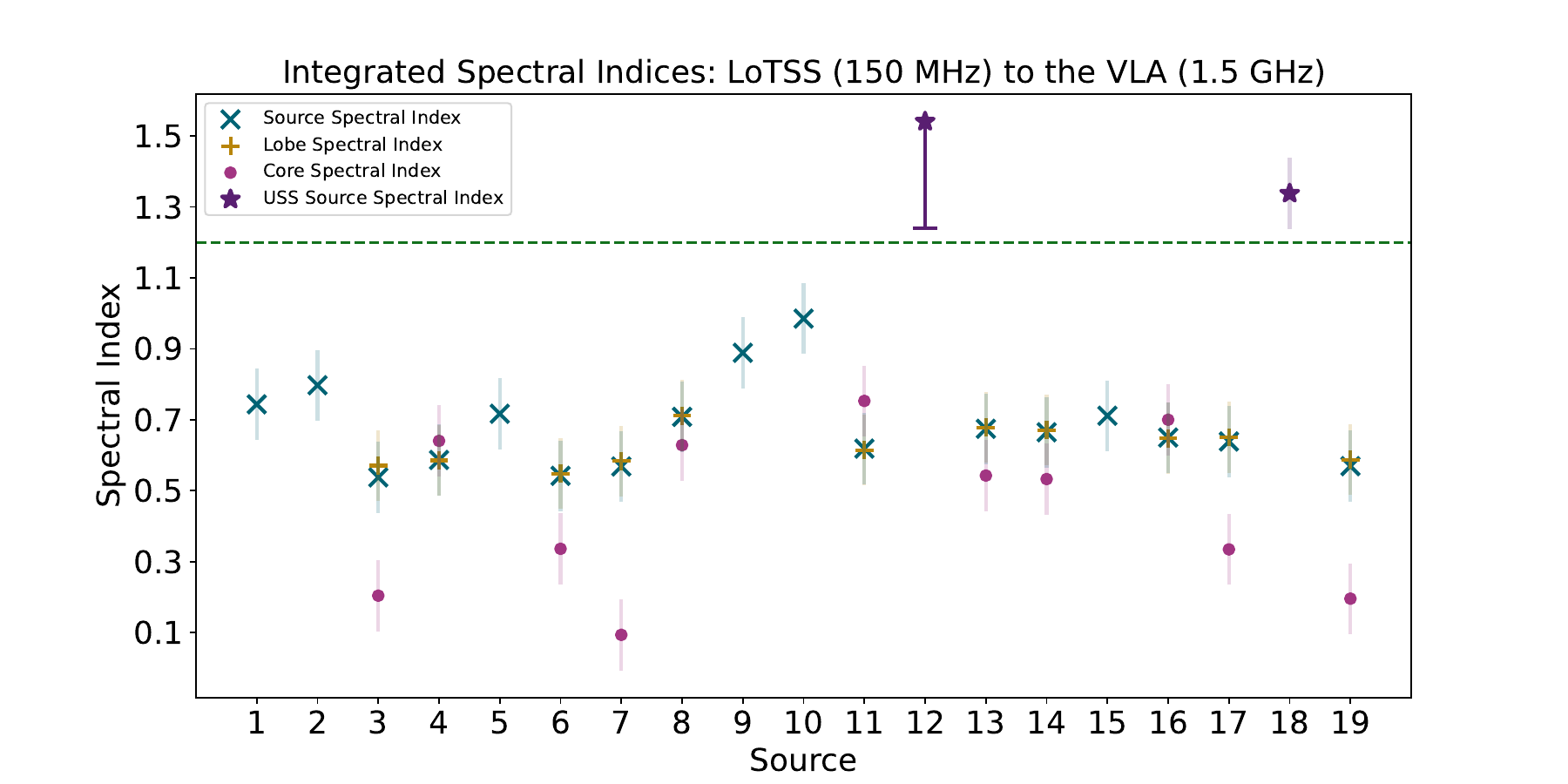}
    \caption{The spectral indices as calculated in \autoref{sec:SpI} for each of the 19 sources. As described in \autoref{sec:SpI} the uncertainty on these values is calculated as $\pm$0.1, and shown with a faint line. The green dashed line is at the value $\alpha = 1.2$, which defines an ultra-steep-spectrum source \citep{komissarov_relic_1994}.  Bounding bars (purple) are given on the \revb{faint detection source} as described in \autoref{sec:SpI}. As these are limits the uncertainties have not been shown on this source. The value of the spectral indices for the cores and lobes are also shown.}
    \label{fig:SpIfig}
\end{figure}

\autoref{fig:SpIfig} shows the total integrated spectral indices for the 19 sample sources, \revb{which are also} listed in \autoref{tab:SpI}. If the USS criterion is taken as $>$1.2 the sources with spectral indices over this boundary are ILTJ134315.98, and ILTJ145936.33. This gives two possible remnant candidates based on the USS criterion.

\subsection{Spectral curvature}
\label{subsec:SC}

The higher energy electrons will radiate their energy away faster, so steeper spectral indices are expected to be found at higher frequencies. The spectral curvature (SPC) criterion looks to find a significant difference between a total integrated spectral index at a higher frequency and a lower one. This would indicate a recent decrease in particle acceleration that is not yet observable at lower frequencies \citep{brienza_search_2017, jurlin_multi-frequency_2021}. 

To explore the possibility of a significant spectral curvature across different frequencies, a set of images at a different, higher frequency was required. We considered use of the VLA Sky Survey (VLASS) quick look images\footnote{\url{http://cutouts.cirada.ca}}. VLASS is a 2 – 4 GHz VLA survey with a resolution of $\sim$2.5 arcsec\footnote{\url{https://science.nrao.edu/vlass}}, and all the images were taken from VLASS Epoch 2. However, as our sources are all larger than 30 arcsec and thus their extended emission is not fully sampled in VLASS, we have chosen to drop this criterion for remnant detection. We believe this will have little impact, as in the results of \citet{jurlin_multi-frequency_2021} none of their selected remnant candidates satisfied their SPC criteria. It is also worth noting that we have used a lower frequency range to calculate the SPC, making it harder to detect curvature.



\subsection{Core prominence}
\label{subsec:CP}

The core prominence (CP) of an extended radio galaxy can act as an indication of the level of activity of the central region of the extended source. If there is a core present there could be a low level of activity, or the source could be a restarter. The CP is the ratio of the higher frequency core flux \revb{density} to the low frequency total flux \revb{density}. Given this, any compact source such as \revb{compact symmetric objects (CSOs) and Giga-Hertz Peaked Spectrum (GPS) sources} will have high CPs. The CP can also be affected by relativistic beaming. This means it is more reliable to consider the CPs of the sample as a whole rather than looking at individual sources. Typical values can range from 0.001 to 0.1 for active radio galaxies with the most powerful ones being as low as $\sim 10^{-4}$, and the highest values $>$ 0.1 being attributed to FR0s \citep{giovannini_radio_1988, mullin_observed_2008, Baldi2015, brienza_search_2017}.

\begin{figure}
    \centering
    \includegraphics[width=\linewidth]{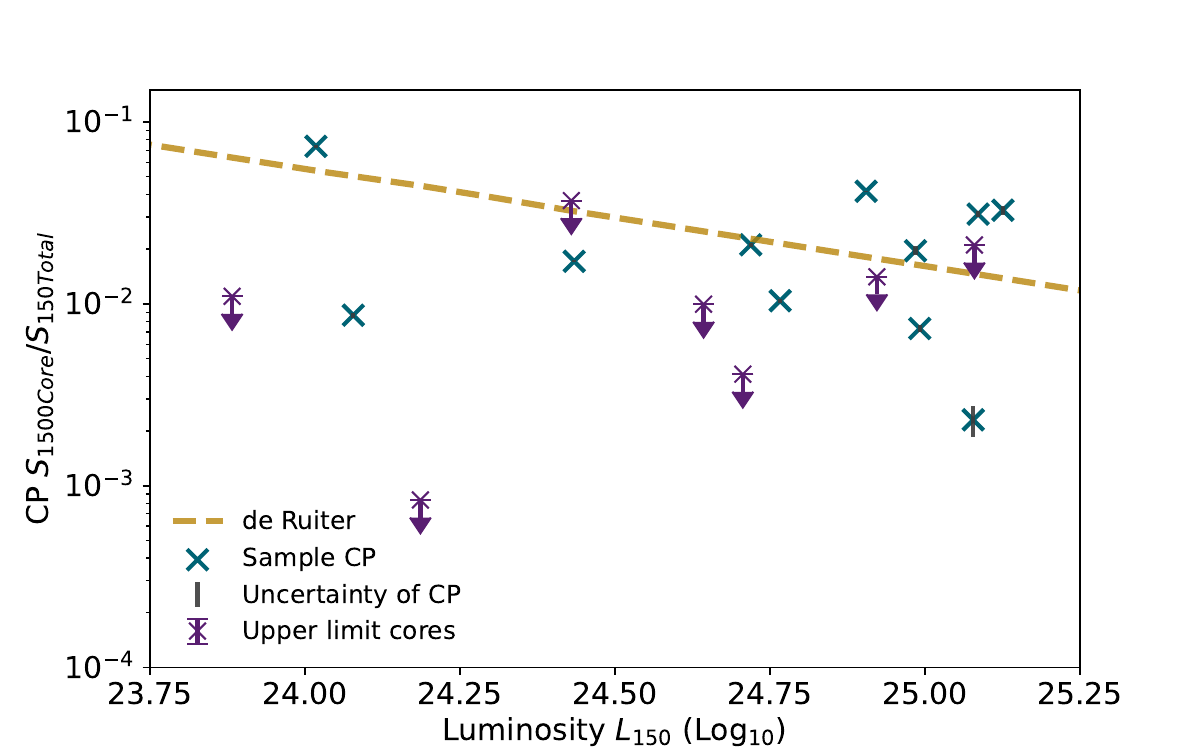}
    \caption{The core prominence of the sources, as defined in \autoref{subsec:CP} plotted against luminosity and the \citet{de_ruiter_vla_1990} relationship. The uncertainties are calculated as described in the text. The yellow dashed line is taken from \citet{jurlin_multi-frequency_2021} and is adjusted for a modern value of the Hubble constant. For those sources without a defined core, an upper limit is plotted as defined in the text.}
    \label{fig:CPfig}
\end{figure}

We have used $\text{CP} = S_{1500\_\text{core}} / S_{150\_\text{source}}$ for all the sources. The total flux \revb{density} for the source was calculated as described in \autoref{sec:SpI}, and the flux \revb{density} for the core was calculated using a 6 arcsec circular region centred on the host location. As in \autoref{sec:SpI} this region was chosen to match the beam size and resolution of the LoTSS images. For some of the sources this matching of size may allow contamination from lobe or jet structures near to the core. For \revb{the seven} sources where a core was not detected in the VLA images, an upper limit was calculated, either using the flux \revb{density} in the defined region or the 3 $\times$ RMS flux \revb{density} if higher. \citet{de_ruiter_vla_1990} showed that there is an inverse relationship between luminosity and CP in the B2 sample. This was adjusted by \citet{jurlin_multi-frequency_2021} for a modern cosmology. Figure \ref{fig:CPfig} shows this adjusted relationship and our calculated CP values for all the sources.

It can clearly be seen that the majority of the sources lie within a reasonable scatter from the line. As the line is a calculated median of the previous data this is to be expected. There is one source \revb{with a core} which is a substantial distance away from this relationship, ILTJ113626.52 (on the lower left), and lies clustered within the upper limits of two \revb{of the} sources with no detected cores. \revb{This group consists of the two sources ILTJ130605.63 and ILTJ133217.44 neither of which has an indication of a core in either of the VLA images or the LOFAR images.  The majority of the source CPs are still within the range [0.001, 0.1] and can thus be considered to be active radio galaxies, with the exception of ILTJ133217.44 where the upper limit lies just outside this range. This indicates that, if cores were present for the sources with upper limits, they should have been detected. A second source, ILTJ145936.33 (on the lower right), can also be considered to lie just outside the main group.}

\subsection{Morphology}
\label{subsec:Morph}

The traditional definition of a remnant radio galaxy is an object that a relaxed morphology with an absence of compact features such as hotspots, cores, and jets \citep{blundell_nature_1999, wang_cosmological_2008}. Determining that a source is a remnant radio galaxy from its morphology can be complicated because of the shape of the radio galaxy during its active phase and the way in which expansion may cause a relaxed morphology to develop. The varying resolution and data quality of images also make it difficult to make clear distinctions about the morphology. However, morphology remains a useful criterion to select remnant candidates \citep{saripalli_atlbs_2012}. \citet{brienza_search_2017} considered remnant candidates to have a relaxed morphology with a low surface brightness in their low frequency images ($SB_\text{{150-MHz}}$ < 50 mJy arcmin$^{-2}$), and an absence of any compact features in both their LOFAR and 5-arcsec FIRST images. As the choice of an appropriate surface brightness threshold is arbitrary, \citet{jurlin_multi-frequency_2021} opted not to use it in their selection process.

\revb{From the VLA observations, all except four have a compact feature such as a hotspot, core, or possible jet. Of the remaining sources, ILTJ133217.44 has bright features that are too large to be defined as hotspots under our definition, but does have classic FRII morphology. The three remaining sources left in the sample meet the criteria of relaxed morphology in the LOFAR images and no compact features at both low or high frequencies: ILTJ112015.05, ILTJ134315.98 and ILTJ145936.33.}

\begin{figure}
    \centering
    \includegraphics[width=\linewidth]{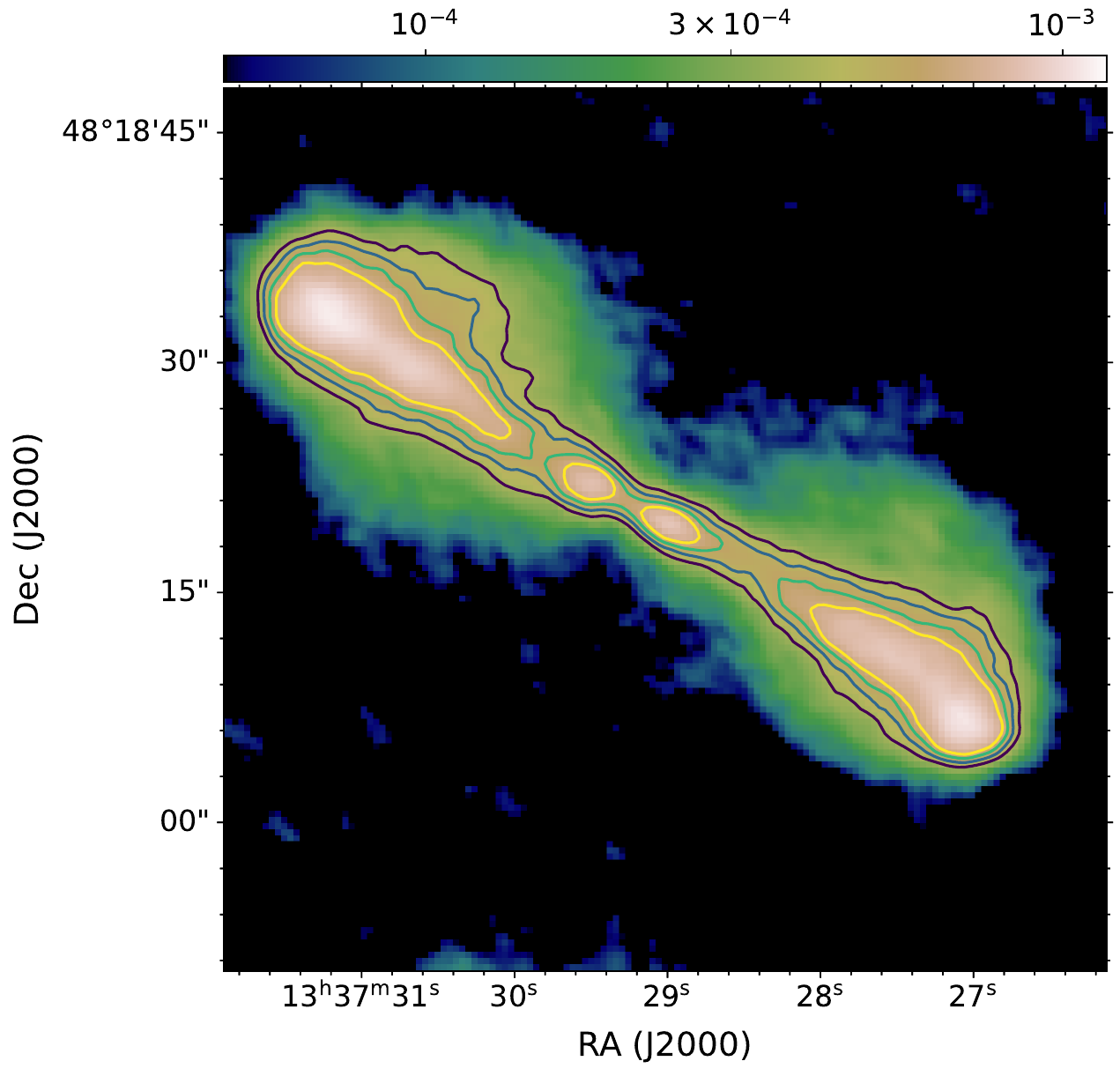}
    \caption{The 1.3-arcsec VLA image of ILTJ133729.25 with increasing sigma contours from 20$\sigma$ up to 35$\sigma$ in steps of 5$\sigma$ to show the presence of a candidate double lobe morphology. \revb{The colour bar shows the flux density in \revbb{mJy/beam.}}}
    \label{fig:35sig}
\end{figure}

Morphology is used to establish restarting activity, especially in the case of double-double radio galaxies (DDRG). In this type of radio sources we see a pair of inner lobes surrounded by an outer pair of faded, extended lobes \citep{schoenmakers_radio_2000, konar_episodic_2013, kuzmicz_optical_2017, mahatma_lotss_2019}. Based on the criteria in \autoref{subsec:LocComFea}, and from its morphology, ILTJ133729.25 can be considered a restarting candidate, due to the double-double morphology that can be identified in the higher resolution VLA images. The original $20\sigma$ contours did not define these regions in the dynamic-range limited image; however as shown in \autoref{fig:35sig} the 35$\sigma$ contours highlight these two putative inner lobes. ILTJ133729.25 is unlikely to be an FRI with inner lobes as this would be unlikely to present, in maps of this resolution, as two very symmetric peaks with a gap in between.

Two other sources, ILTJ113626.52 and ILTJ144650.49, also demonstrate a brighter, elongated core and potentially separated hotspots or jet features, which could morphologically indicate restarting sources.
These two sources, together with ILTJ133729.25, discussed in the previous paragraph, comprise the three sources with bold highlights in \autoref{tab:SpI}, which are those systems in which the core spectral index is steeper than that of both the source and the lobes. These are the highest core spectral indices in the sample, and unusually high for the cores of active FRIIs. The steeper spectral indices are a signature of recent activity at the core, since they indicate optically thin, and therefore physically larger, emitting regions, providing further evidence that these sources are restarting candidates.

\subsection{Summary of remnant and restarting results}
\label{RRSum}

As noted above, classifying a radio galaxy as a remnant or restarting source is challenging, and there are many different criteria by which radio sources can be assessed.

\begin{table*}
    \centering
    \renewcommand{\arraystretch}{1.5} 
    \caption{Summary of the individual sources meeting each criteria for remnant and restarting classification.  The sources discussed in each subsection are listed in the first column, and the criteria they satisfy are given in the second column.  Any key point related to the sources and discussed in each subsection are listed in the final column.}
    \begin{tabular}{p{3.5cm} p{3.5cm} p{7cm}}
    \hline
    \textbf{Source Name} & \textbf{Criteria} & \textbf{Key Points} \\
    \hline \hline
    ILTJ112015.05\revb{+503254.9}     &    Morphology                 &     Amorphous morphology.                          \\
    ILTJ113626.52\revb{+501320.3}     &    Morphology \newline SpI \newline CP &  Morphology indicates a possible restarter. \newline Spectral index of the core is greater. \newline CP still in range for an active radio galaxy.      \\
    ILTJ133729.25\revb{+481822.3}     &    Morphology \newline SpI    &     DDRG - Two clear inner lobes. \newline Spectral index of the core is greater.             \\
    ILTJ134315.98\revb{+553139.6}     &    Morphology \newline USS    &     Amorphous morphology.\newline Integrated spectral index is a lower limit.                        \\
    ILTJ144650.49\revb{+514625.3}     &    Morphology \newline SpI    &     Morphology indicates a possible restarter.\newline Spectral index of the core is greater.          \\
    ILTJ145936.33\revb{+484219.8}     &    Morphology \newline USS \newline CP    &     Amorphous morphology. \newline Integrated spectral index is a lower limit. \newline CP lies slightly outside the group range.                        \\
    \hline
    \end{tabular}
    \label{tab:R&Rs}
\end{table*}

From the summary in \autoref{tab:R&Rs} it is clear that there are \revb{three} sources which can be considered remnant candidates: ILTJ112015.05, ILTJ134315.98, and ILTJ145936.33. Two of these sources meet the USS criterion, and all of them can be classified as having a relaxed or amorphous morphology. There are also three possible restarting candidates: ILTJ113626.52, which was noted for its low outlying CP and possible restarting morphology; ILTJ133729.25 which has a double-double morphology; and ILTJ144650.49 which has a possible restarting morphology. Supporting this conclusion, these three sources are also noted in \autoref{sec:SpI} to have high core spectral indices, which are comparatively steep (within the typical active galaxy range) and steeper than their total spectral index. Therefore, approximately \revb{$15.8_{-11.1}^{+22.6}$ per cent (three)} are possible remnant candidates and $15.8_{-11.1}^{+22.6}$ per cent (three) of the population are restarting candidates. The remaining objects must be considered to be normal, active sources.

\section{Discussion and Conclusions}
\label{sec:Conc}

\cite{mingo_revisiting_2019} discovered a population of low luminosity FRIIs in LoTSS DR1. In this paper we have explored the properties of this population using a sample of 19 sources at least one order of magnitude below the traditional FR luminosity break. These sources were observed with the VLA in the L-Band using the A, B, and C configurations, resulting in the first systematic high frequency (1.5 GHz), higher resolution (1.3 arcsec) study of this population. We compared the radio morphologies of the FRII-low sample in the new high-resolution, higher frequency images with their LOFAR morphologies, and with a representative sample of LOFAR-selected FRII-highs to provide a comparison sample of the traditional FRII population. With these images we have looked at the presence of compact features, their spectral properties and used these to draw the following conclusions:

\begin{itemize}
    \item Compact features (cores) are very common in the FRII-lows, and more common than in the FRII-highs.
    \item Hotspots appear to be \revb{equally} prevalent in the FRII-lows and the FRII-highs.
    \item FRII-lows have a higher prevalence of cores compared to the FRII-highs, and the FRII-lows have jet features, whereas there are none in the FRII-highs. However, we do not have comparable sensitivity in our FRII-high images for a rigorous comparison of prevalence.
    \item Roughly 50 per cent of the FRII-low sources have total spectral indices of 0.7 - 1.0, which are typical for active galaxies; in other words, the FRII-low population does not consist exclusively of remnant sources.
    \item We have found \revb{$\sim16$ per cent (three)} FRII-low sources which are remnant candidates and $\sim16$ per cent (three) sources which are restarting candidates. The remaining objects are most likely standard, active sources.
    \item Along with the three restarting candidates, $\sim32$ per cent (six) of the FRII-low sources have clear FRII morphology at higher frequency and resolution (ILTJ115011.27, ILTJ121623.58, ILTJ130605.63, ILTJ133217.44, ILTJ135152.95, and ILTJ150827.77).
\end{itemize}

To summarize, our new higher-frequency, higher-resolution images for a sample of 19 RLAGN classed as low-luminosity FRIIs via LoTSS have demonstrated a diverse population that includes active FRIIs, restarting and remnant sources, as well as some sources that appear more FRI-like at higher frequency. Our results highlight that (not unexpectedly) frequency and resolution do affect the morphological classification of a source: edge-brightened sources at low frequency and comparatively low resolution may appear as hybrid or FRI sources when observed at a higher frequency and/or resolution \citep{rudnick_radio_2021-1}, such as ILTJ112654.44. However, we find that there remains a significant proportion of RLAGN below the traditional FR luminosity break that have morphologies strongly indicative of FRII dynamics: they have compact hotspots and demonstrate integrated spectral indices consistent with those of an active galaxy. This population could be the result of lower powered jets in a minimally dense environment that allows them to remain relativistic to larger distances, thus producing the edge-brightened morphology of the FRIIs \citep[e.g.][]{mingo_accretion_2022, clews_radio-loud_2025}. As \cite{mingo_revisiting_2019} argued, it is no longer safe to assume that only high-luminosity radio sources have FRII-like jet physics. 



\section*{Acknowledgements}

BB acknowledges support from the UK Research and Innovation Science and Technology Facilities Council (UKRI STFC) under the grant ST/X002543/1. JHC and BM acknowledge support from the UKRI STFC under grants ST/X001164/1 and ST/T000295/1. BM further acknowledges support from UKRI STFC for an Ernest Rutherford Fellowship [grant number ST/Z510257/1]. MJH acknowledges support from the UKRI STFC under grant ST/Y001249/1.

The National Radio Astronomy Observatory and Green Bank Observatory are facilities of the U.S. National Science Foundation operated under cooperative agreement by Associated Universities, Inc.

LOFAR is the Low Frequency Array designed and constructed by ASTRON. It has observing, data processing, and data storage facilities in several countries, which are owned by various parties (each with their own funding sources), and which are collectively operated by the LOFAR ERIC under a joint scientific policy. The LOFAR resources have benefited from the following recent major funding sources: CNRS-INSU, Observatoire de Paris and Université d'Orléans, France; BMBF, MIWF-NRW, MPG, Germany; Science Foundation Ireland (SFI), Department of Business, Enterprise and Innovation (DBEI), Ireland; NWO, The Netherlands; The Science and Technology Facilities Council, UK; Ministry of Science and Higher Education, Poland; The Istituto Nazionale di Astrofisica (INAF), Italy.

This research made use of the Dutch national e-infrastructure with support of the SURF Cooperative (e-infra 180169) and the LOFAR e-infra group. The Jülich LOFAR Long Term Archive and the German LOFAR network are both coordinated and operated by the Jülich Supercomputing Centre (JSC), and computing resources on the supercomputer JUWELS at JSC were provided by the Gauss Centre for Supercomputing e.V. (grant CHTB00) through the John von Neumann Institute for Computing (NIC).

This research made use of the University of Hertfordshire high-performance computing facility and the LOFAR-UK computing facility located at the University of Hertfordshire and supported by STFC [ST/P000096/1], and of the Italian LOFAR IT computing infrastructure supported and operated by INAF, and by the Physics Department of Turin university (under an agreement with Consorzio Interuniversitario per la Fisica Spaziale) at the C3S Supercomputing Centre, Italy.

This research is part of the project LOFAR Data Valorization (LDV) [project numbers 2020.031, 2022.033, and 2024.047] of the research programme Computing Time on National Computer Facilities using SPIDER that is (co-)funded by the Dutch Research Council (NWO), hosted by SURF through the call for proposals of Computing Time on National Computer Facilities. 
\section*{Data Availability}

The LOFAR data underlying this work \revb{are} part of the LoTSS DR2 survey, which can be accessed from \url{https://lofar-surveys.org/dr2_release.html}. The VLA data used in this work \revb{are} publicly available in the National Radio Astronomy Observatory (NRAO) Science Archive under proposal code [19A-151]. The image maps are available on request from the authors.



\bibliographystyle{mnras}
\bibliography{FRIILowsPaper.bib} 




\appendix

\section{Features}
\label{App:Features}

\begin{figure*}

    \begin{minipage}{0.24\textwidth}
    \centering
        \includegraphics[width=\linewidth]{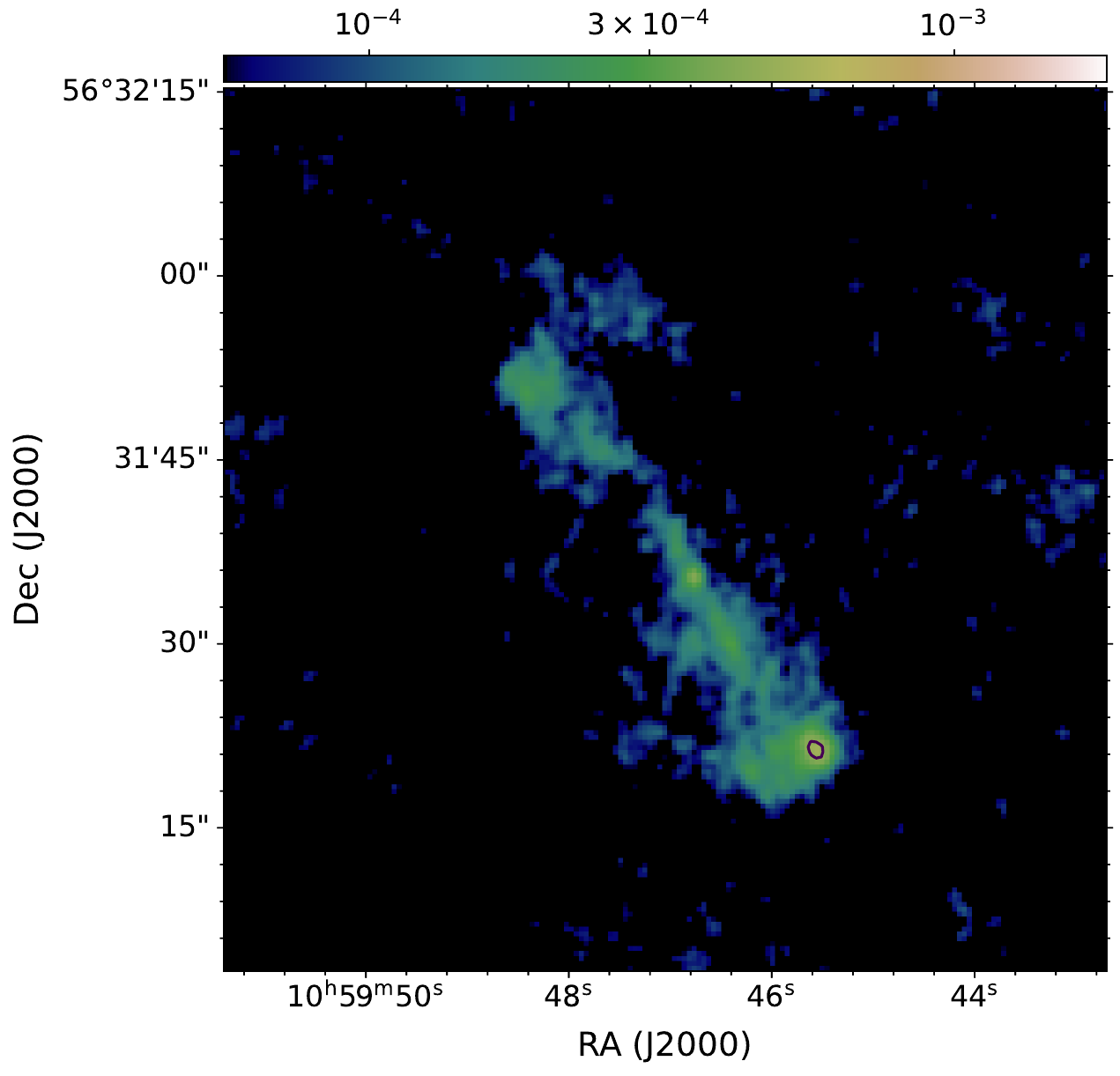}
        \subcaption{ILTJ105946.75+563136.4}
    \end{minipage}
    \begin{minipage}{0.24\textwidth}
    \centering
        \includegraphics[width=\linewidth]{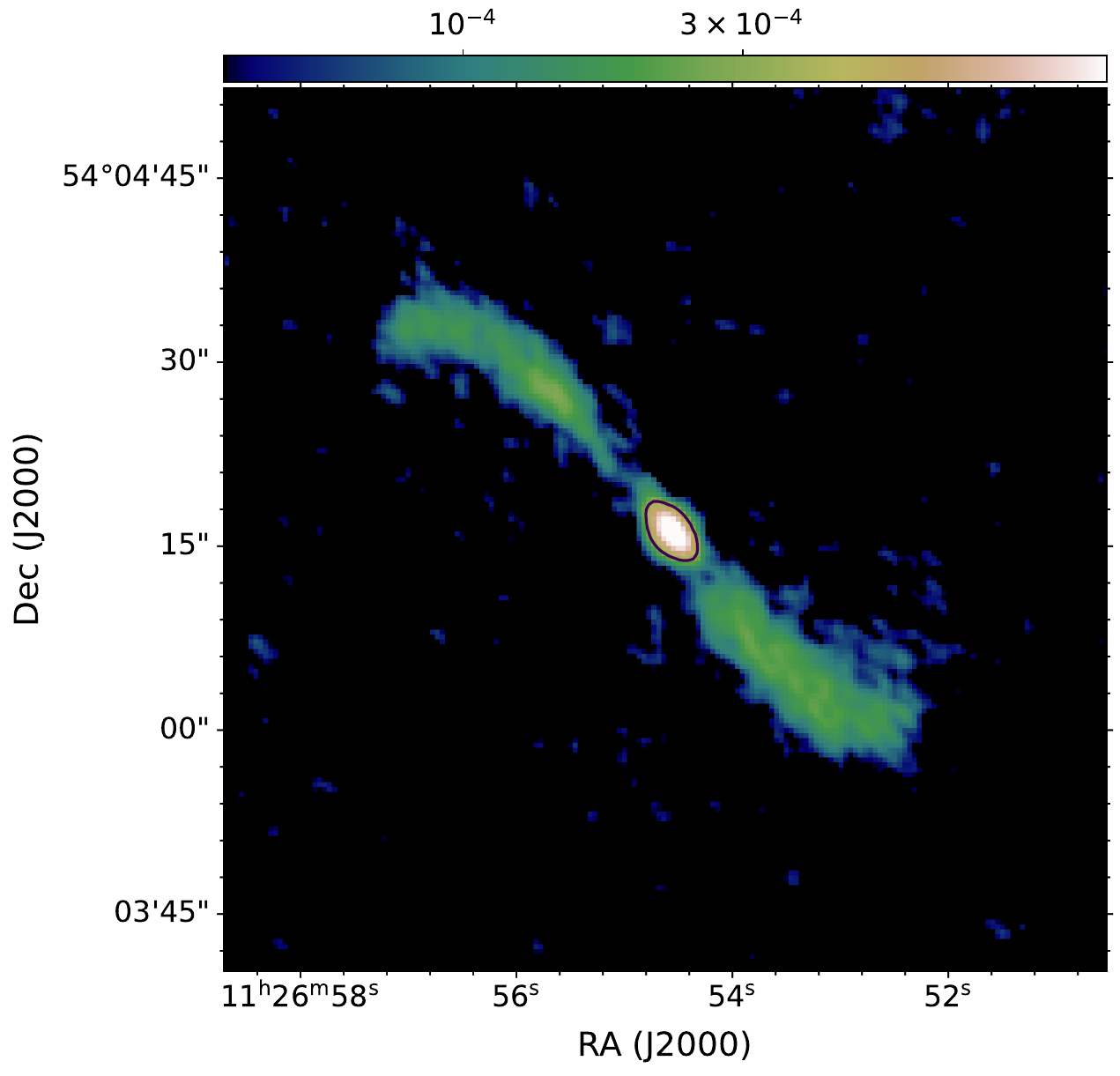}
        \subcaption{ILTJ112654.44+540415.3}
    \end{minipage} 
    \begin{minipage}{0.24\textwidth}
    \centering
        \includegraphics[width=\linewidth]{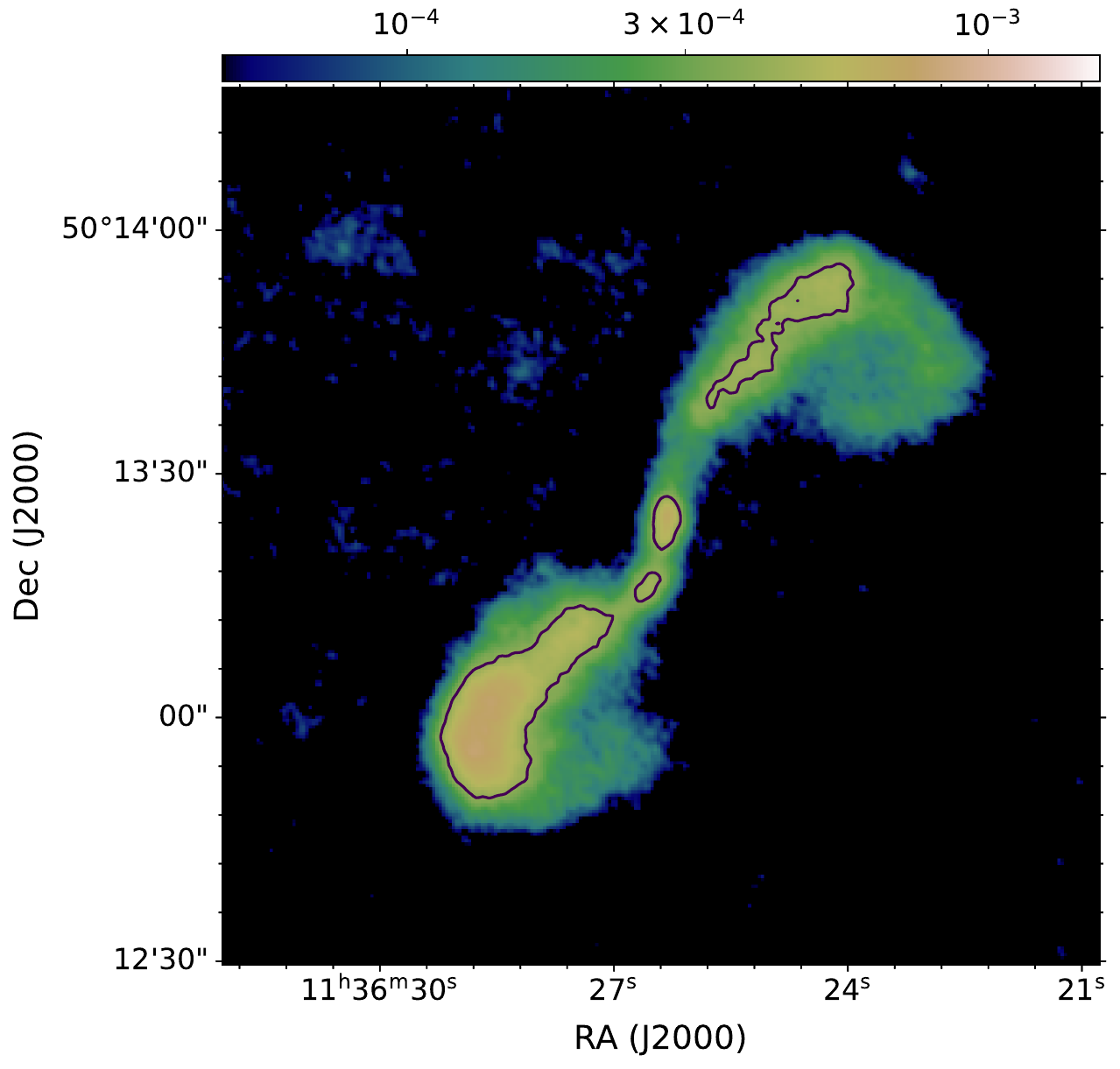}
        \subcaption{ILTJ113626.52+501320.3}    
    \end{minipage}   
    \begin{minipage}{0.24\textwidth}
    \centering
        \includegraphics[width=\linewidth]{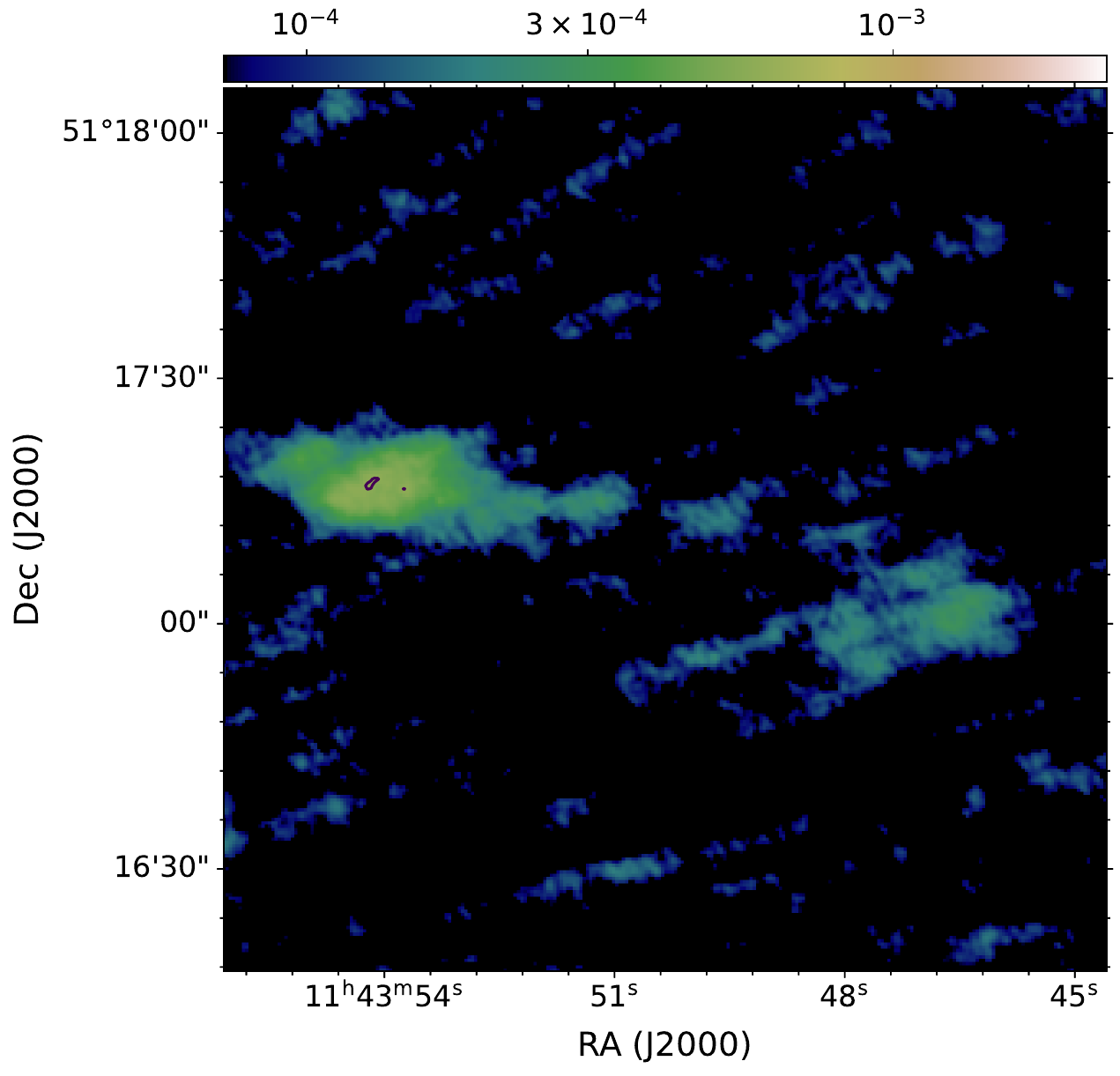}
        \subcaption{ILTJ114351.49+511712.6}
    \end{minipage}  
    
    \begin{minipage}{0.24\textwidth}
    \centering
        \includegraphics[width=\linewidth]{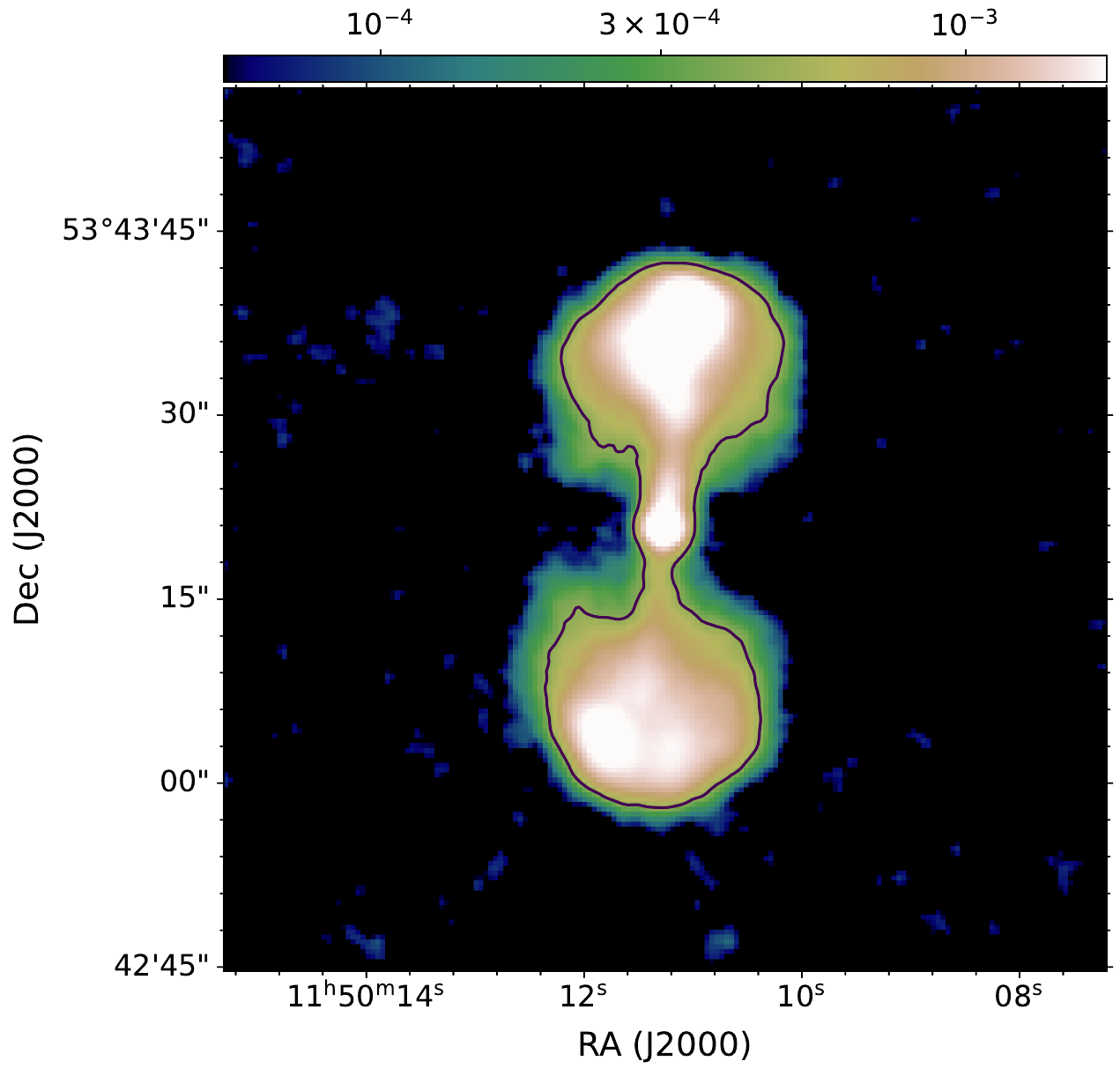}
        \subcaption{ILTJ115011.27+534320.9}    
    \end{minipage} 
    \begin{minipage}{0.24\textwidth}
    \centering
        \includegraphics[width=\linewidth]{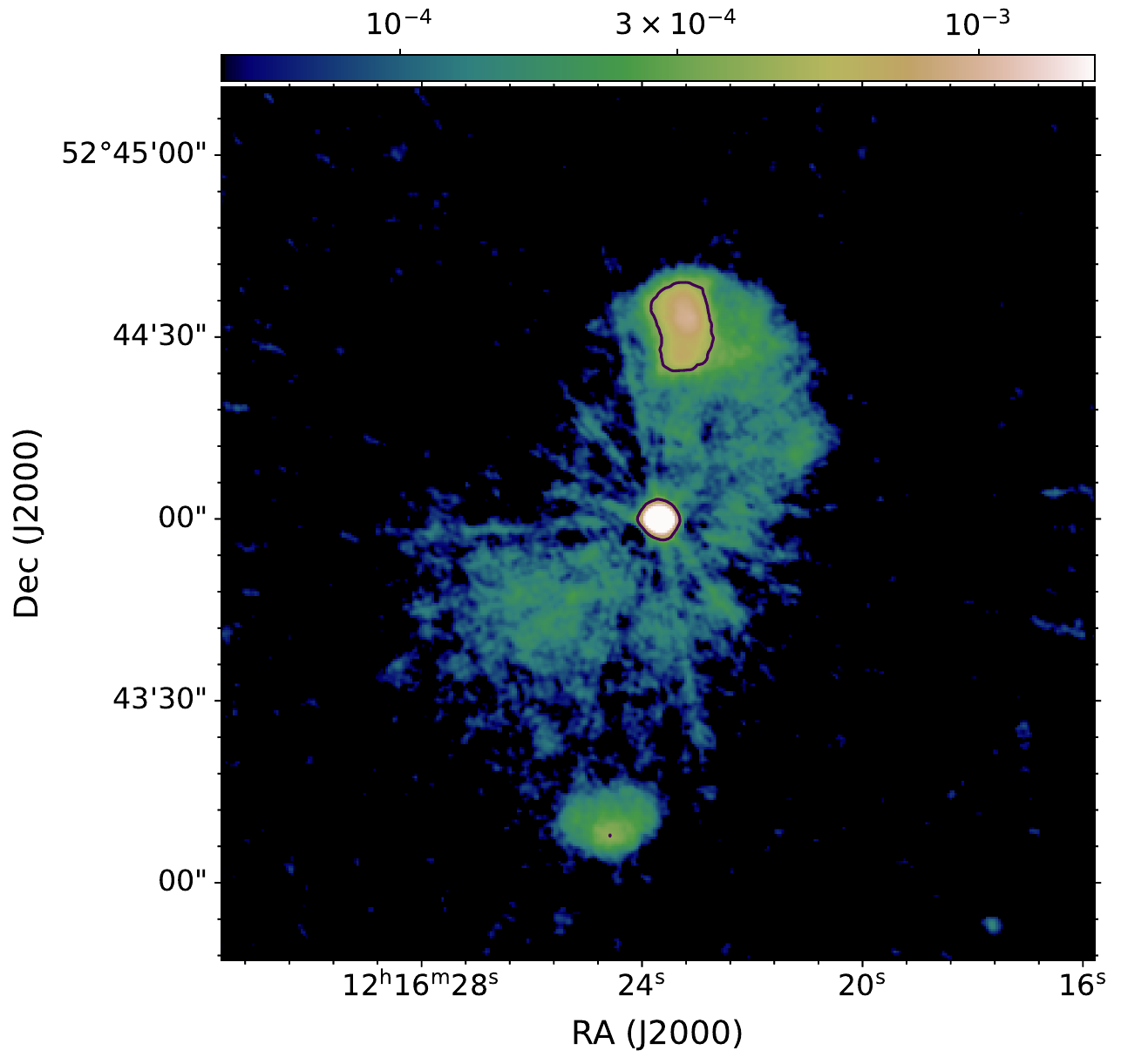}
        \subcaption{ILTJ121623.58+524409.4}
    \end{minipage}  
    \begin{minipage}{0.24\textwidth}
    \centering
        \includegraphics[width=\linewidth]{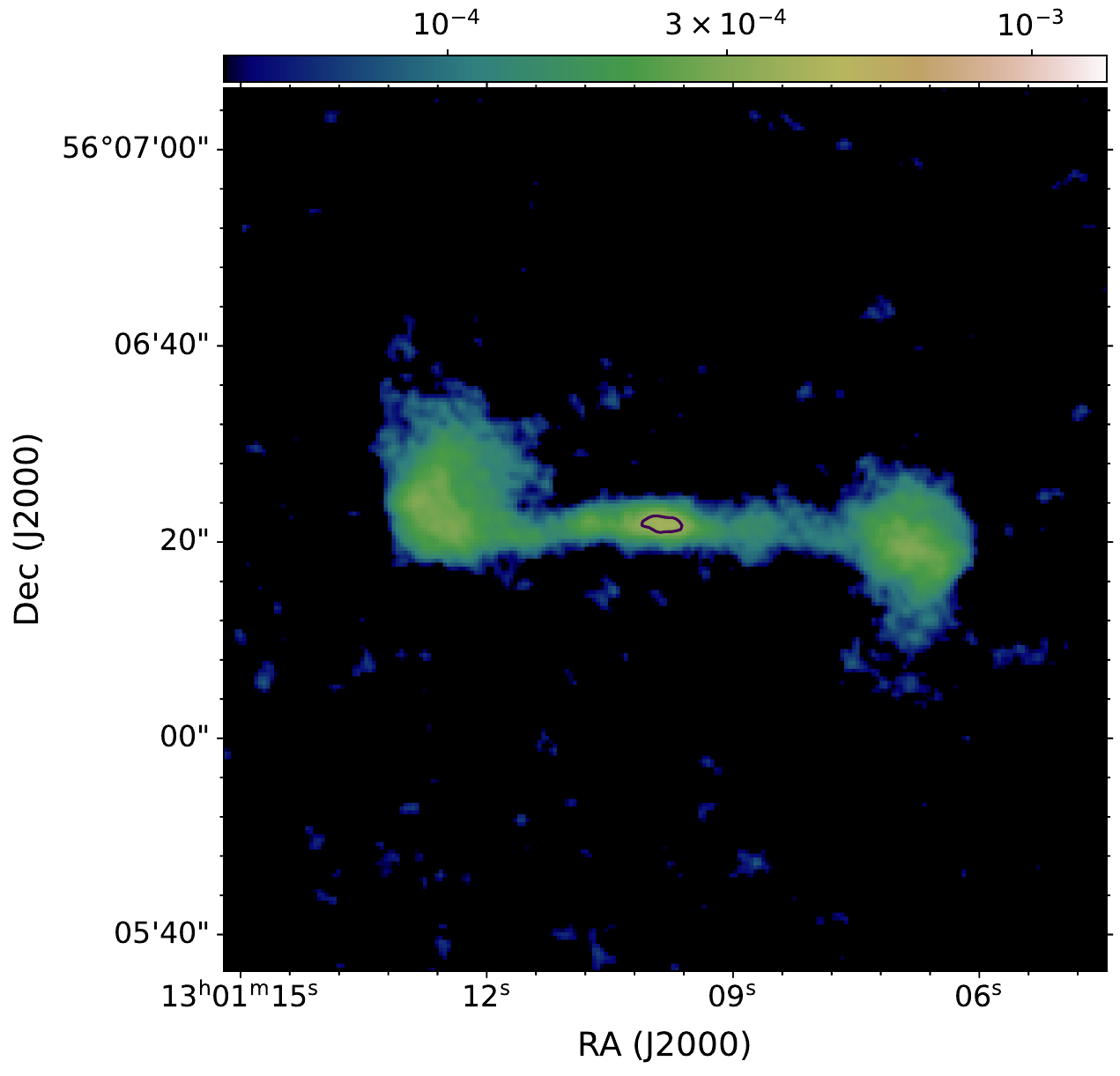}
        \subcaption{ILTJ130109.83+560623.4}
    \end{minipage}
    \begin{minipage}{0.24\textwidth}
    \centering
        \includegraphics[width=\linewidth]{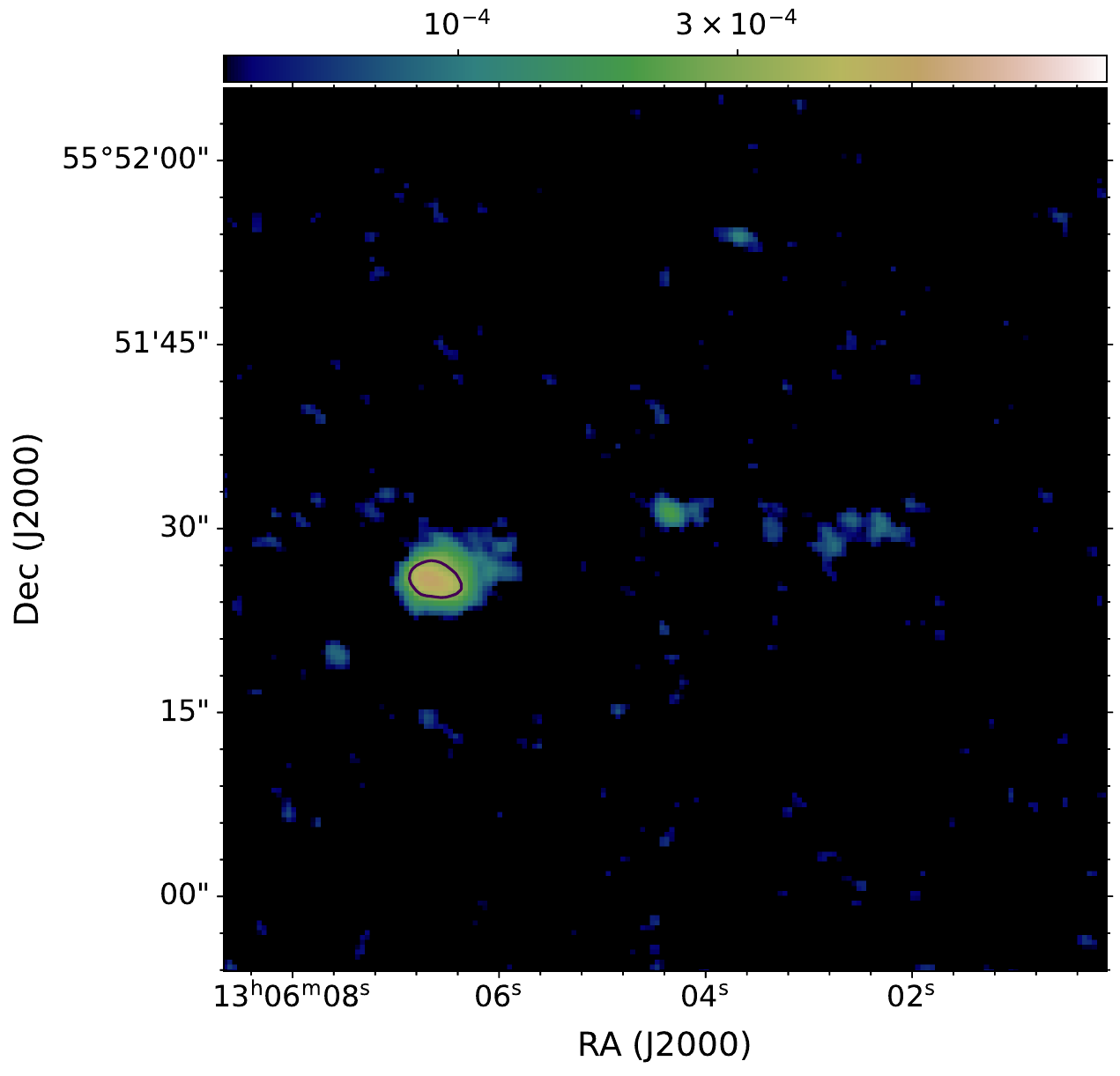}
        \subcaption{ILTJ130605.63+555127.6}
    \end{minipage} 
    
    \begin{minipage}{0.24\textwidth}
    \centering
        \includegraphics[width=\linewidth]{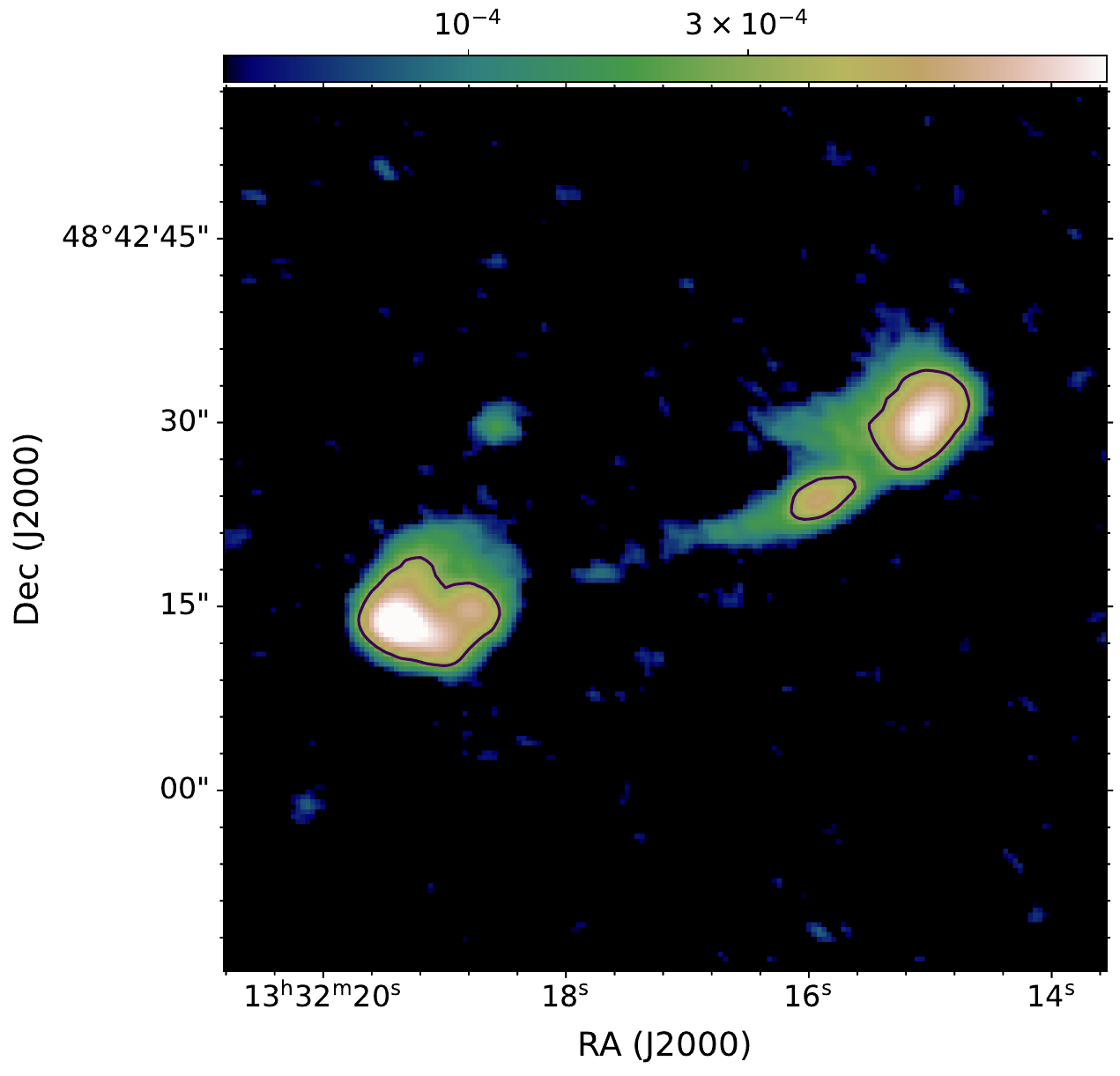}
        \subcaption{ILTJ133217.44+484221.7}
    \end{minipage}   
    \begin{minipage}{0.24\textwidth}
    \centering
        \includegraphics[width=\linewidth]{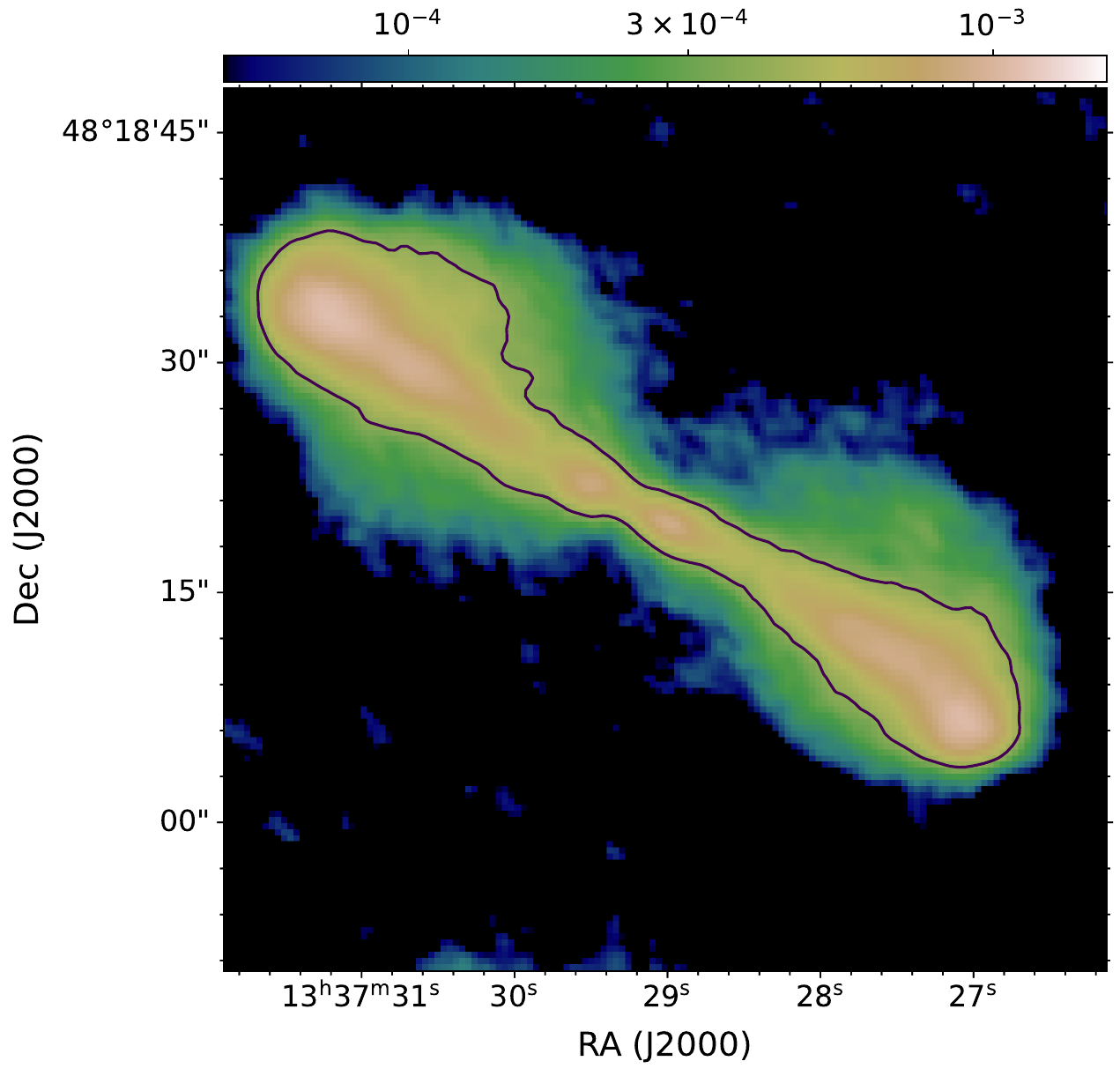}
        \subcaption{ILTJ133729.25+481822.3}
    \end{minipage}    
    \begin{minipage}{0.24\textwidth} 
    \centering
        \includegraphics[width=\linewidth]{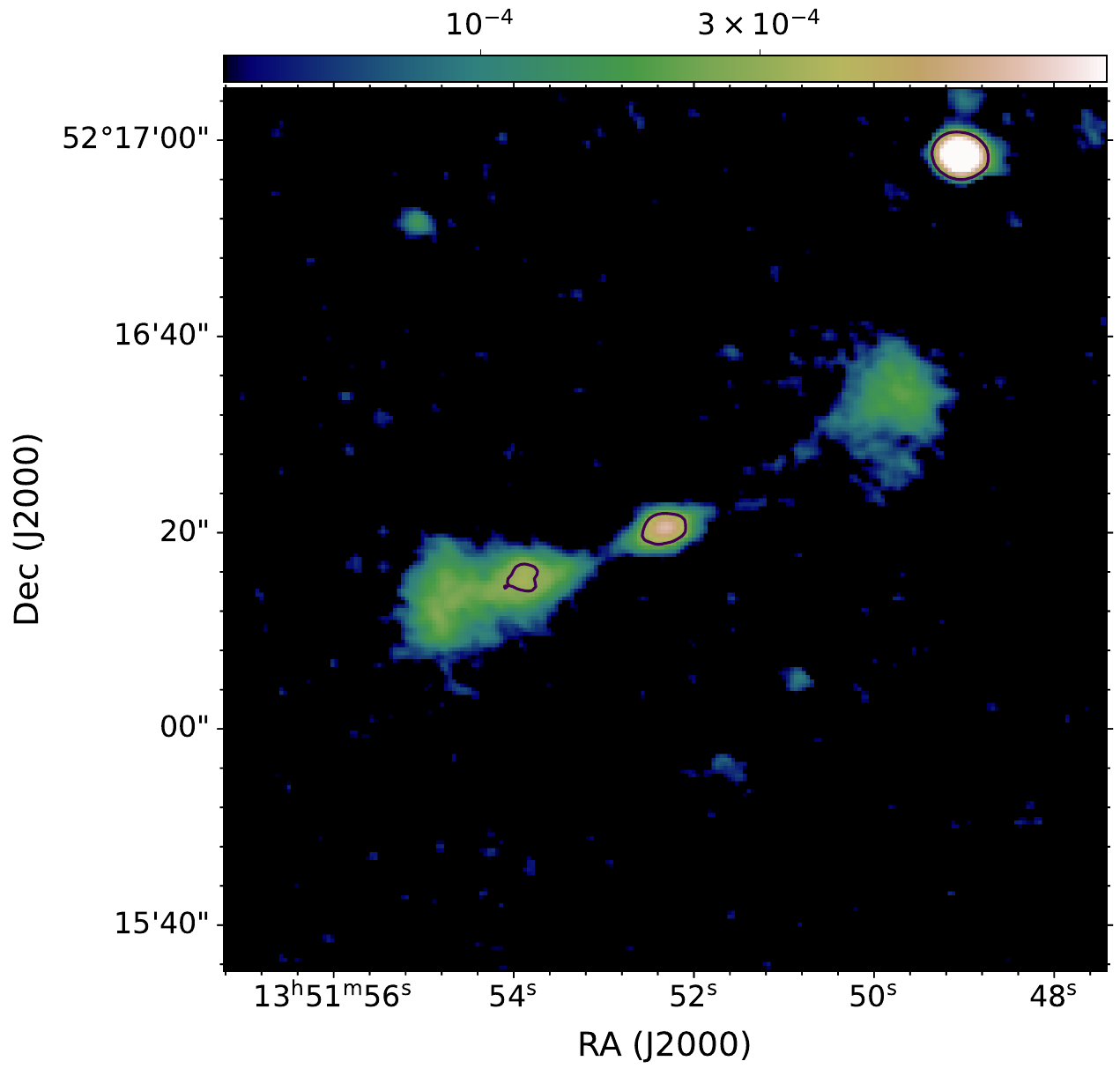}
        \subcaption{ILTJ135152.95+521618.8}
    \end{minipage}
    \begin{minipage}{0.24\textwidth}
    \centering
        \includegraphics[width=\linewidth]{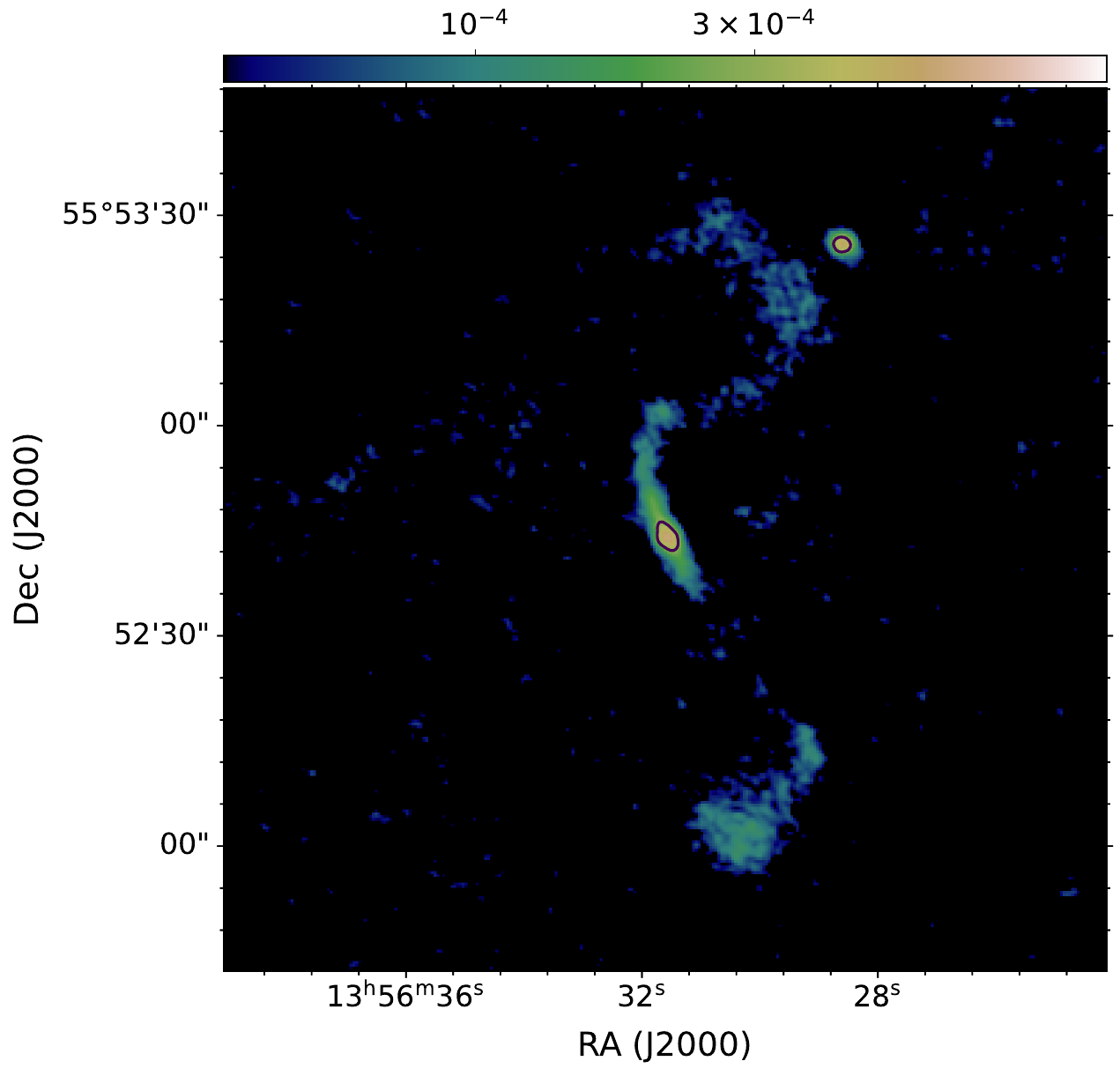} 
        \subcaption{ILTJ135630.51+555245.1}
    \end{minipage}
    
    \begin{minipage}{0.24\textwidth}
    \centering
        \includegraphics[width=\linewidth]{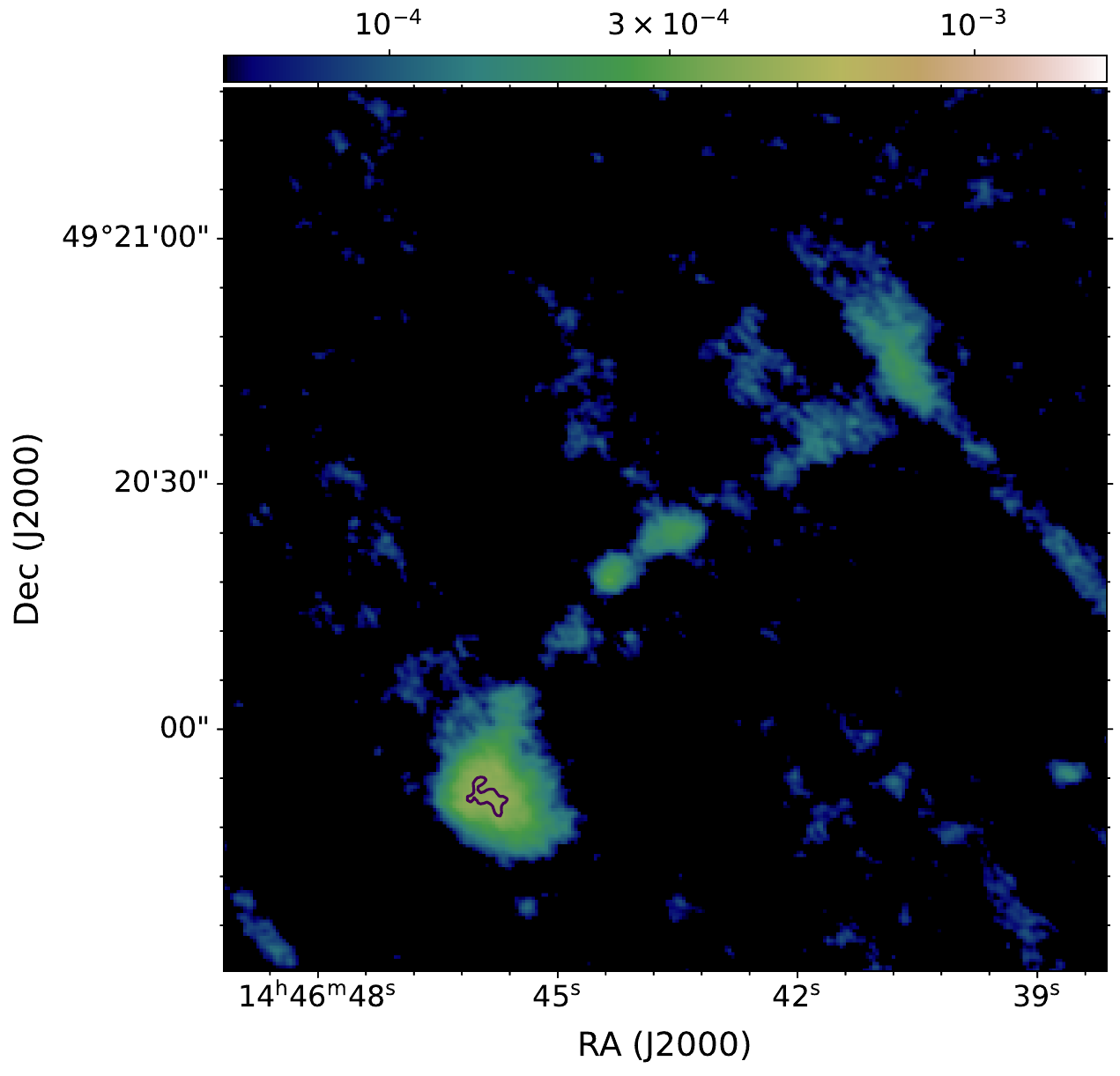} 
        \subcaption{ILTJ144644.12+492012.3}
    \end{minipage}
    \begin{minipage}{0.24\textwidth} 
    \centering
        \includegraphics[width=\linewidth]{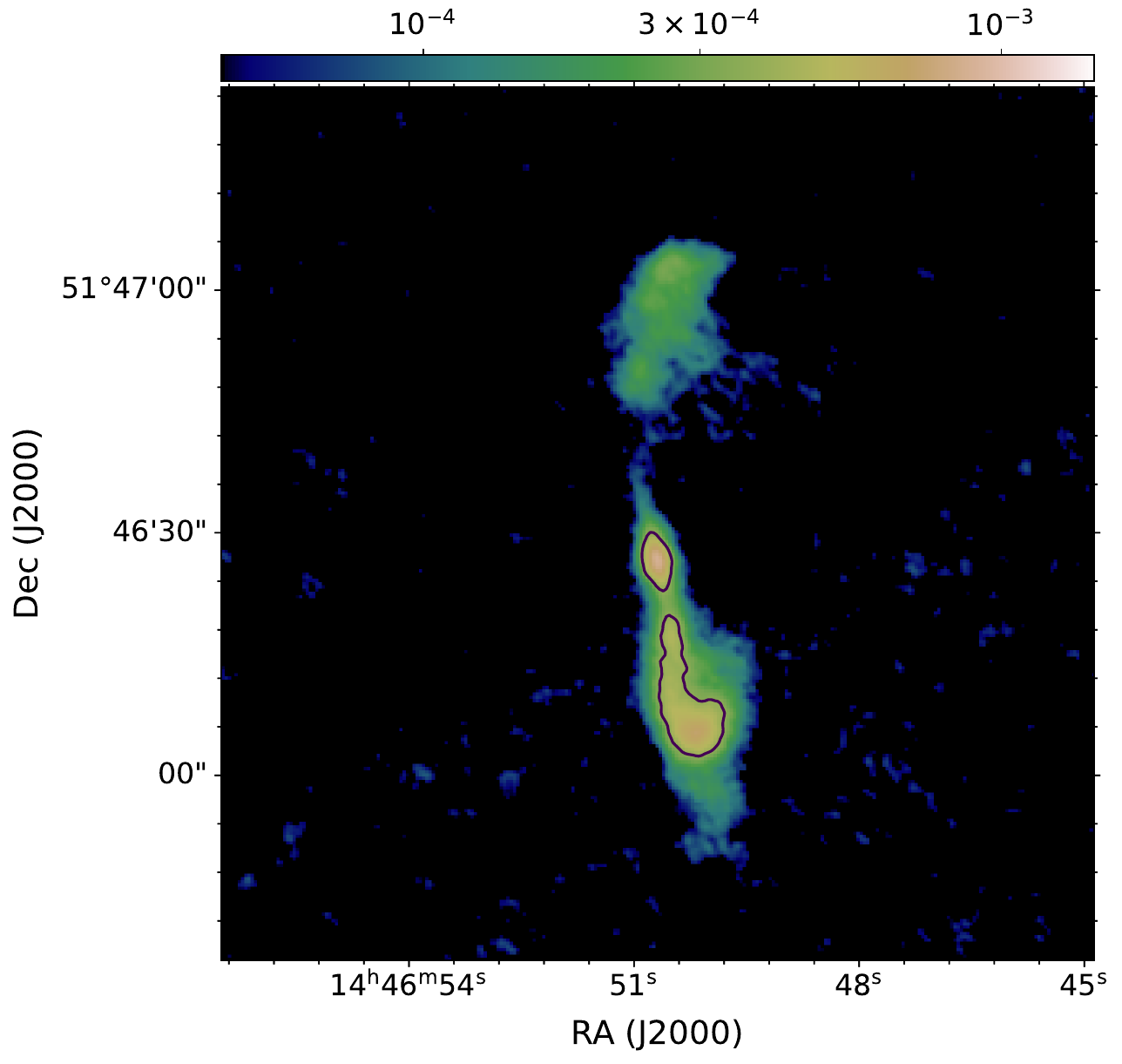} 
        \subcaption{ILTJ144650.49+514625.3}
    \end{minipage}
    \begin{minipage}{0.24\textwidth}
    \centering
        \includegraphics[width=\linewidth]{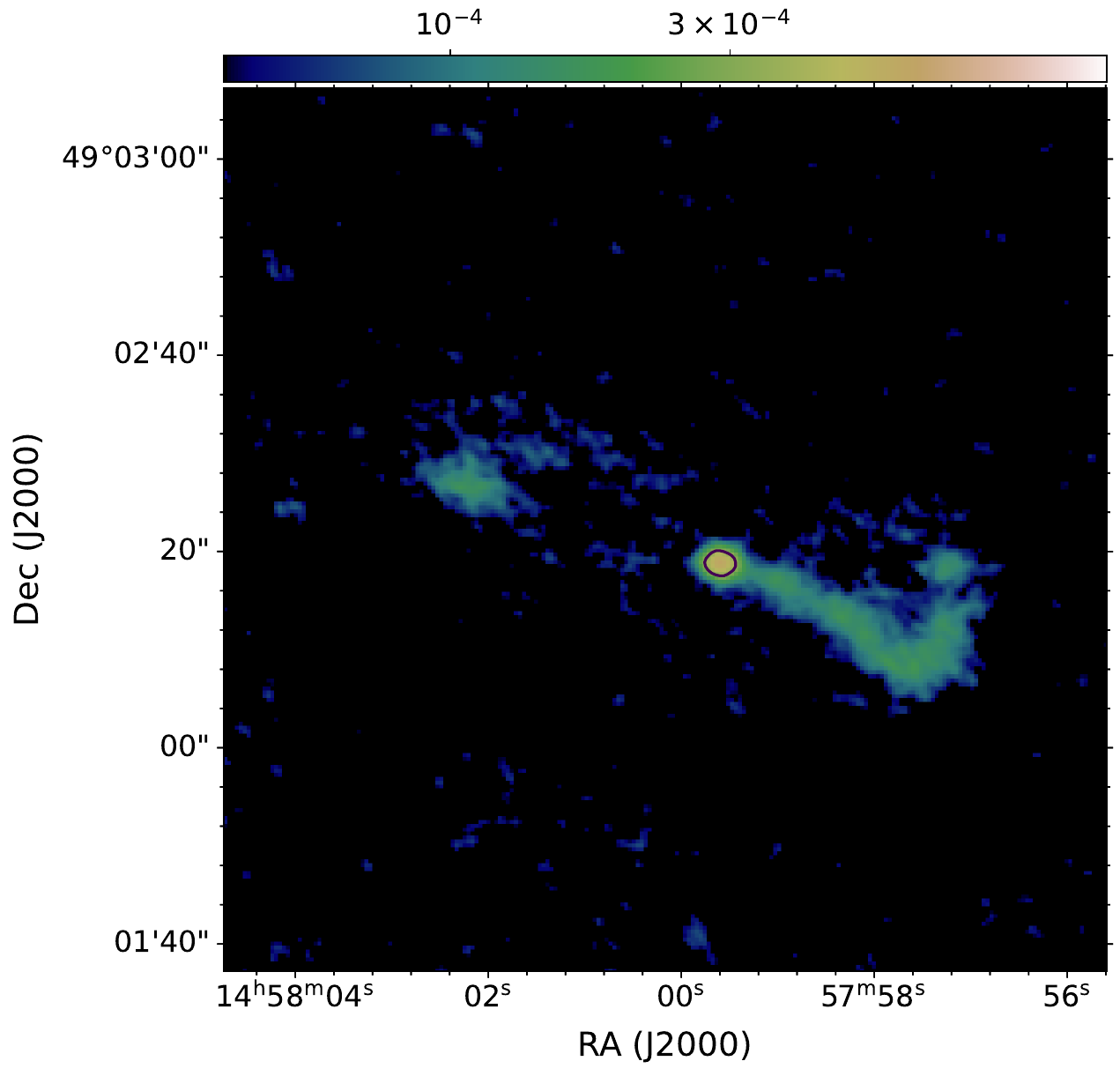} 
        \subcaption{ILTJ145759.29+490219.2}
    \end{minipage} 
    \begin{minipage}{0.24\textwidth}
    \centering
        \includegraphics[width=\linewidth]{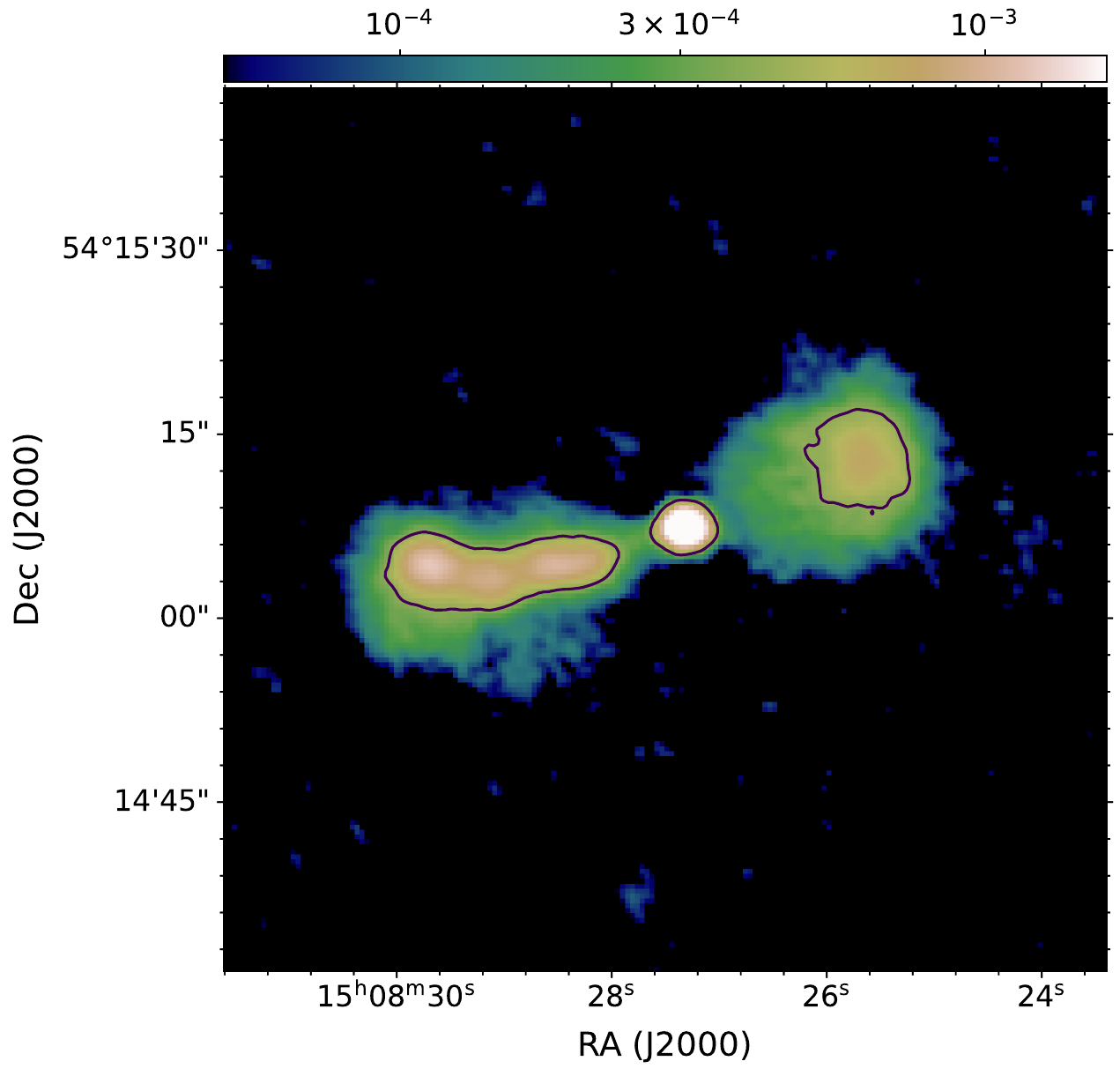} 
        \subcaption{ILTJ150827.77+541507.1}
    \end{minipage}
    \caption{The VLA images of the FRII-low sample with \revb{2}0-sigma contours. These are the \revb{sources which have detections over twenty} times the rms level. These contours have been used to determine whether the sources contain hotspots, cores, or jets as described in \autoref{subsec:Morph}. \revb{The colour bars show the flux density in \revbb{mJy/beam.}}}
    \label{fig:VLAContours} 
\end{figure*}

We present \revbb{below, in \autoref{fig:VLAContours}} the FRII-low VLA images, \revbb{described in \autoref{subsec:FRII-lows}} with the \revb{20$\sigma$} contours on them. These contours have been used to define the compact features within each sample, \revbb{as described in \autoref{sec:SDyn}}.

\begin{figure*}
    \centering
        \begin{minipage}{0.47\linewidth}
        \centering
            \includegraphics[width=\linewidth]{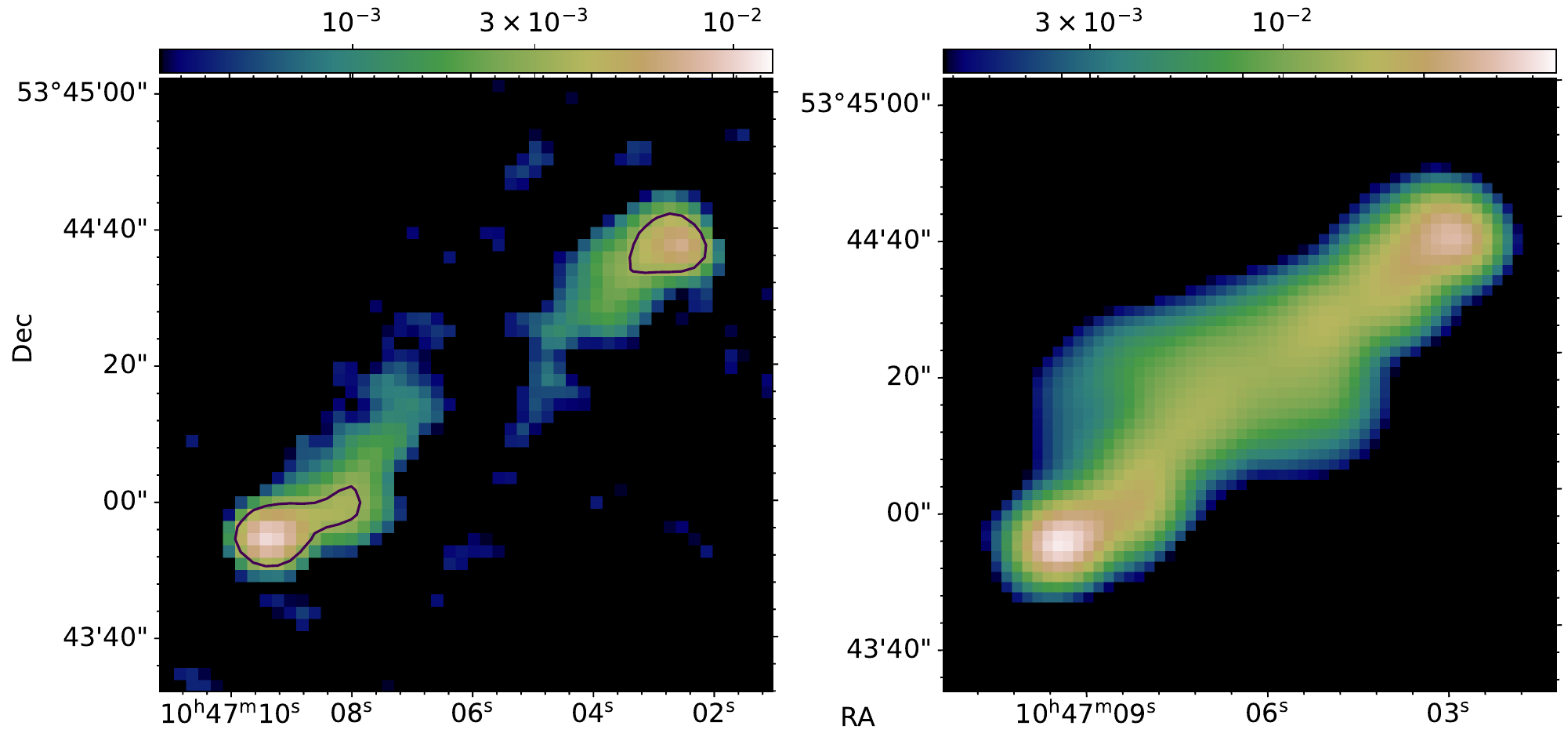}
        \subcaption{J104706+534419}
        \label{fig:enter-label}
        \end{minipage}
        \hfill
        \begin{minipage}{0.47\linewidth}
        \centering
            \includegraphics[width=\linewidth]{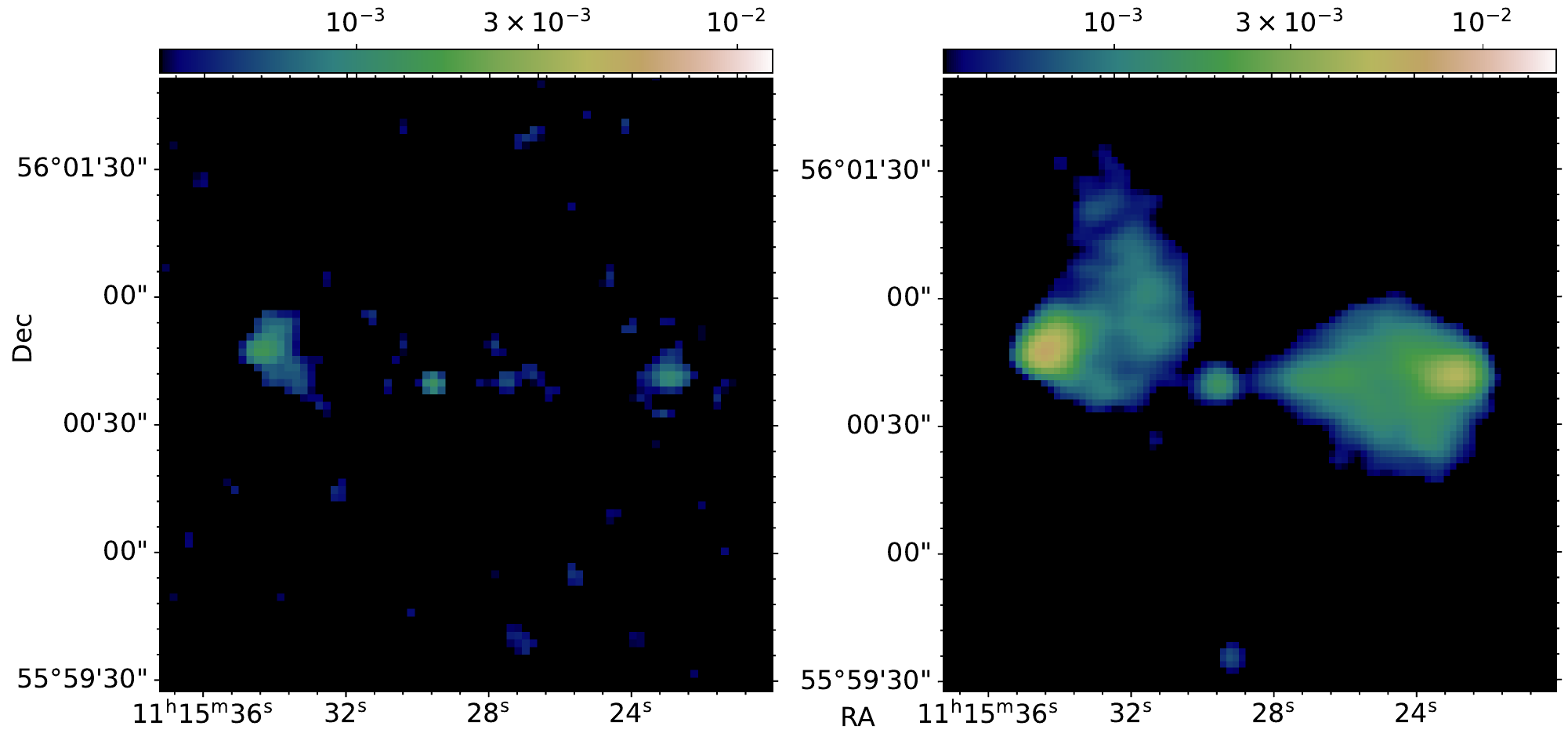}
        \subcaption{J111529+560039}
        \label{fig:enter-label} 
        \end{minipage}

        \begin{minipage}{0.47\linewidth}
        \centering
            \includegraphics[width=\linewidth]{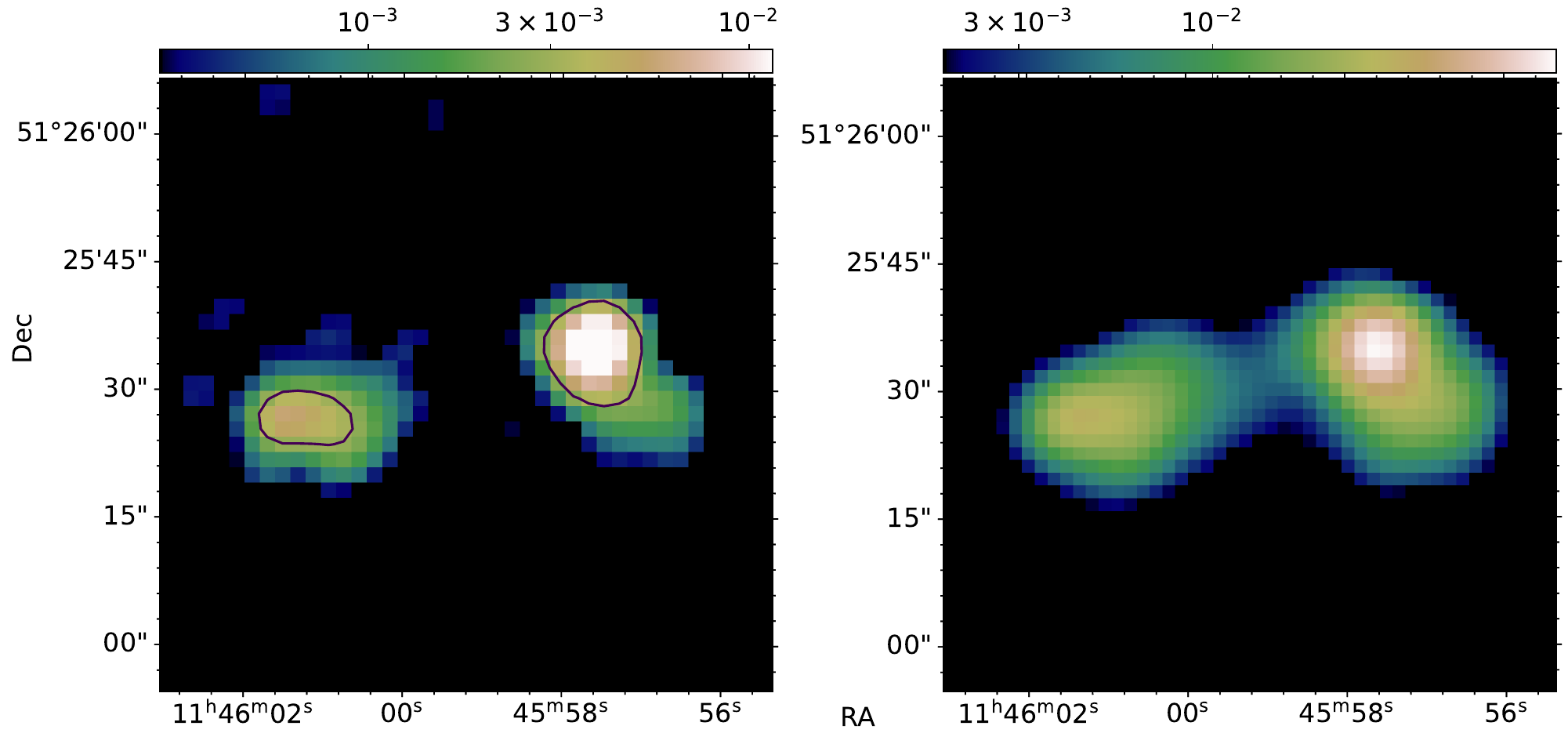}
        \subcaption{J114559+512528}
        \label{fig:enter-label} 
        \end{minipage}
        \hfill
        \begin{minipage}{0.47\linewidth}
        \centering
            \includegraphics[width=\linewidth]{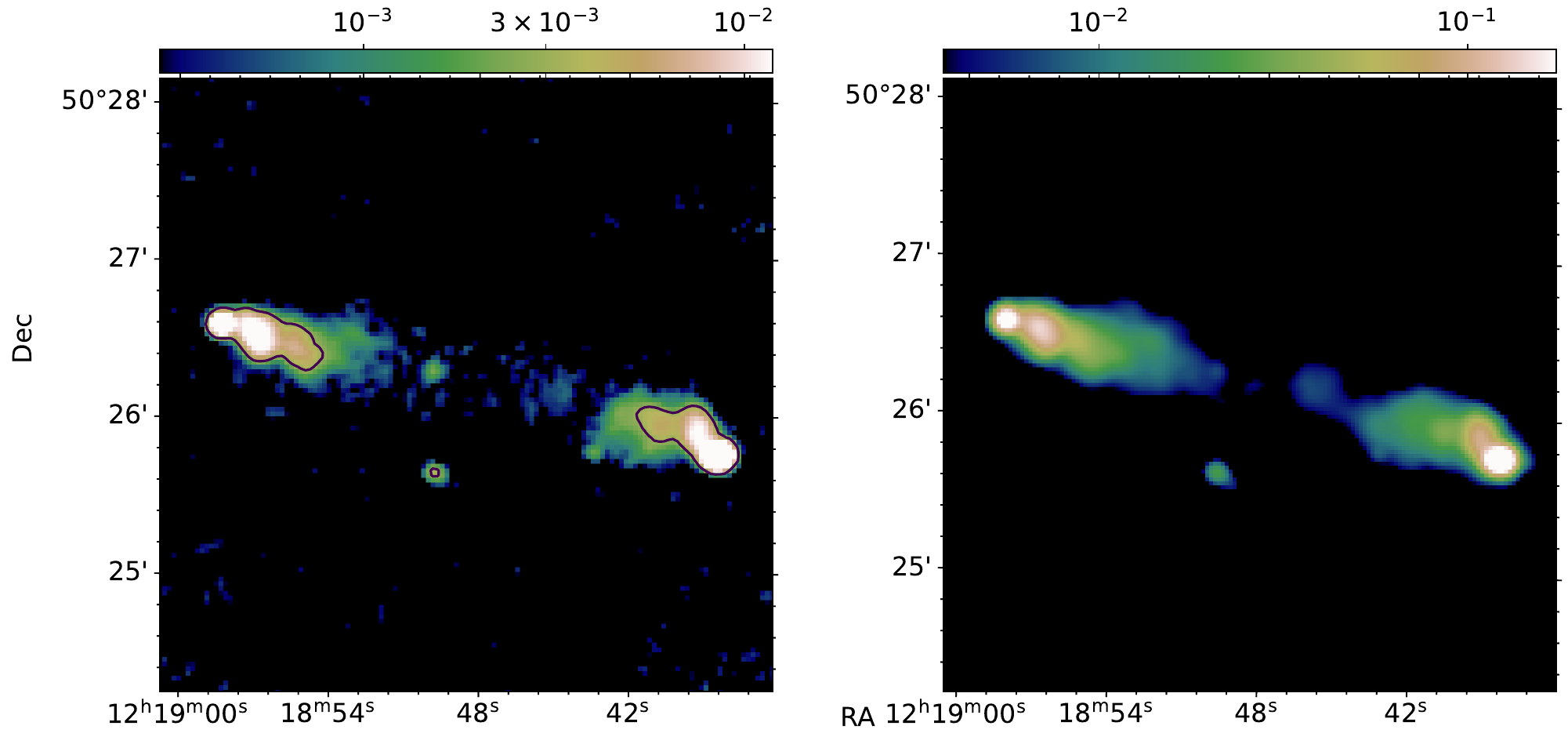}
        \subcaption{J121849+502617}
        \label{fig:enter-label} 
        \end{minipage}

        \begin{minipage}{0.47\linewidth}
        \centering
            \includegraphics[width=\linewidth]{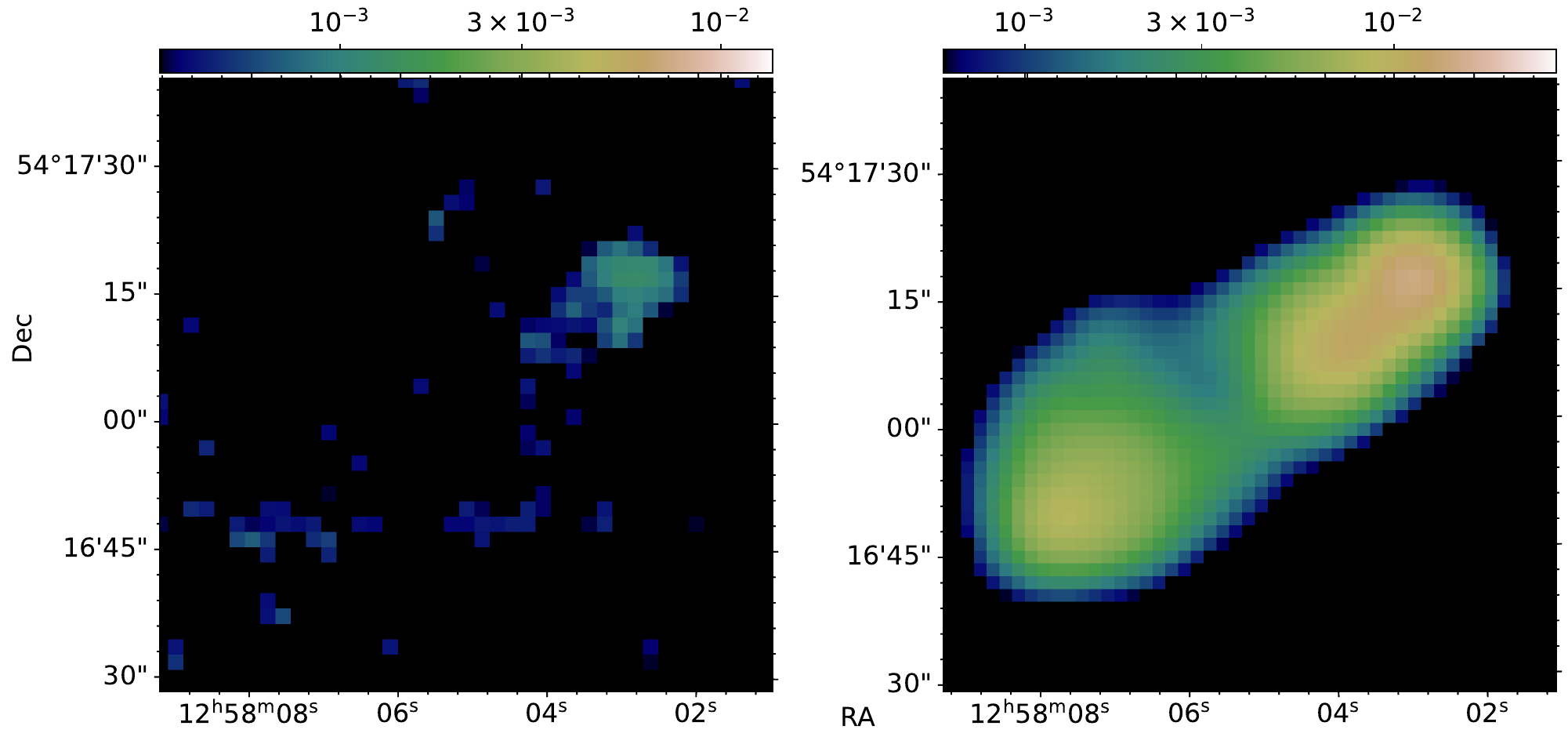}
        \subcaption{J125804+541702}
        \label{fig:enter-label} 
        \end{minipage}
        \hfill
        \begin{minipage}{0.47\linewidth}
        \centering
            \includegraphics[width=\linewidth]{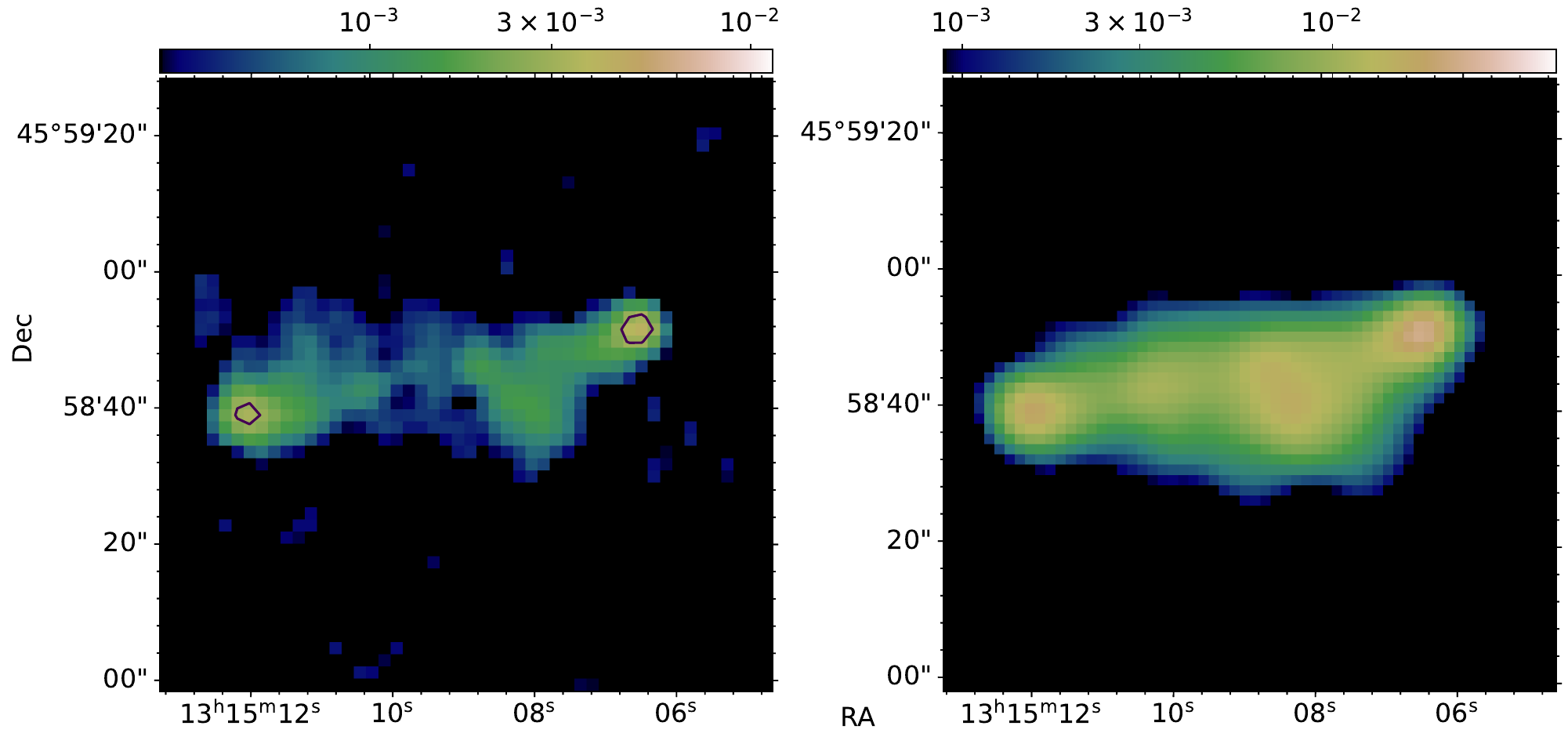}
        \subcaption{J131508+455846}
        \label{fig:enter-label} 
        \end{minipage}

        \begin{minipage}{0.47\linewidth}
        \centering
            \includegraphics[width=\linewidth]{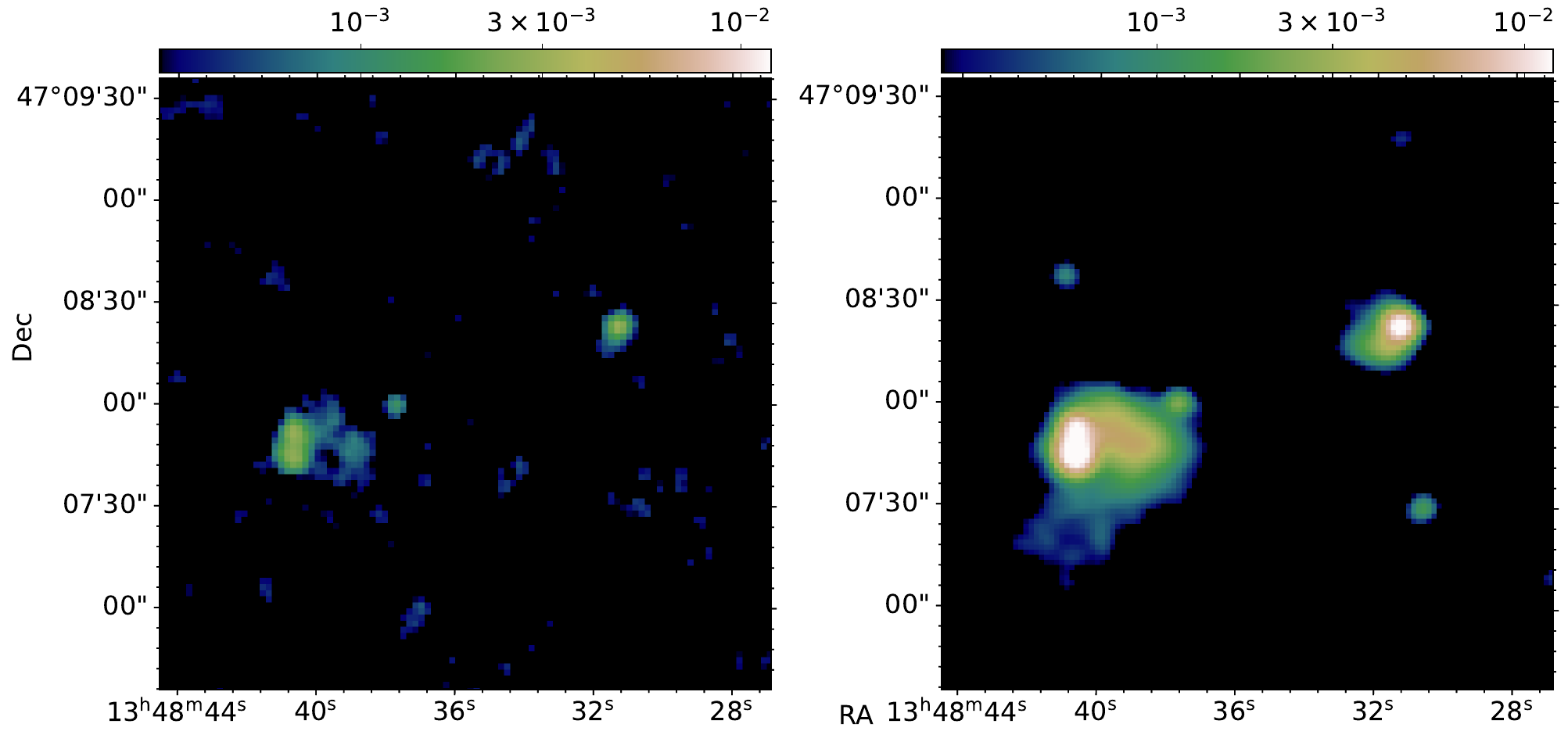}
        \subcaption{J134837+470800}
        \label{fig:enter-label} 
        \end{minipage}
        \hfill
        \begin{minipage}{0.47\linewidth}
        \centering
            \includegraphics[width=\linewidth]{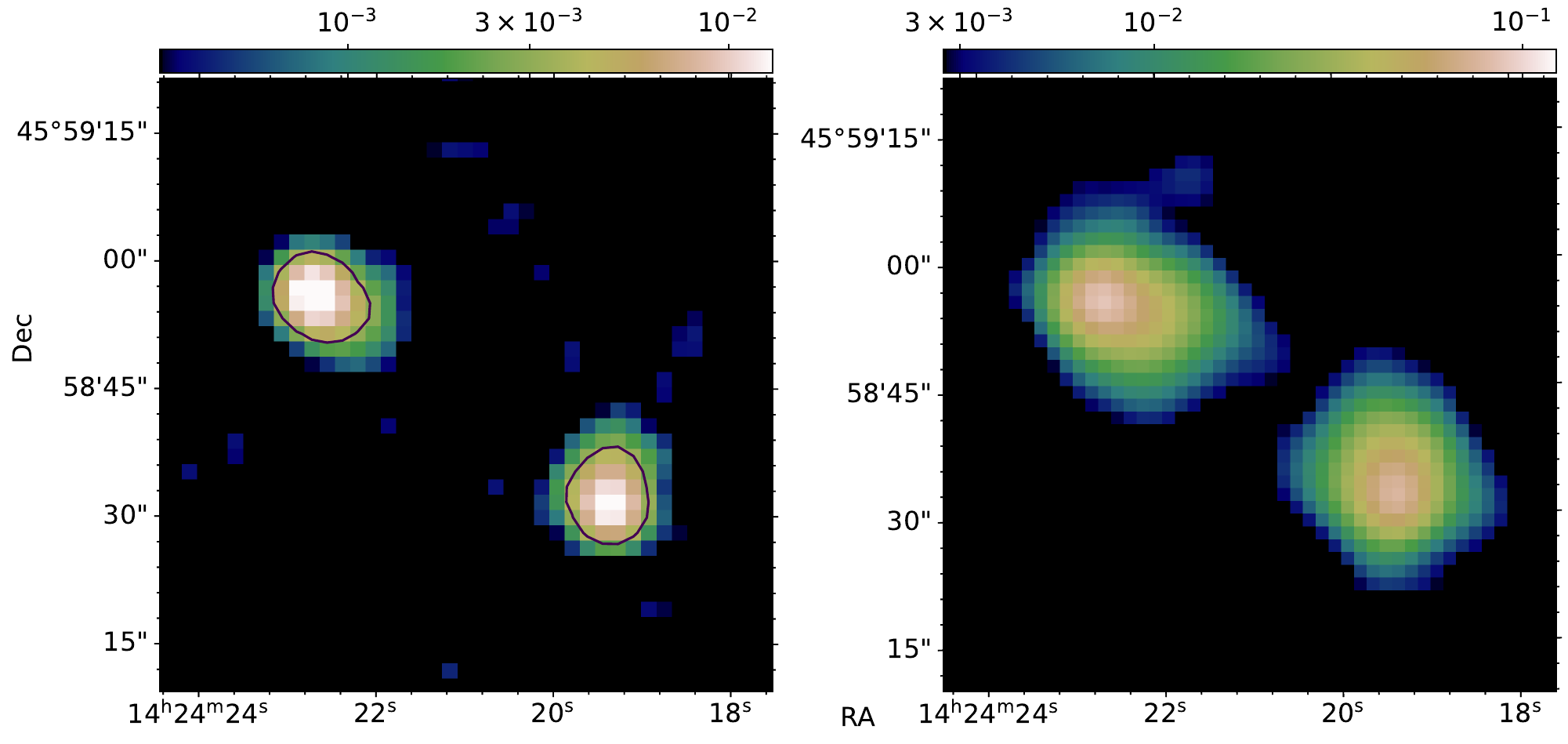}
        \subcaption{J142420+455837}
        \label{fig:enter-label} 
        \end{minipage}

        \begin{minipage}{0.47\linewidth}
        \centering
            \includegraphics[width=\linewidth]{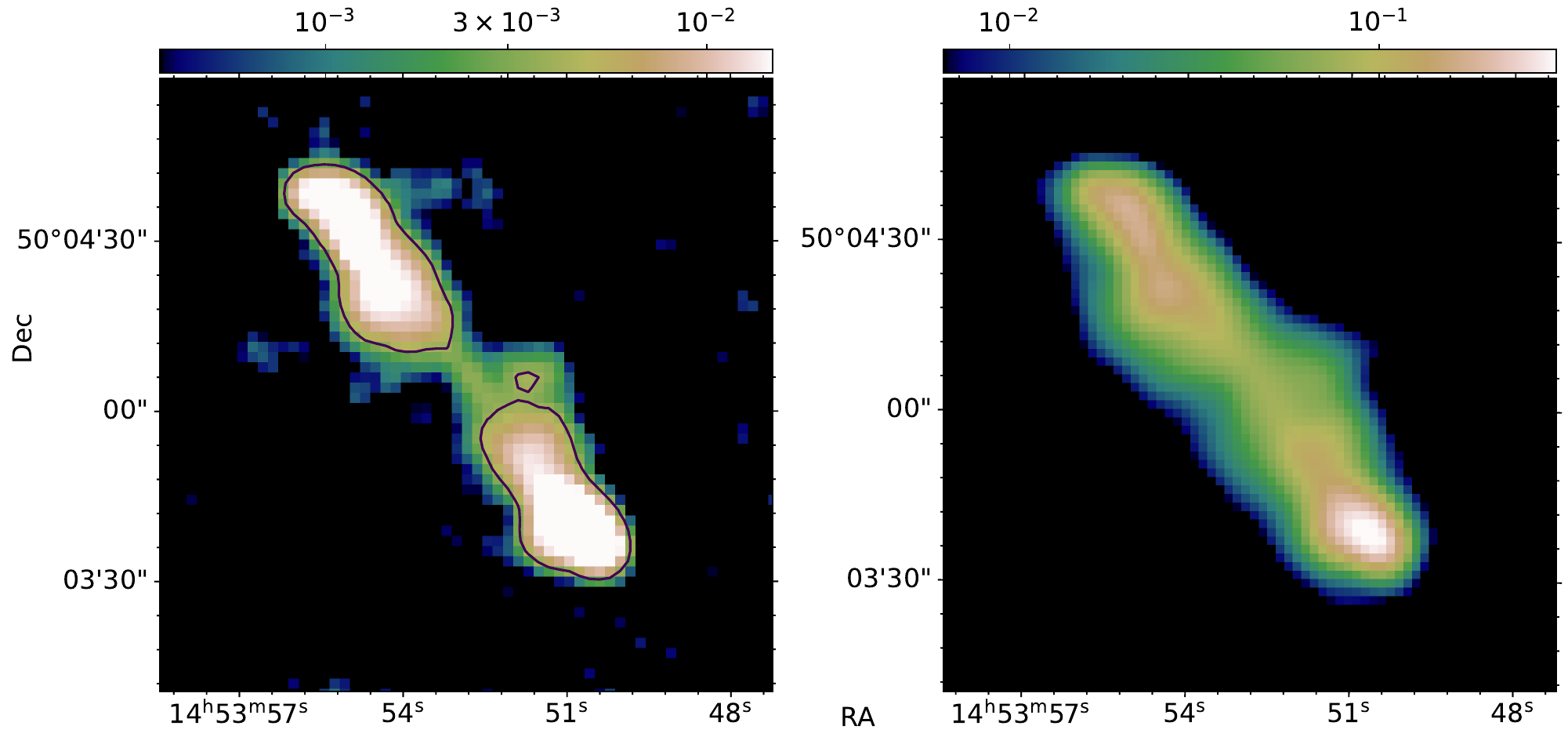}
        \subcaption{J145352+500406}
        \label{fig:enter-label} 
        \end{minipage}
        \hfill
        \begin{minipage}{0.47\linewidth}
        \centering
            \includegraphics[width=\linewidth]{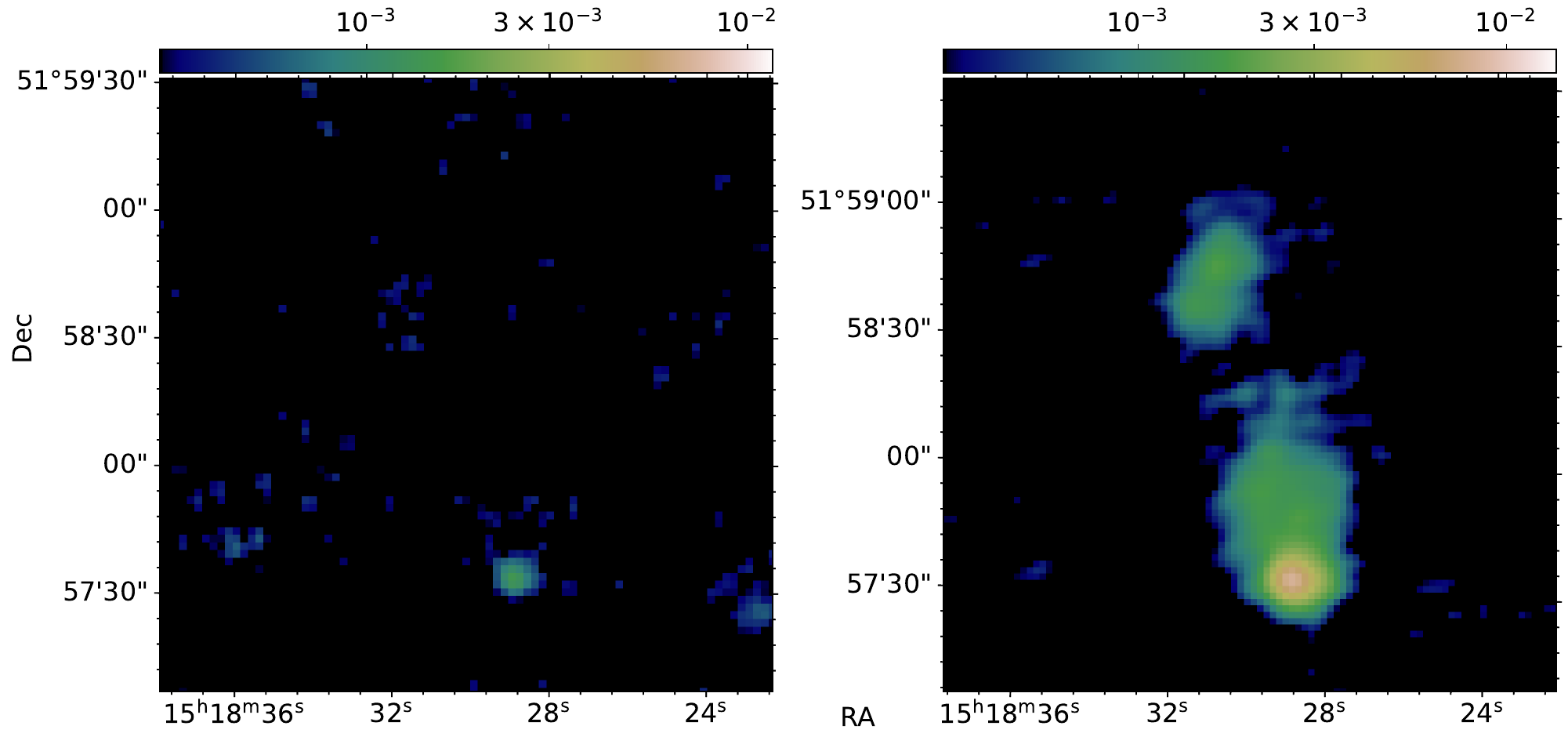}
        \subcaption{J151830+515817}
        \label{fig:enter-label} 
        \end{minipage}
    \caption{The FRII-high images from LoTSS and FIRST and described in \autoref{subsec:FRII-highs}. The images are matching pairs; those on the right are the FIRST 1.4-GHz, 5-arcsec images, and the ones on the left are the LoTSS 150-MHz, 6-arcsec images. The \revb{2}0$\sigma$ contours have been added to the FIRST images where possible. \revb{The colour bars show the flux density in \revbb{mJy/beam.}}}
\label{fig:HighImages}
\end{figure*}

Also included \revbb{in \autoref{fig:HighImages}} are the images of the sample of FRII-highs; both the LoTSS images and the equivalent FIRST images, \revbb{described in \autoref{subsec:FRII-highs}}. Overlaid on the FIRST images are the \revb{2}0$\sigma$ contours, where it is possible to plot them \revbb{ \citep{becker_first_1995}}.

\bsp	
\label{lastpage}
\end{document}